\documentclass[journal=jacsat,manuscript=article,layout=twocolumn]{achemso}

\usepackage{geometry}

\setkeys{acs}{articletitle=true}
\usepackage{lettrine}

\usepackage[version=3]{mhchem} 
\usepackage[T1]{fontenc}       

\usepackage[fontsize=10pt]{scrextend}

\usepackage{newtxtext}
\usepackage{bm} 
\usepackage[labelfont=bf]{caption} 

\usepackage[hyperfootnotes=false]{hyperref}	 
\hypersetup{
  colorlinks,
  citecolor=blue,
  linkcolor=blue,
  urlcolor=blue
}

\usepackage{titlesec} 
\titleformat{\subsection}[runin]
{\normalfont\bfseries}{\thesubsection{.}}{1em}{}[.]

\titleformat{\section}
{\normalfont\sffamily\large\bfseries}{\thesection{.}}{1em}{\MakeUppercase}

\usepackage[table]{xcolor}
\usepackage{graphicx}
\usepackage{amsmath,fullpage,amssymb}
\usepackage{graphics,epsfig}
\usepackage{makeidx}
\usepackage{gensymb}
\usepackage{siunitx}

\usepackage{soul}	

\def\rmd{{\mathrm{d}}}

\def\rme{{\mathrm{e}}}

\def\Eq{eq}
\def\Eqs{eqs}

\def\ie{{\em i.e.}}
\def\eg{{\em e.g.}}

\newcommand{\Fig}{Fig.~\ref}

\newcommand{\kB}{k_\textrm{B}}

\newcommand{\cmt}[1]{\textcolor{red} {[#1]}} 

\newcommand{\trm}[1]{{\textrm{#1}}}

\SectionNumbersOn
\let\oldmaketitle\maketitle
\let\maketitle\relax

\title{Tuning contact angles of aqueous droplets on hydrophilic and hydrophobic surfaces by surfactants
}

\author{Fabio Staniscia}
\affiliation{\rm\small Department of Theoretical Physics, Jo{\v z}ef Stefan Institute, SI-1000 Ljubljana, Slovenia}
\author{Horacio V. Guzman}
\affiliation{\rm\small Department of Theoretical Physics, Jo{\v z}ef Stefan Institute, SI-1000 Ljubljana, Slovenia}
\author{Matej Kandu\v{c}}
\affiliation{\rm\small Department of Theoretical Physics, Jo{\v z}ef Stefan Institute, SI-1000 Ljubljana, Slovenia}
\email{matej.kanduc@ijs.si}

\begin{document}
\pagenumbering{arabic}
\noindent

\parindent=0cm
\setlength\arraycolsep{2pt}

\twocolumn[	
\begin{@twocolumnfalse}
\oldmaketitle

\begin{abstract}\small
Adsorption of small amphiphilic molecules occurs in various biological and technological processes, sometimes desired, the other times unwanted (\eg, contamination). Surface-active molecules preferentially bind to interfaces and affect their wetting properties. We study the adsorption of short-chained alcohols (simple surfactants) to the water--vapor interface and solid surfaces of various polarities using molecular dynamics simulations. 
The analysis enables us to establish a theoretical expression for the adsorption coefficient, which exponentially scales with the molecular surface area and the surface wetting coefficient, and which is in good agreement with the simulation results. The competition of the adsorptions to both interfaces of a sessile droplet alters its contact angle in a nontrivial way. The influence of surfactants is strongest on very hydrophilic and very hydrophobic surfaces, whereas droplets on surfaces of moderate hydrophilicity are much less affected.
\vspace{5ex}
\end{abstract}
\end{@twocolumnfalse}]

\section{Introduction}
Adsorption of dissolved molecules from an aqueous phase onto interfaces with air and solids is a ubiquitous phenomenon in natural and technological processes. 
For instance, the adsorption of organic material (\eg, microorganisms and pollen) plays a prominent role in several aspects of atmospheric and oceanic environments~\cite{bfLangmuir2004,li2017aqueous,salta2013marine,ButtACSApM2020bf}.
Adsorption is essential in many applications, ranging from detergency, printing, surface catalysis, dialysis, filtration~\cite{SinghJMaterChem2012} to petrochemical processes~\cite{BERA13} and removal of water pollutants~\cite{BUSCA2008}.
Yet, adsorption is a process often challenging to predict and control.
Uncontrolled adsorption contributes to surface contamination, biofouling (\ie, unwanted bacterial adhesion), loss of product to vessel surfaces, clogging of small constrictions in coronary stents~\cite{thierry2002nitinol, rosenhahn2010role} or microfluidic devices~\cite{tegenfeldt2004micro}, and deterioration of biosensors~\cite{niedzwiecki2010single}. 

It is known that many small molecules and proteins tend to adsorb better onto hydrophobic surfaces, while hydrophilic surfaces are generally more resistant to adsorption~\cite{prime1991self,sigal1998effect, vogler1998structure}, making them suitable self-cleaning materials against biofouling~\cite{rosenhahn2010role}.
Using water contact angle as a proxy for the hydrophobicity of the surface became attractive, even to explain such complex phenomena as cellular responses to synthetic surfaces in culture media or simulated medical device service environments~\cite{polacosMaterials2020}. However, in many complex biological scenarios, other factors become important as well \cite{alexander2017water}.
Unfortunately, the surfactant adsorption processes are challenging to study experimentally, in particular, because the adsorbing layers are typically below a few nanometers in thickness, often comprising a single molecular monolayer~\cite{manne1995molecular, Paria04,WatanabeLangmuir2002, sendner2009interfacial}.


An important effect of adsorbed molecules is that they reduce the surface tension of the interface to which they adsorb~\cite{CHANG19951,VelardeJCIS2000}, which is why surfactants are often used to enhance the wetting ability of aqueous solutions~\cite{von1999spreading} and to suppress  hydrophobic cavitation~\cite{diamant1996kinetics, israelachvili2006}.
Surface-active molecules can dramatically alter the substrate wettability, and therewith leading to phenomena such as superspreading~\cite{rafai2002superspreading} or autophobing (spontaneous retraction of a drop after initial spreading)~\cite{bera2016surfactant,afsar2004dewetting}. 
Determining the relationship between the surface tension and the structures of surfactant additives at different temperatures, pressures, salinities, and pH regimes is critical for the design in many industry sectors, ranging from consumer chemicals to oil and gas extraction~\cite{sresht2017combined,TutejaPTRSA2019}.
In recent years, we witnessed an enormous interest in surfactant-containing droplets, where the surfactant's adsorption to the solid--water and air--water interface can render wetting in a nontrivial way~\cite{eckmann2001wetting, bera2018counteracting, thiele2018equilibrium, tadmor2019drops, kwiecinski2019evaporation, tadmor2021drops, sun2021ion,ButtSM2016surfButt, bera2021antisurfactant}.


Among the vast number of additives, alcohols hold a special place, being by far the most frequently used~\cite{Zana95}. Short-chained alcohols are the simplest molecules that contain both hydrophobic and hydrophilic groups and are therefore excellent model systems in studies of interfaces~\cite{Mounir96,Stewart03,Wilson97,JanczukJCIS2006,hub2012organic}. 
Alcohols are the most common cosurfuctants added to surfactant and oil systems, for instance, in microemulsions. Alcohol adsorption is also relevant to distillation~\cite{Karlsson2017}, biofuels~\cite{LEVARIO2012},
biomass transformation~\cite{Corma07}, pharmacological processes (binding to membranes and proteins)~\cite{SEEMAN71,Bull78,Peoples96}, and in aerosol science \cite{Tarbuck06,Szori2011}.

In this work, we employ molecular dynamics (MD) simulations to study how short-chained alcohols (\ie, methanol, 1-propanol, and 1-pentanol) adsorb to two kinds of interfaces: water--vapor and solid--water. For the latter we used surfaces with various levels of hydrophilicity, characterized by different wetting coefficients
(the cosine of the water contact angle)~\cite{KapplBook2003,KapplJPCM2005}. 
The three linear alcohols are very well soluble in water~\cite{VanDerSpoel06}, which enables studying the effect of chain length directly.
Since they adsorb to both interfaces and lower their surface tension, we will refer to them also as surfactants~\cite{IUPACgoldbook} in this study.
We compute the adsorption of alcohols onto the interfaces and analyze their dependence on the chain length and the surface contact angle, $\theta$, expressed in terms of the wetting coefficient, $\cos\theta$. 
We invoke a continuum-level approach to rationalize the observed relationship between the adsorption and the wetting coefficient.
Furthermore, using the Gibbs adsorption-isotherm formalism, we relate the surfactant adsorption to the decrease in the surface tensions. This enables us to analyze the variation of droplet contact angles in the dependence of the surfactant concentration.





\section{Methods}

\subsection{Atomistic models}
We used the simple point charge/extended (SPC/E) model for water~\cite{spce} combined with the GROMOS force field~\cite{oostenbrink_biomolecular_2004} for simulating alcohols and the solid surface. The all-atom structures and topology files for alcohols were obtained from the ATB repository~\cite{malde2011automated}. 

\begin{figure}[h]
  \centering\includegraphics[width=0.4\textwidth]{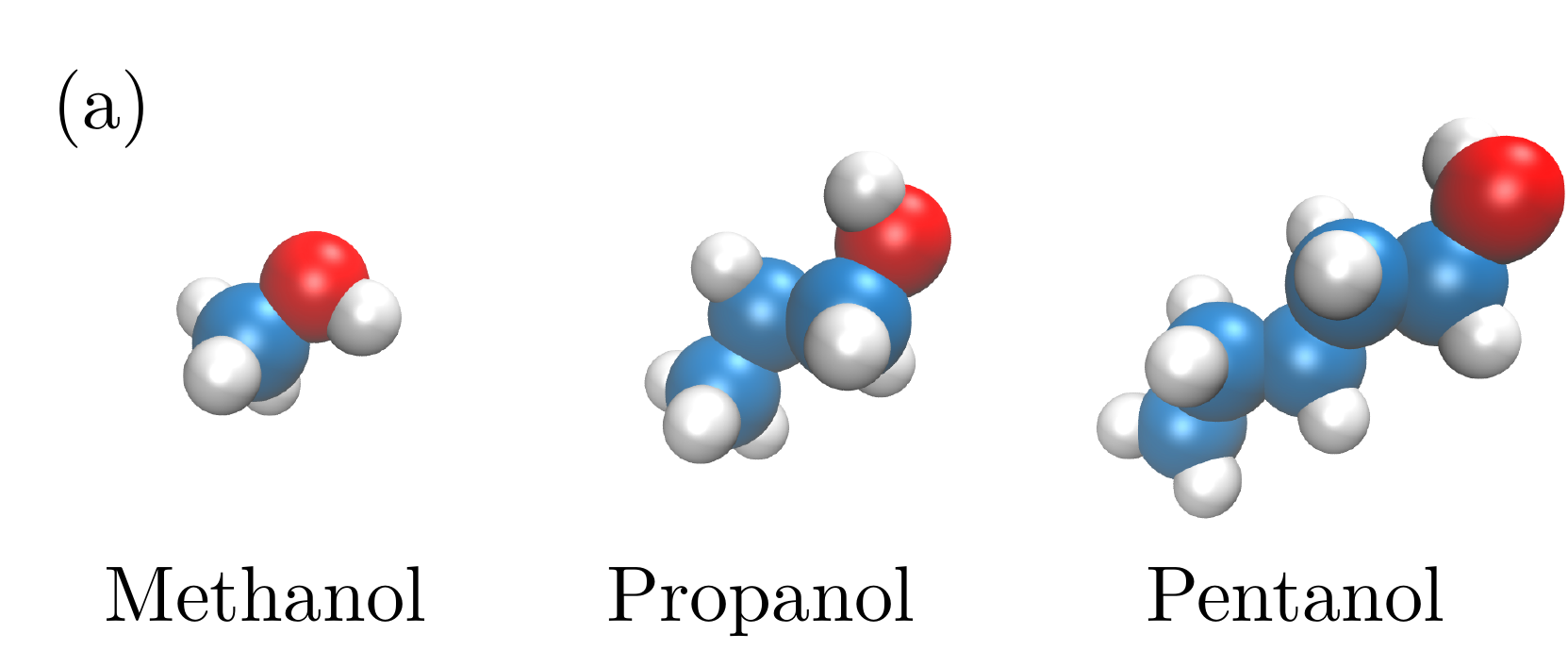}\\
\includegraphics[width=0.4\textwidth]{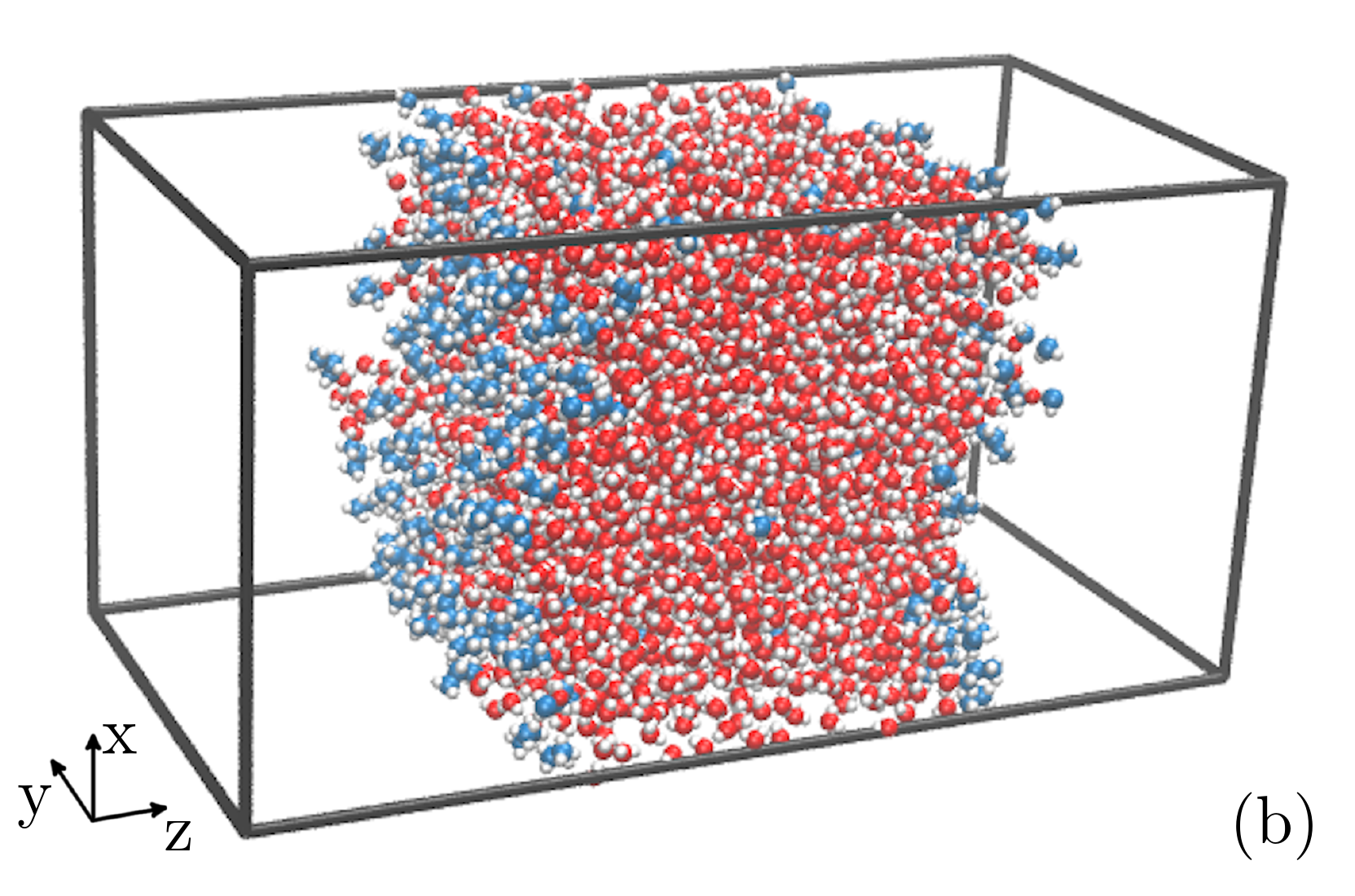}\\
\includegraphics[width=0.42\textwidth]{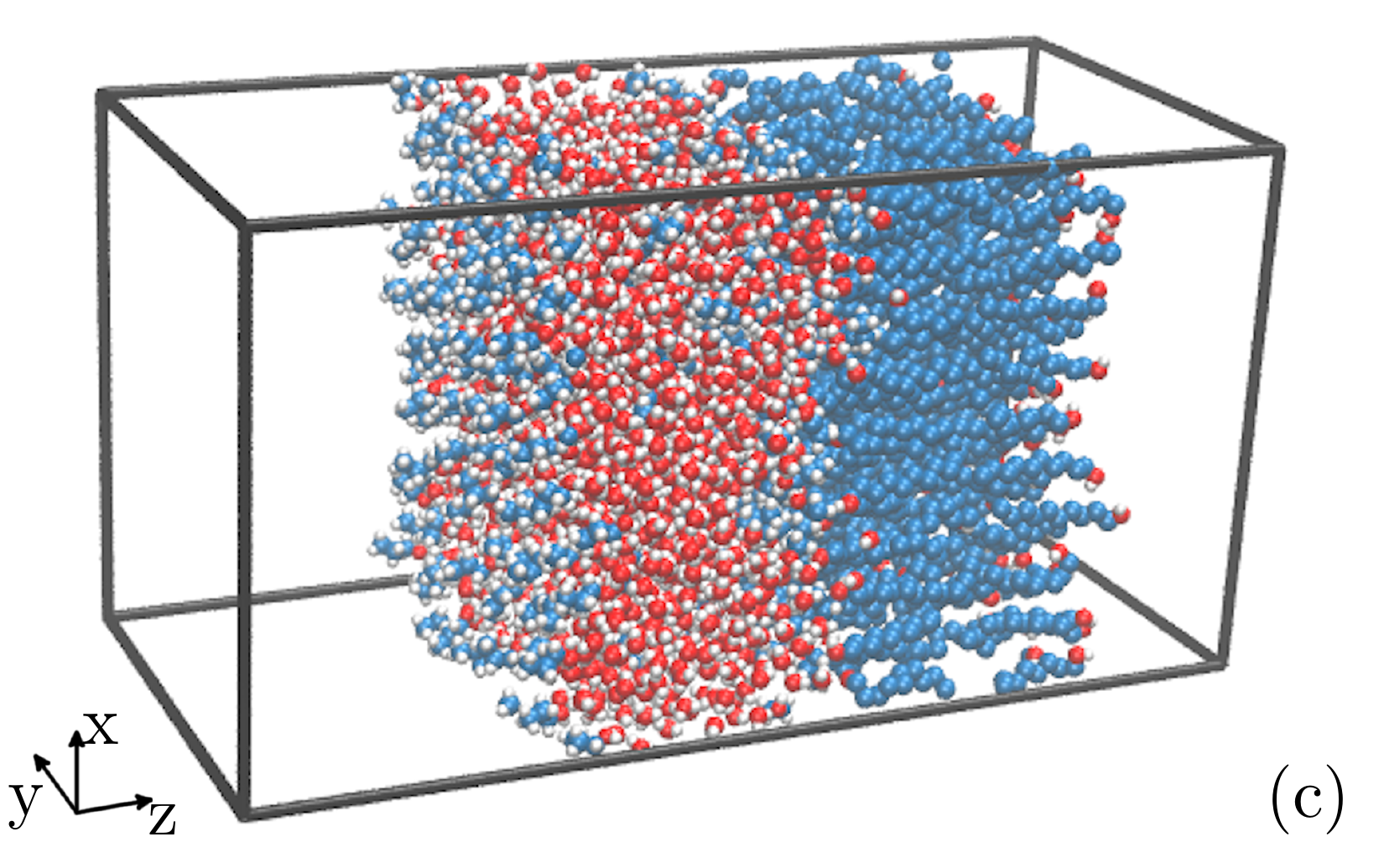}
\caption{Simulation models. (a) Surfactant molecules in this work: methanol, 1-propanol, and 1-pentanol. (b) Simulation box of a water slab containing surfactant molecules, used to study the water--vapor adsorption.
(c) Simulation box of a water slab in contact with the planar surface.
\label{fig:molecules}}
\end{figure}

To simulate the adsorption at the water--vapor interface, we set up an NVT (fixed number of particles, volume, and temperature) simulation with box dimensions of 5~nm $\times$ 5~nm $\times$ 10~nm with a water slab (containing various number of alcohol molecules) of thickness 5~nm in the middle (see \Fig{fig:molecules}b).
Periodic boundary conditions were applied in all three directions. 
The vapor layer (of thickness 5~nm) was large enough such that the water slab did not interfere with its periodic images along the $z$ direction.



For the planar solid surface, we adopted an atomistic model introduced before~\cite{KANDUC14, KANDUC16, kanduc2017going}, which mimics a self-assembled monolayer. The surface was composed of restrained, hexagonally packed aliphatic chains terminated by  hydroxyl (OH) head groups with the area density of 4.3 nm$^{-2}$. For the aliphatic chains, the united-atom representation was used.
To generate different hydrophilicities of the surface, we rescaled the original partial charges in the OH groups by the factors 0, 0.4, 0.6, 0.7, and 0.8, which results in the water contact angles of the surface of $\theta=\>$135$^\circ$, 120$^\circ$, 97$^\circ$, 76$^\circ$, and 45$^\circ$, respectively, as determined previously~\cite{kanduc2017going}. 
A 5-nm-thick water slab with added surfactant molecules was placed in contact with the surface.
The simulation box (of lateral dimensions 5.2 nm $\times$ 4.5 nm and height 10 nm) was replicated in all three directions {\em via} periodic boundary conditions (see \Fig{fig:molecules}c).

\subsection{Simulations and data analysis details}
The MD simulations were performed with the GROMACS 2019 simulation package \cite{gromacs}. The temperature was maintained at 300 K using the velocity-rescaling thermostat~\cite{Bussi07} with a time constant of 0.1 ps.
In NPT (fixed number of particles, pressure, and temperature) simulations (used for the Kirkwood--Buff integrals), the pressure was controlled with the Parrinello--Rahman barostat~\cite{Parrinello81,Nose83} of time constant 1.0 ps. Electrostatics was treated using particle-mesh-Ewald methods \cite{PME1, PME2} with a real-space cutoff of 0.9 nm. The LJ potentials were cut off at 0.9 nm in order to be compatible with the previous studies~\cite{KANDUC14, KANDUC16, kanduc2017going}.
The simulation times spanned up to 300~ns (3 independent realizations of 100~ns) for the water--vapor systems and 100~ns for the surface systems. 

When performing fits of our data with a given function we use an orthogonal distance regression algorithm~\cite{Boggs90}, which allows to include the uncertainty of the data in both the $x$ and $y$-coordinate.
This is necessary since, for some set of data, the relative uncertainty is much larger for the $x$-coordinate.


\section{Results and discussions}

\subsection{Adsorption at the water--vapor interface}\label{sec:adsorption_wat-wap}

\begin{figure*}[t!]
  \centering
	\includegraphics[width=0.32\textwidth]{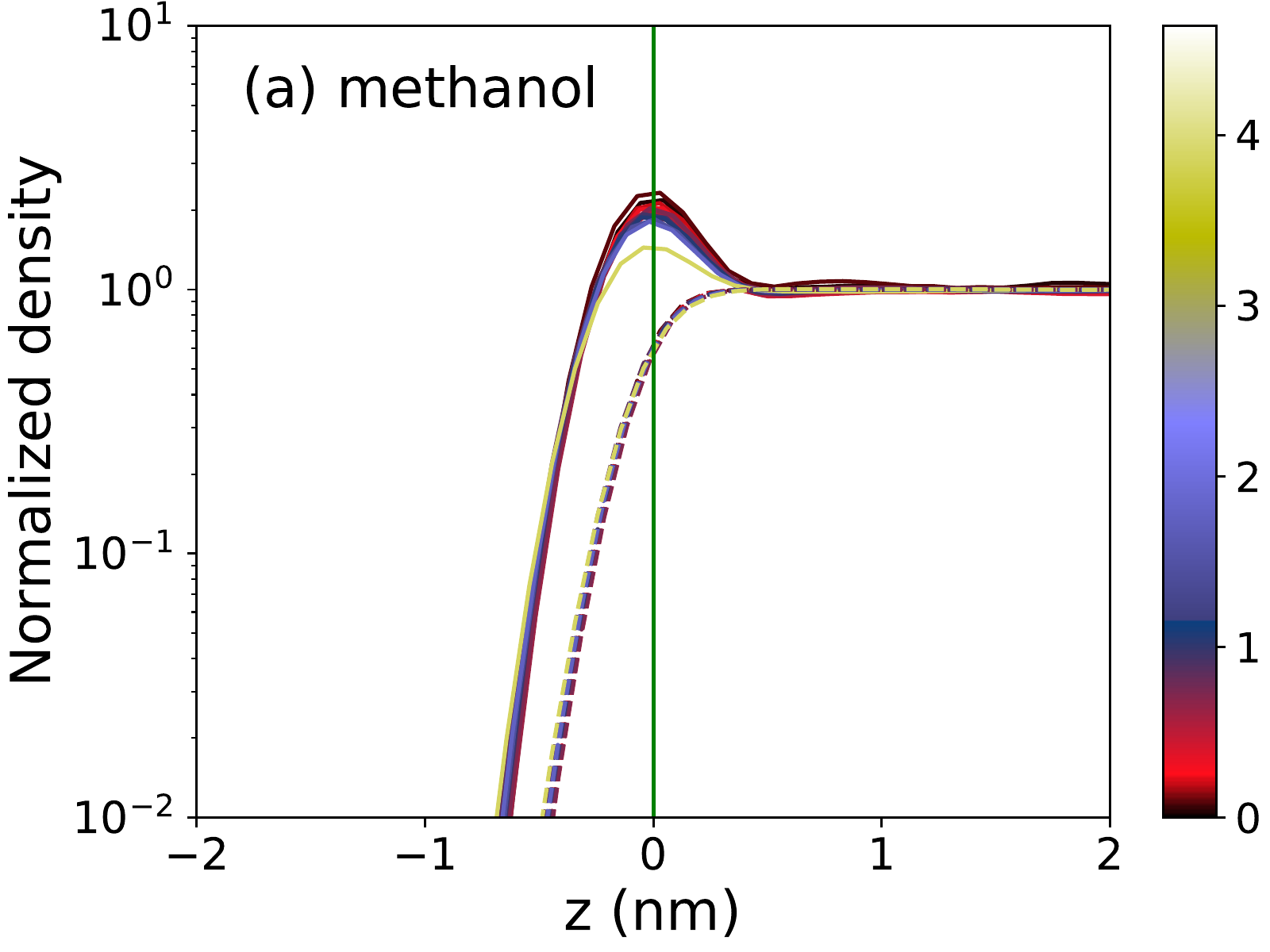}
\includegraphics[width=0.32\textwidth]{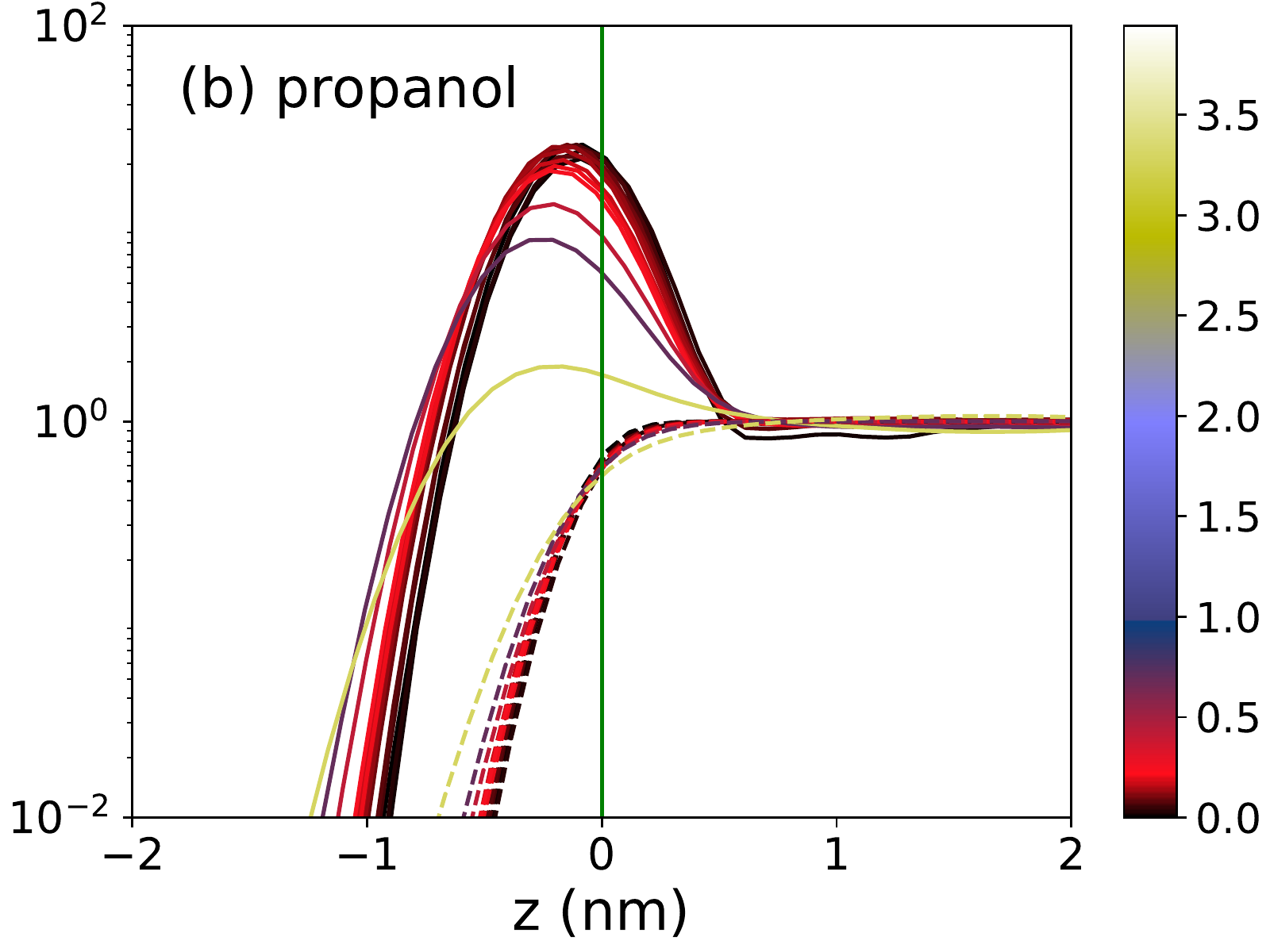}
\includegraphics[width=0.32\textwidth]{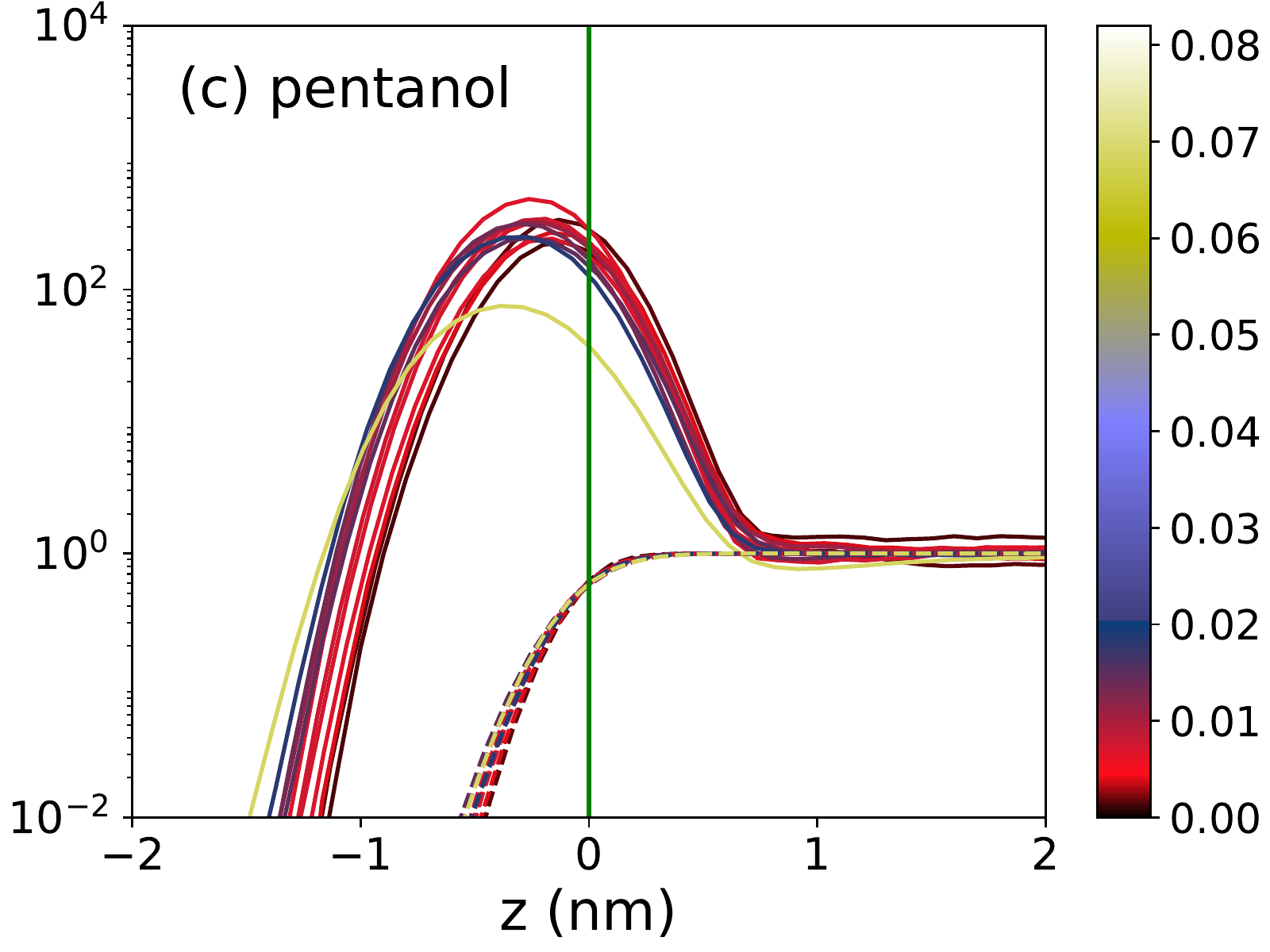}
	\caption{Normalized water (dashed lines) and surfactant (solid lines) density profiles (in logarithmic plot) at the water--vapor interface for different concentrations of (a)~methanol, (b)~propanol, and (c)~pentanol. 
	Different colors correspond to different bulk concentrations $c_0$ of the surfactant shown by the color bar on the right in the unit of nm$^{-3}$.
The green vertical lines indicate the Gibbs dividing surface of the water phase.}
\label{fig:dens_prof_slab}
\end{figure*}

We start by examining the adsorption behavior at the water--vapor interface (a proxy for the air--water interface), which is one of the most studied interfaces~\cite{Karlsson2017, VanDerSpoel06, hub2012organic}.
Figure~\ref{fig:dens_prof_slab} shows normalized density profiles of water [$c_\trm{w}(z)/c_\trm{w0}$; dashed lines] with various concentrations of surfactant [$c(z)/c_\trm{0}$; solid lines] in the proximity of the liquid--vapor interface. 
The pronounced density peaks of surfactants at the interface indicate preferential adsorption.

The adsorption is commonly quantified as the surface excess number density $\Gamma$ of the surfactant across the effective water--vapor boundary,
\begin{equation}
 \Gamma = \int_{-\infty}^{z_\trm{d}} \rmd z \, c (z) +\int_{z_\trm{d}}^{\infty} \rmd z \, [c (z)-c_0] , 
\end{equation}
where $c_0$ is the bulk surfactant concentration and $z_\trm{d}$ denotes the position of the Gibbs dividing surface of water (\ie, the position at which the excess water adsorption vanishes).
\begin{figure*}[t!]
  \centering
\includegraphics[width=0.32\textwidth]{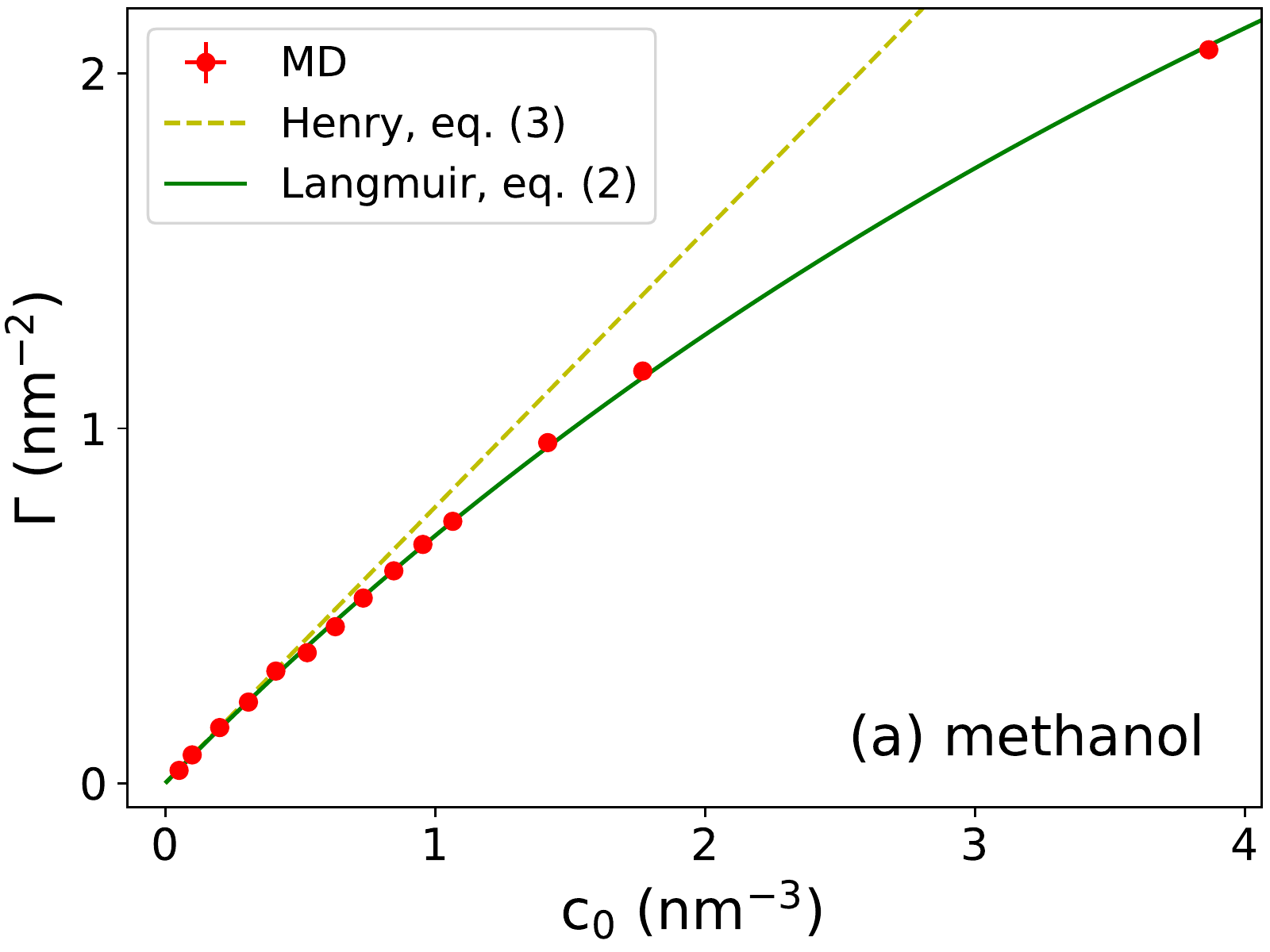}
\includegraphics[width=0.32\textwidth]{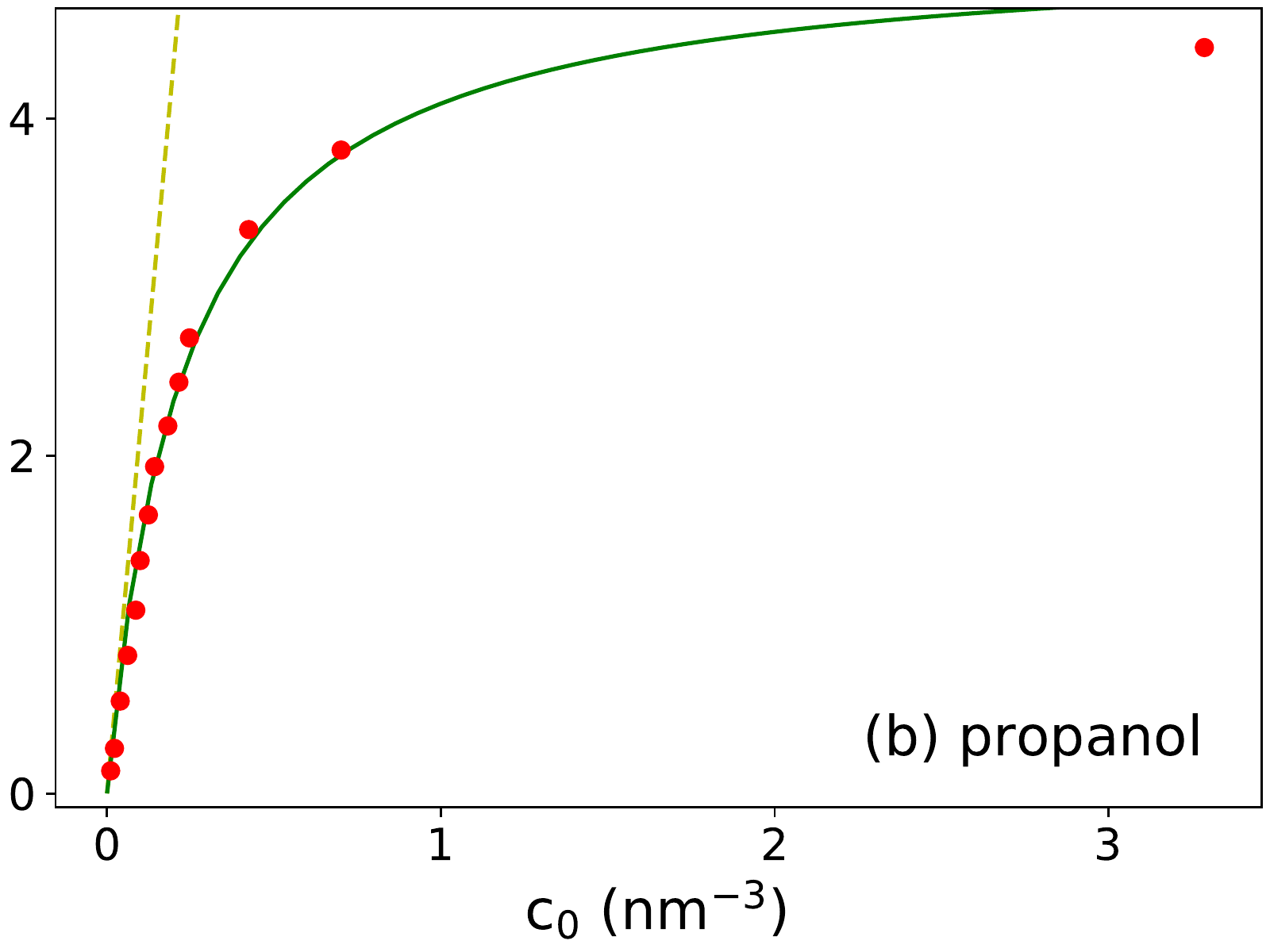}
\includegraphics[width=0.32\textwidth]{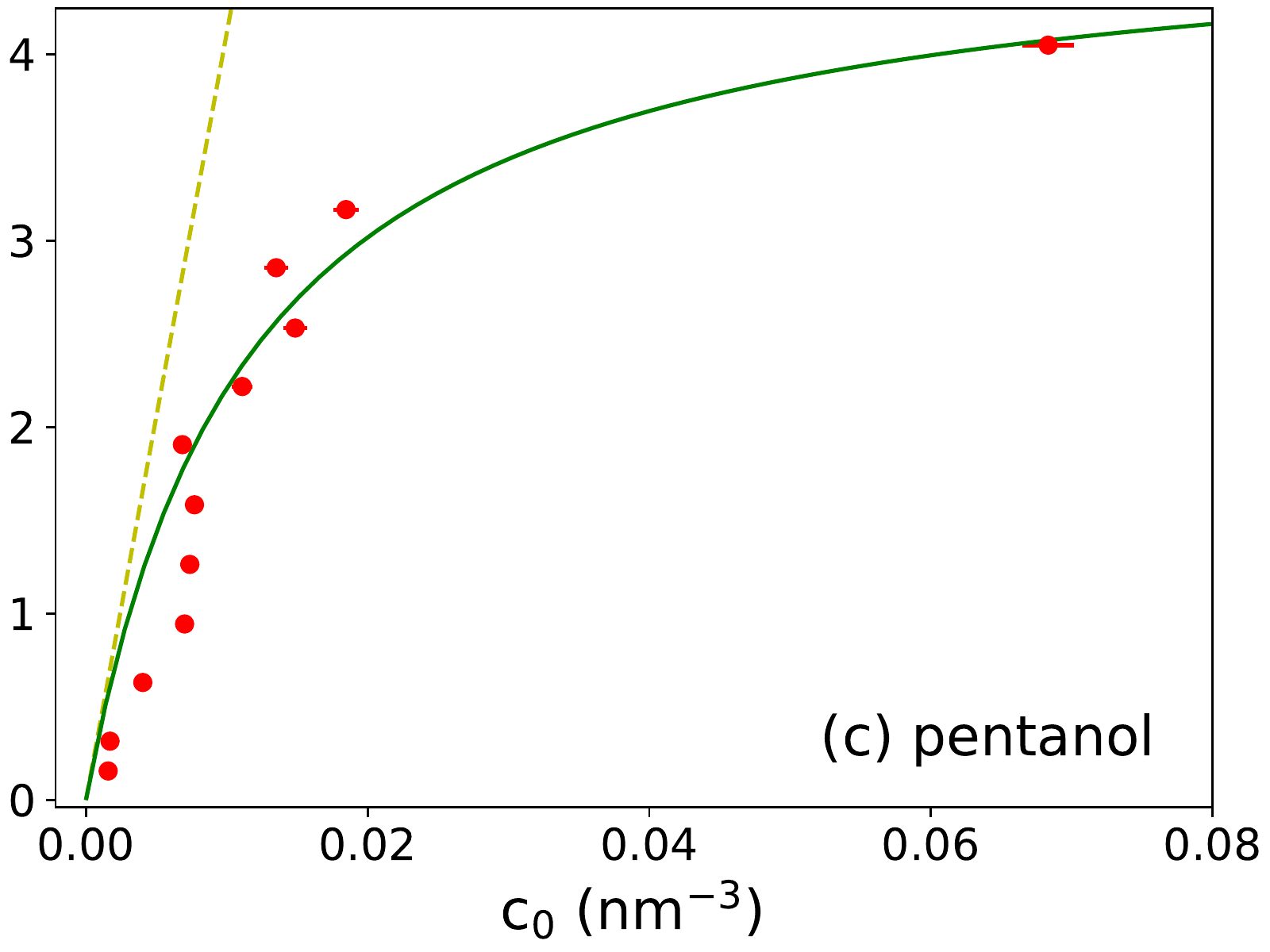}
	\caption{Adsorption $\Gamma$ at the water--vapor interface as a function of the bulk concentration of (a) methanol, (b) propanol, and (c) pentanol. The MD data (red symbols) are fitted with the Langmuir isotherm, \Eq~\ref{eq:langmuir} (green line). Yellow dashed lines correspond to Henry's law (\Eq~\ref{eq:Henry}), for which the coefficient $K_\trm{v}$ is obtained from the Langmuir fit ($K_\trm{v}=k_\trm{c}\Gamma_\infty$).   \label{fig:gamma_c0_slab}}
\end{figure*}

In \Fig{fig:gamma_c0_slab}, we plot evaluated adsorptions $\Gamma$ as a function of the bulk concentration $c_0$ for all three alcohol surfactants. Generally, at first a linear trend for low concentrations starts leveling off at higher concentrations, which can be well described by the Langmuir adsorption isotherm~\cite{CHANG19951,Bleys85,EASTOE2000,Swenson2019}
\begin{equation}
\Gamma = \Gamma_{\infty} \frac{k_\trm{c} \, c_0}{1 + k_\trm{c} \,  c_0} \, ,
\label{eq:langmuir}
\end{equation}
shown in \Fig{fig:gamma_c0_slab} and where $k_\trm{c}$ and $\Gamma_{\infty}$ are fitting parameters.
For low concentrations, \Eq~\eqref{eq:langmuir} reduces to Henry's law, \begin{equation}
    \Gamma = K c_0
    \label{eq:Henry}
\end{equation}
where $K = k_\trm{c}\,\Gamma_{\infty}$ is the adsorption coefficient. We denote the latter as $K_\trm{v}$ when representing the adsorption coefficient to the water--vapor interface and $K_\trm{s}$ to the solid--water interface. Henry's law is also shown in \Fig{fig:gamma_c0_slab} for comparison, with the adsorption coefficient $K_\trm{v}$ as obtained from the fit of the Langmuir isotherm.  
The adsorption coefficient $K_\trm{v}$, as well as $k_\trm{c}$, grows rapidly with molecular size: starting from $K_\trm{v}=$~0.8~nm for methanol, 21~nm for propanol, and 410~nm for pentanol.
Experimental values, obtained from literature~\cite{Bleys85,Joos89,CHANG19951}  (and references therein), for propanol and pentanol are $K_\trm{v}=$\,32~nm and 270~nm, respectively, which is in good agreement with the MD result given the high sensitivity on size, as we will see later on.

In contrast, the resulting saturation values of $\Gamma_{\infty}$ are comparable for the three alcohol species (6.52~nm$^{-2}$ for methanol, 5.06~nm$^{-2}$ for propanol and 4.80~nm$^{-2}$ for pentanol) because the adsorbed molecules occupy similar areas.
    Experimental data give $\Gamma_{\infty} \simeq 3.5\, \mbox{nm}^{-2}$  for propanol and pentanol~\cite{Bleys85,CHANG19951} (and references therein), which also compares reasonably well to our MD results. 
    Note that the accuracy of $\Gamma_\infty$ is not very high because of very few data points at high concentrations.

A notable effect of surfactant adsorption at the water--vapor interface is that it reduces the surface tension, $\gamma$. 
The reduction is calculated by the Gibbs adsorption equation, $\rmd\gamma=-\Gamma\rmd\mu$,
where $\mu$ is the surfactant chemical potential.
Both $\Gamma$ and $\mu$ depend on the concentration of surfactants, $c_0$. Whereas for $\Gamma(c_0)$, we assume the Langmuir isotherm (\Eq~\ref{eq:langmuir}), for the chemical potential we invoke the Kirkwood--Buff (KB) relation~\cite{KirkwoodBuff51,BenNaim77,Smith06},
\begin{equation}
\left( \frac{\partial c_0}{\partial \mu}\right)_{T,P} =  \frac{c_0 + c_0^2 ({\cal G_\trm{mm}-{\cal G}_\trm{wm}})}{\kB\,T}
\label{eq:KBmu}
\end{equation}
where $\kB$ is the Boltzmann constant, $T$ the temperature, and ${\cal G}_\trm{mm}$ and ${\cal G}_\trm{wm}$ are the molecule--molecule and water--molecule KB integrals, defined as
\begin{equation}
{\cal G}_{ij} =  \int_0^{\infty} \left[ g_{ij}(r) - 1 \right] 4 \pi r^2 \, \mbox{d}r \, \label{eq:g_ij_corr}
\end{equation}
where $ g_{ij}(r) $ is the radial distribution function between species $i$ and $j$ in bulk. Evaluated $ g_{ij}(r) $ in bulk solutions and calculated KB integrals ${\cal G}_\trm{mm}$ and ${\cal G}_\trm{wm}$ are shown in sec.~S-1 of the Supplementary Information (SI). Both KB integrals are roughly constant for low concentrations. Therefore,
we can combine \Eqs~\ref{eq:langmuir} and \ref{eq:KBmu} with the Gibbs adsorption equation. After integration, we obtain the relation between the surface tension reduction $\Delta \gamma$ and the adsorption $\Gamma$

\begin{equation}
\Delta \gamma =   \frac{\kB T\, \Gamma_\infty}{\xi} \, \ln \left(1 -\xi \, \frac{\Gamma}{\Gamma_{\infty}} \right)
\label{eq:gamma_vs_gamma}
\end{equation}
with the correction factor
\begin{equation}
    \xi=1-\frac{\Gamma_\infty ({\cal G_\trm{mm}-{\cal G}_\trm{wm}})}{K_\trm{v}}
    \label{eq:dgamma}
\end{equation}

The the correction factor $\xi$ amounts to $\sim 0.650$ for methanol, $\sim 0.950$ for propanol, and $\sim 0.998$ for pentanol. Let us briefly discuss the dependence of $\xi$ on surfactant size. Denoting the linear size of the molecule as $l$, $\Gamma_\infty$ roughly scales as $\sim l^{-2}$ (\ie, corresponding to the tightly packed monolayer of surfactants) and for non-aggregating molecules ${\cal G_\trm{mm}-{\cal G}_\trm{wm}}\sim l^3$ (\ie, corresponding to the volume of the surfactant). The product of the two scales with the size of the molecule, $\Gamma_\infty ({\cal G_\trm{mm}-{\cal G}_\trm{wm}})\sim l$. As we will see, the adsorption coefficient $K_\trm{v}$ increases exponentially with molecular size, therefore the correction $\xi$ is important for small molecules, whereas for larger molecules, it tends to unity, $\xi\to 1$; consistent with the simulation results.

\begin{figure}[h]
  \centering
\includegraphics[width=0.45\textwidth]{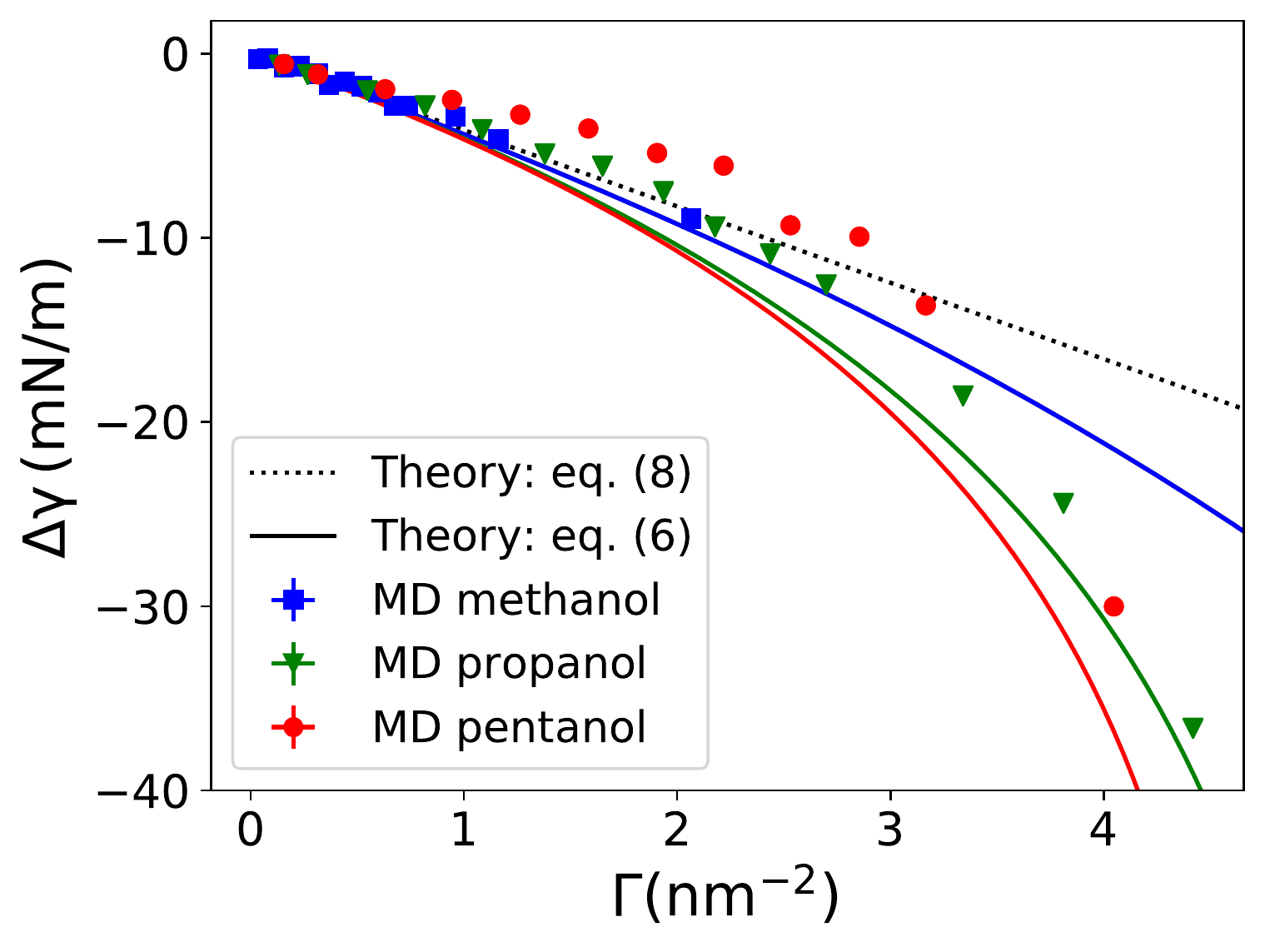}
	\caption{Reduction of the water--vapor surface tension {\em versus} adsorption as obtained from MD simulations (symbols) and theoretical predictions: 
	\Eq~\ref{eq:gamma_vs_gamma} (solid lines) and
    its linear expansion \Eq~\ref{eq:gamma_lin} (dotted line).}
\label{fig:gama_gamma_slab}
\end{figure}

In the limit of low adsorption (\ie, $\Gamma\ll\Gamma_\infty$, relevant at low concentrations), \Eq~\ref{eq:gamma_vs_gamma} simplifies to a linear form
\begin{equation}
\Delta \gamma \simeq -  \kB T\, \Gamma 
\label{eq:gamma_lin}
\end{equation}
which follows directly from Henry's law~\cite{Bonn09} and by assuming ideal behavior of the chemical potential. 
The second-order term in the above expansion is $-(\kB T\xi/2\Gamma_\infty)\Gamma^2$, from which it follows that \Eq~\ref{eq:gamma_lin} is expected to be valid for $\Gamma\ll\xi^{-1}\Gamma_\infty$ (\ie, when the second-order term is much smaller than the first term).
Figure \ref{fig:gama_gamma_slab} shows the relation between the surface-tension reduction $\Delta\gamma$ and the surfactant adsorption $\Gamma$ as obtained from the simulations (calculated from the diagonal pressure-tensor components~\cite{nijmeijer1990wetting}) and theory (\Eqs~\ref{eq:gamma_vs_gamma} and \ref{eq:gamma_lin}).
For small adsorption, the simple linear relation given by \Eq~\ref{eq:gamma_lin} (dotted line) matches very well the MD data. For higher adsorptions, the surface tension progressively sinks with adsorption, which is considerably well captured by the nonlinear relation (\Eq~\ref{eq:gamma_vs_gamma}).
However, some deviations are observed for intermediate values of $\Gamma$ for propanol and pentanol. Clearly, the underlying theoretical assumptions have limitation, one of which is the use of the Langmuir isotherm, especially for fitting the pentanol data (\Fig{fig:dens_prof_slab}c). The agreement could probably be improved by considering more complex isotherms with more fitting parameters (\eg, Frumkin isotherm)~\cite{diamant1996kinetics, CHANG19951, EASTOE2000}, but this is beyond the scope of this study, which focuses rather on low-adsorption regimes.


\subsection{Adsorption onto solid surfaces}
We now turn our attention to solid surfaces and investigate how changing the polarity, manifesting in different contact angles ($\theta\simeq45^\circ$--$135^\circ$) affects the adsorption of the three surfactants.
More details are provided in the Methods section and in refs.~\citenum{KANDUC14, KANDUC16, kanduc2017going}.

\begin{figure}[h!]
  \centering
    \includegraphics[width=0.2\textwidth]{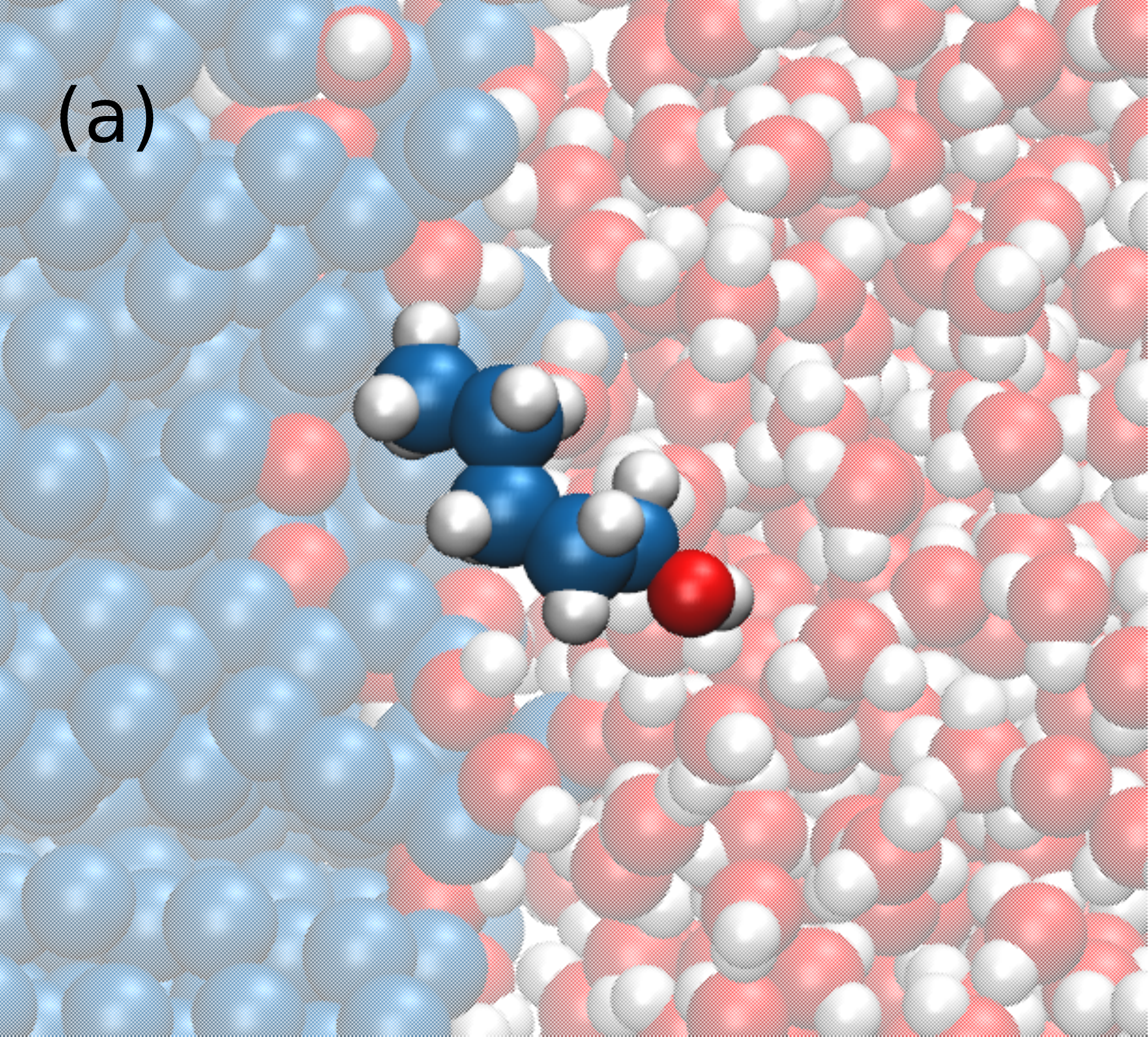}
    \includegraphics[width=0.25\textwidth]{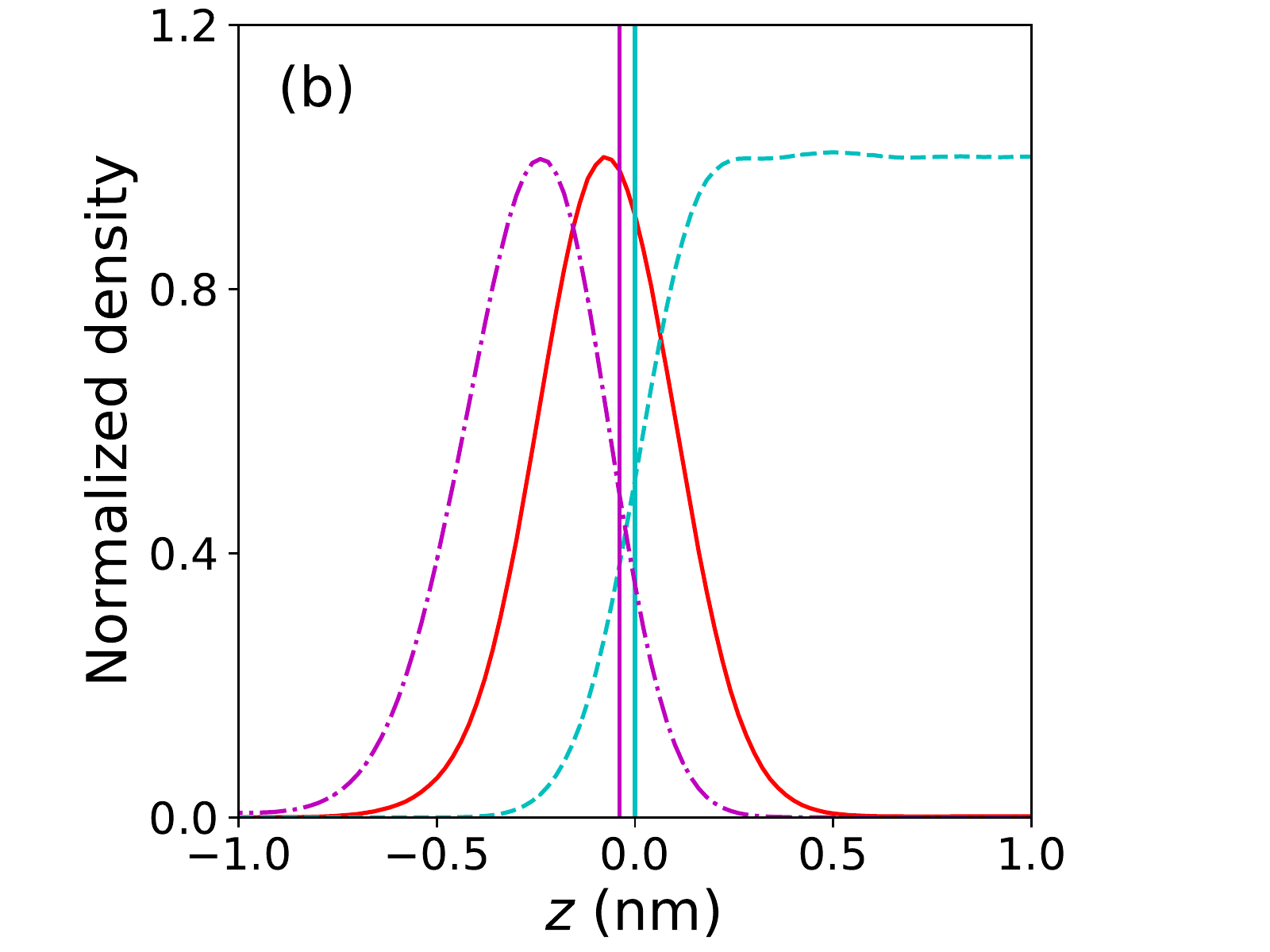}
   	\caption{(a) Snapshot of an adsorbed pentanol molecule on the nonpolar surface with $\theta=135^\circ$.
   	    (b) Corresponding rescaled density profiles of pentanol (red solid line) with bulk concentration of $c_0=0.001$~nm$^{-3}$, the surface OH groups (magenta dash-dotted line), and water (cyan dashed line). 
   	        Effective phase boundaries are depicted by the Gibbs dividing surface for water (cyan) and the position at half height on the water side of the OH group (magenta).
	\label{fig:snap}}
\end{figure}

We first take a look at some details of adsorption. 
Figure~\ref{fig:snap}a is a snapshot of a pentanol molecule adsorbed on the hydrophobic surface with $\theta=135^\circ$. The molecule partially penetrates into the surface interior by locally deforming the neighboring surface molecules. From the  density profiles of this scenario, shown in \Fig{fig:snap}b, we estimate that the molecule penetrates into the surface's interior roughly by a half of its size. Similar behavior is also found for the other two alcohols and other surface polarities; see Fig.~S-3 in the SI. 

\begin{figure*}[t!]
  \centering
\includegraphics[width=0.32\textwidth]{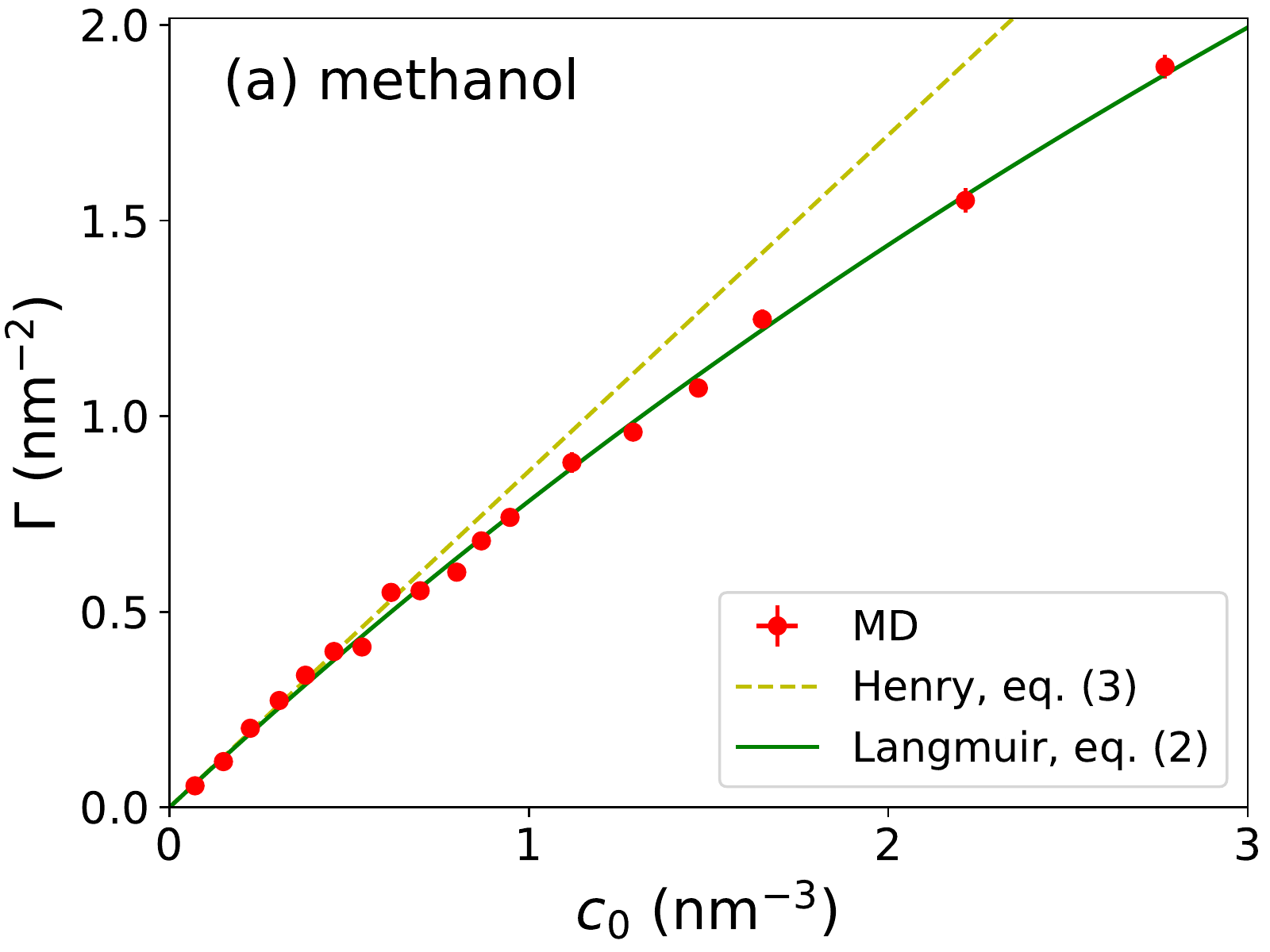} 
\includegraphics[width=0.32\textwidth]{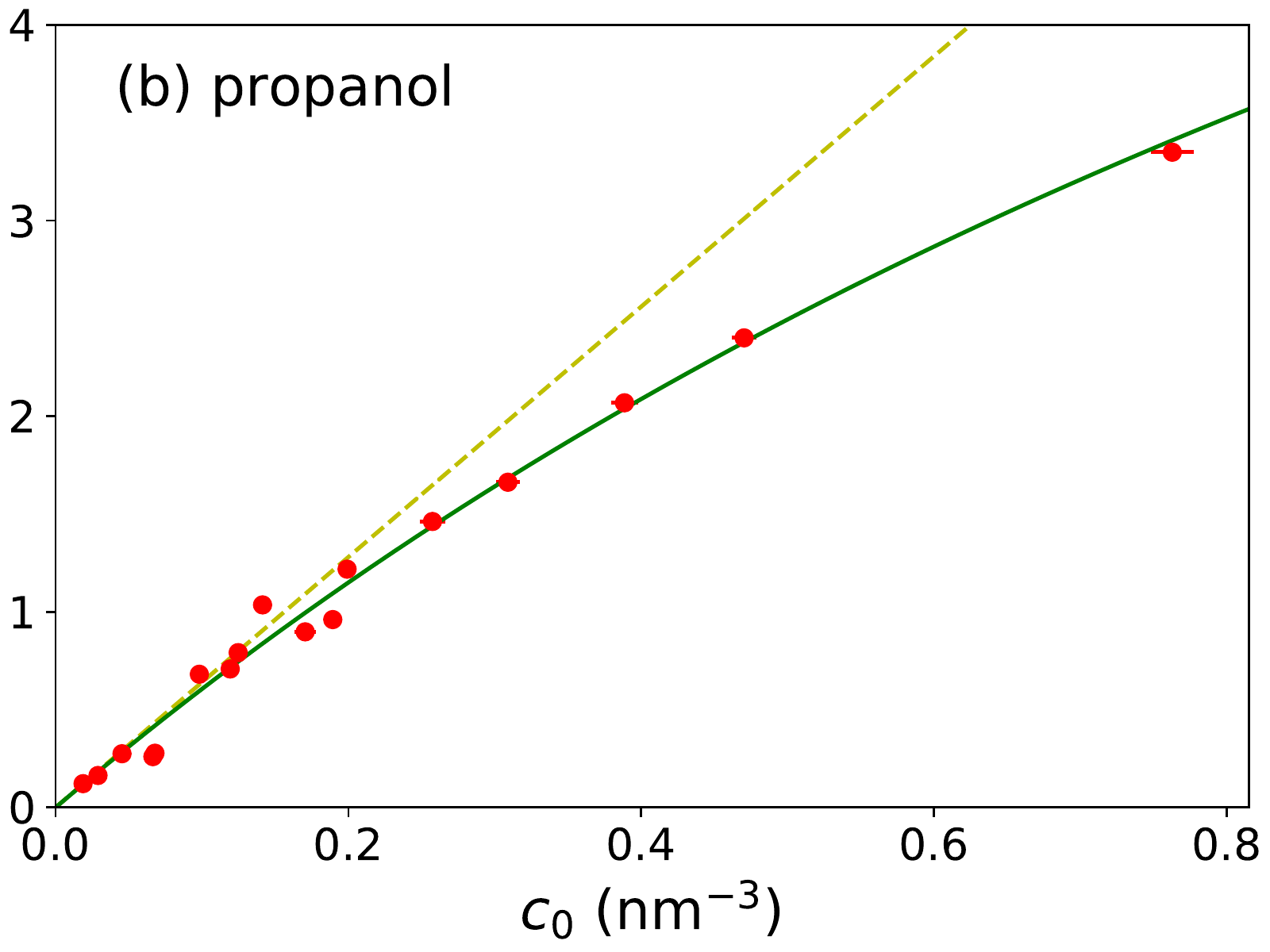} 
\includegraphics[width=0.32\textwidth]{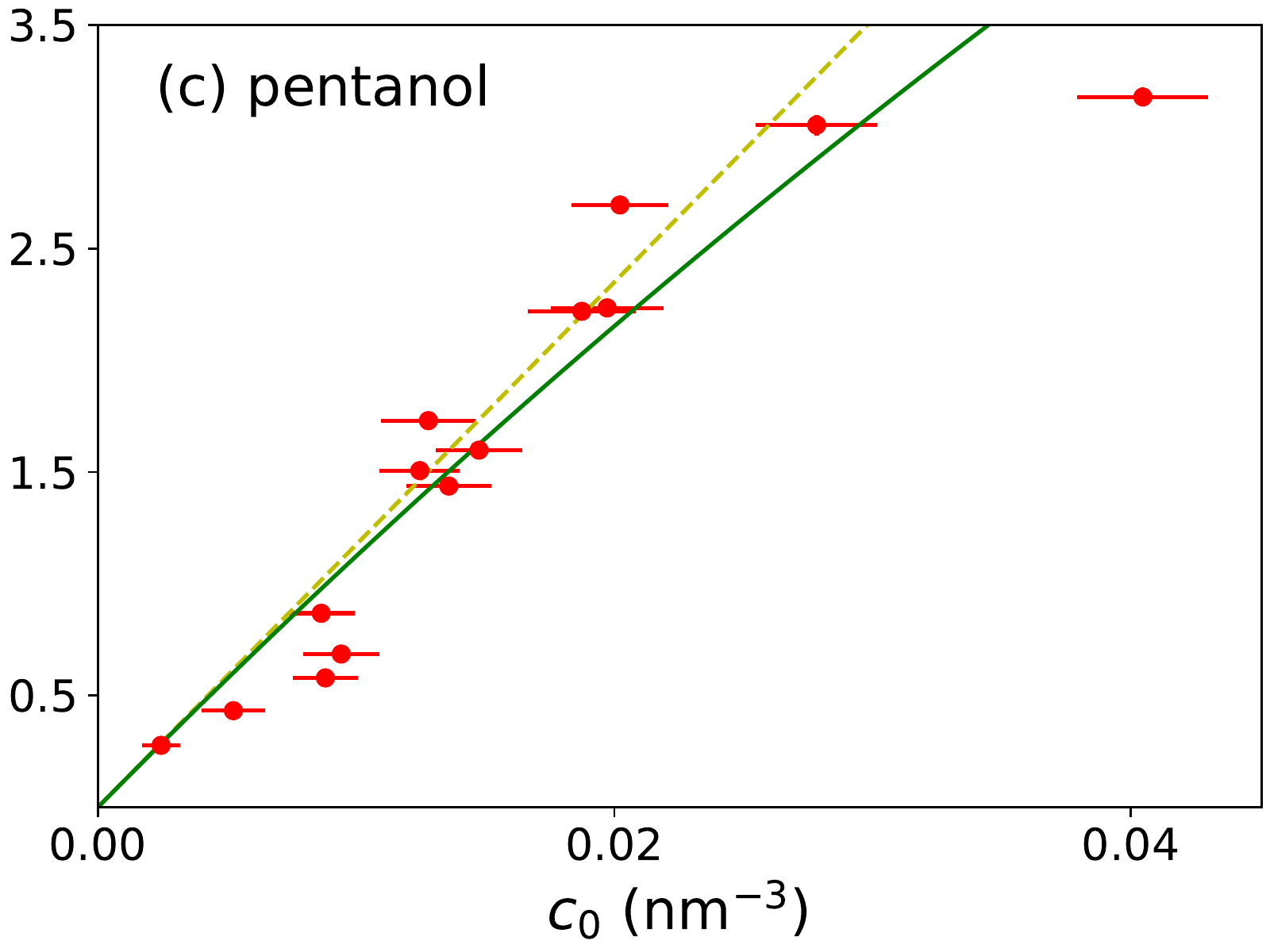} 
	\caption{
    Adsorption onto the surface with a wetting contact angle of $\theta=97^\circ$ as a function of the bulk concentration of (a) methanol, (b) propanol, and (c) pentanol. MD values are shown by red circles, whereas solid green lines show the fits of the Langmuir isotherm. Yellow dashed lines correspond to Henry's law (\Eq~\ref{eq:Henry}), for which the coefficient $K_\trm{v}$ is taken from the Langmuir fit.
   }
	\label{fig:gama_c0_surf_06}
\end{figure*}

Following the same procedure as for the water--vapor adsorption, we evaluate the adsorption--concentration relations, a few representative examples of which are shown
in \Fig{fig:gama_c0_surf_06} for a mildly hydrophobic surface with $\theta=97^\circ$ (the rest can be found in sec. S-2 of the SI).
The overall qualitative behavior is the same as at the water--vapor interface and it can be likewise well described by the Langmuir isotherm (shown by solid lines in \Fig{fig:gama_c0_surf_06}). 
The values of  $\Gamma_\infty$ are shown in Fig.~S-7 in the SI.

\begin{figure*}[t!]
  \centering
    \includegraphics[width=0.323\textwidth]{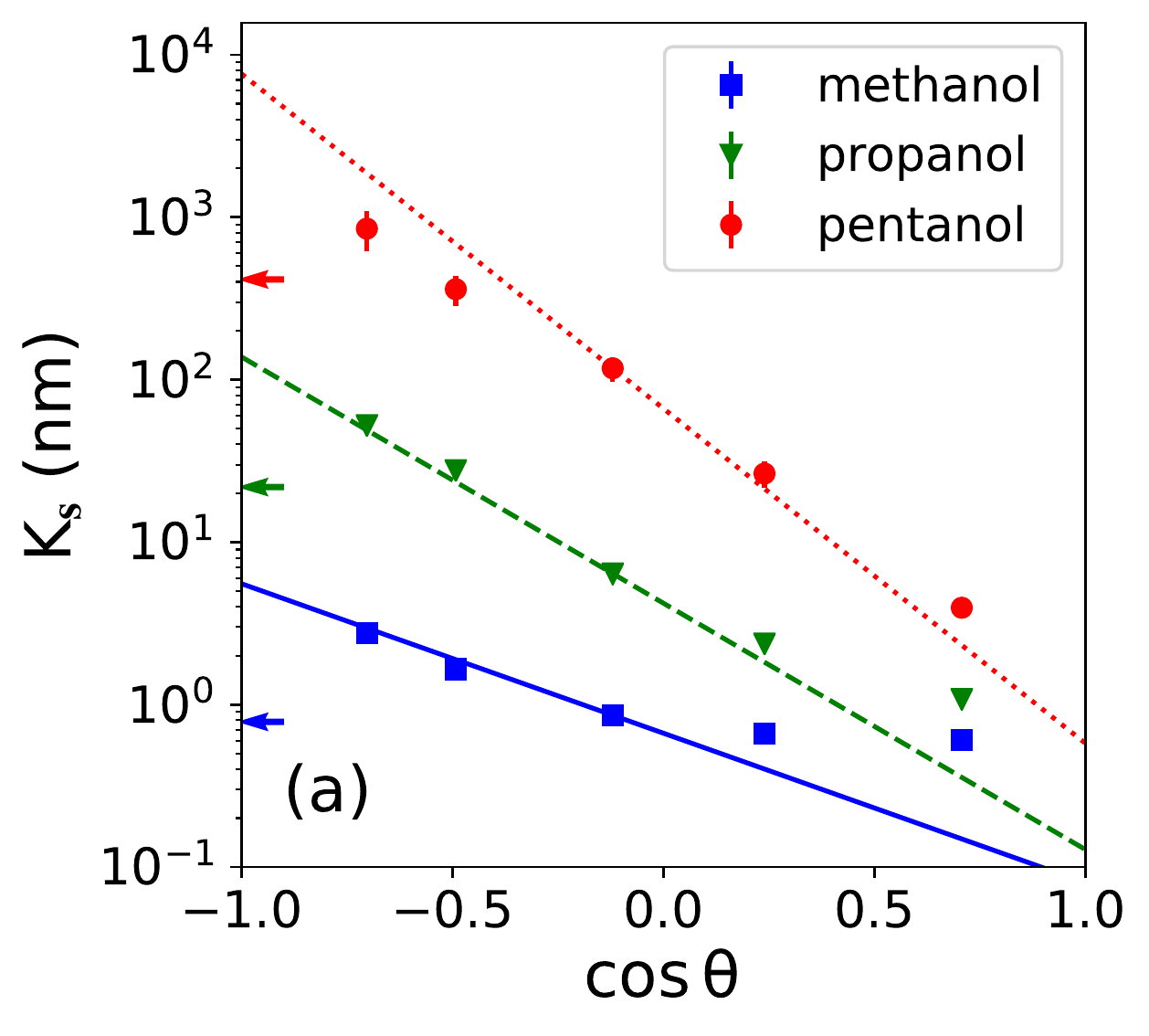}
    \includegraphics[width=0.30\textwidth]{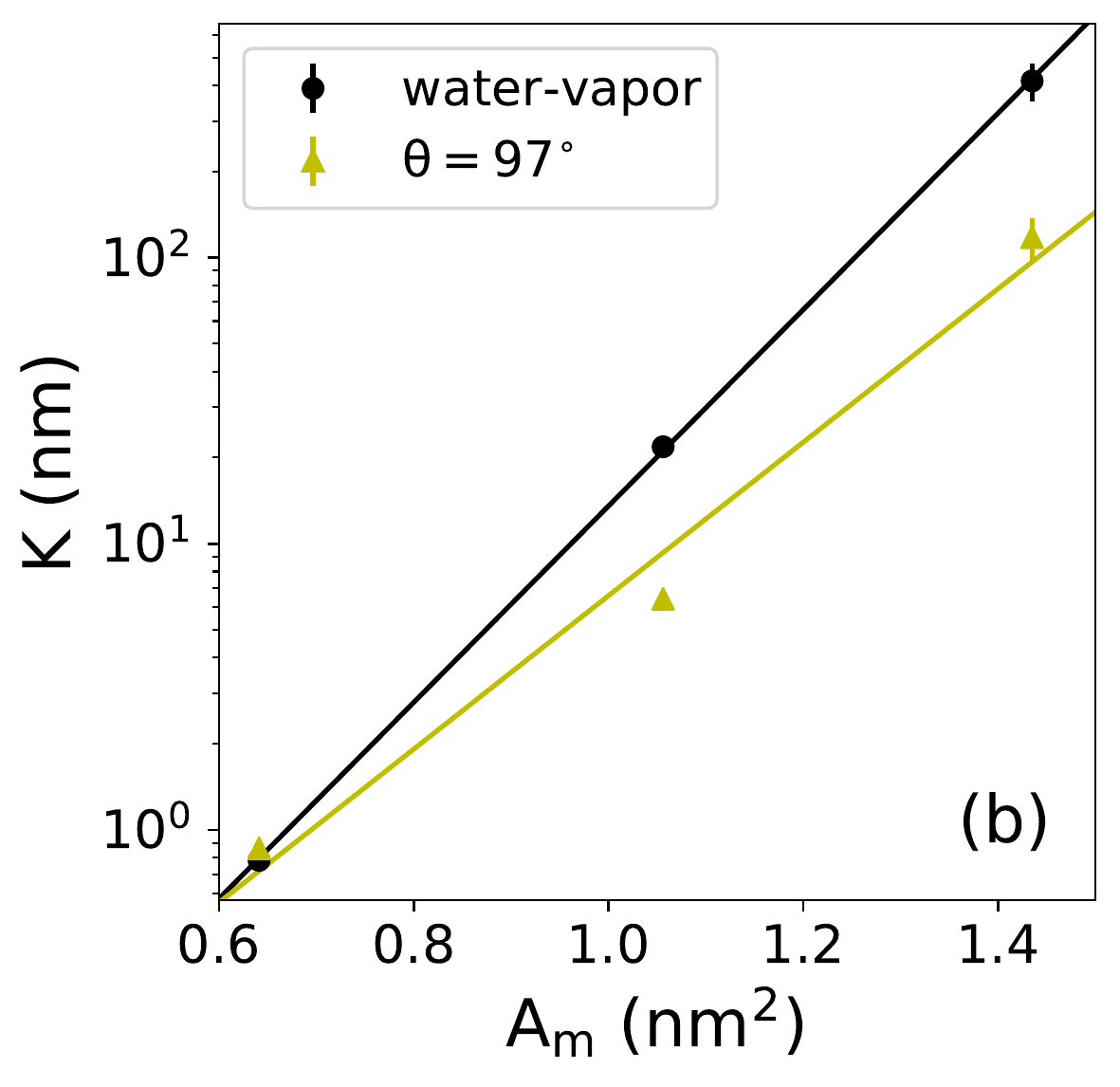}
    \includegraphics[width=0.31\textwidth]{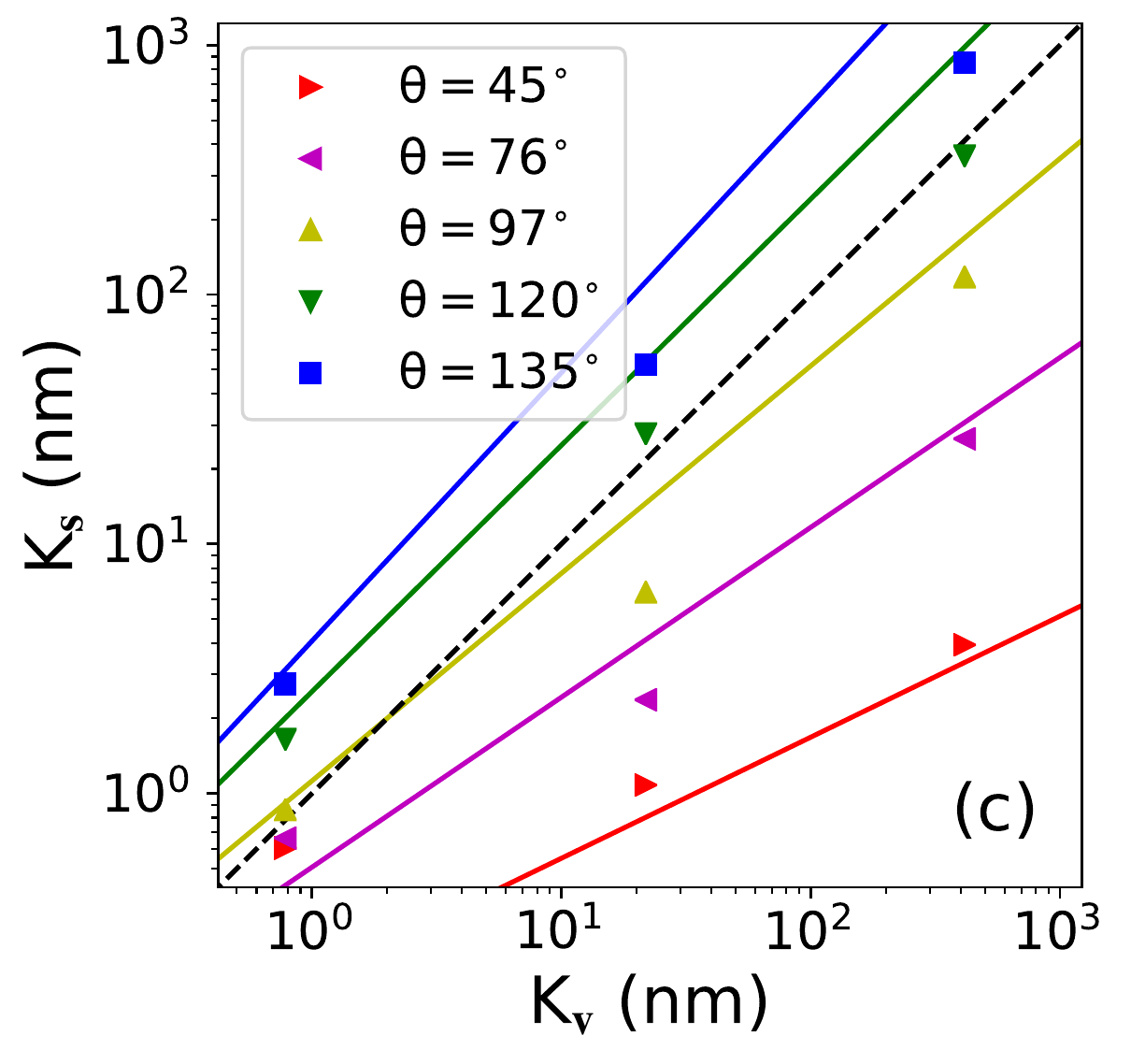}
\caption{Adsorption coefficients. (a)~Adsorption coefficient {\em versus} wetting coefficient for all three alcohols.
	The lines are predictions of \Eq~\ref{eq:acostheta}, whereby the coefficient $K_\trm{s}^{(0)}$ (controlling the offset) was used as a fitting parameter to the middle data points, with the wetting coefficient closest to zero ($\cos\theta=-0.12$). The arrows indicate the adsorption coefficients to the water--vapor interface, $K_\trm{v}$.
	(b)~Adsorption coefficient $K_\trm{s}^{(0)}$ for vanishing wetting coefficient {\em versus} molecular surface area $A_\trm{m}$. A comparison with the the water--vapor interface $K_\trm{v}$ is also shown. The solid lines are fitted exponential functions (\Eqs~\ref{eq:KsAm} and \ref{eq:KvAm}), which give $\tilde \gamma_\trm{s}\simeq 25.6$ mN/m and $\tilde \gamma_\trm{v}\simeq 32.7$ mN/m. 
	(c)~Correlation between adsorption coefficient to the solid surface ($K_\trm{s}$) and to the water--vapor interface ($K_\trm{v}$). The symbols are MD results, solid lines are predictions of \Eq~\ref{eq:logK}. The dashed diagonal line denotes the symmetric case $K_\trm{s}=K_\trm{v}$.
	\label{fig:gamma_gc_a_xi}}
\end{figure*}

In \Fig{fig:gamma_gc_a_xi}a, we plot the adsorption coefficients onto the surface, $K_\trm{s}$, against the surface wetting coefficient, $\cos\theta$. The outcomes clearly show that the hydrophobic surfaces have a much higher propensity to molecular adsorption than hydrophilic surfaces, which is consistent with the overall adsorption correlation with the contact angle found in various contexts~\cite{vogler1998structure, ishida1999, rosenhahn2010role,schwierz2012relationship}. 
Moreover, the results even suggests an approximate quantitative relation of the form $\log K_\trm{s} \sim \cos\theta$, which we will rationalize in the following.

\begin{figure}[]
  \centering
    \includegraphics[width=0.45\textwidth]{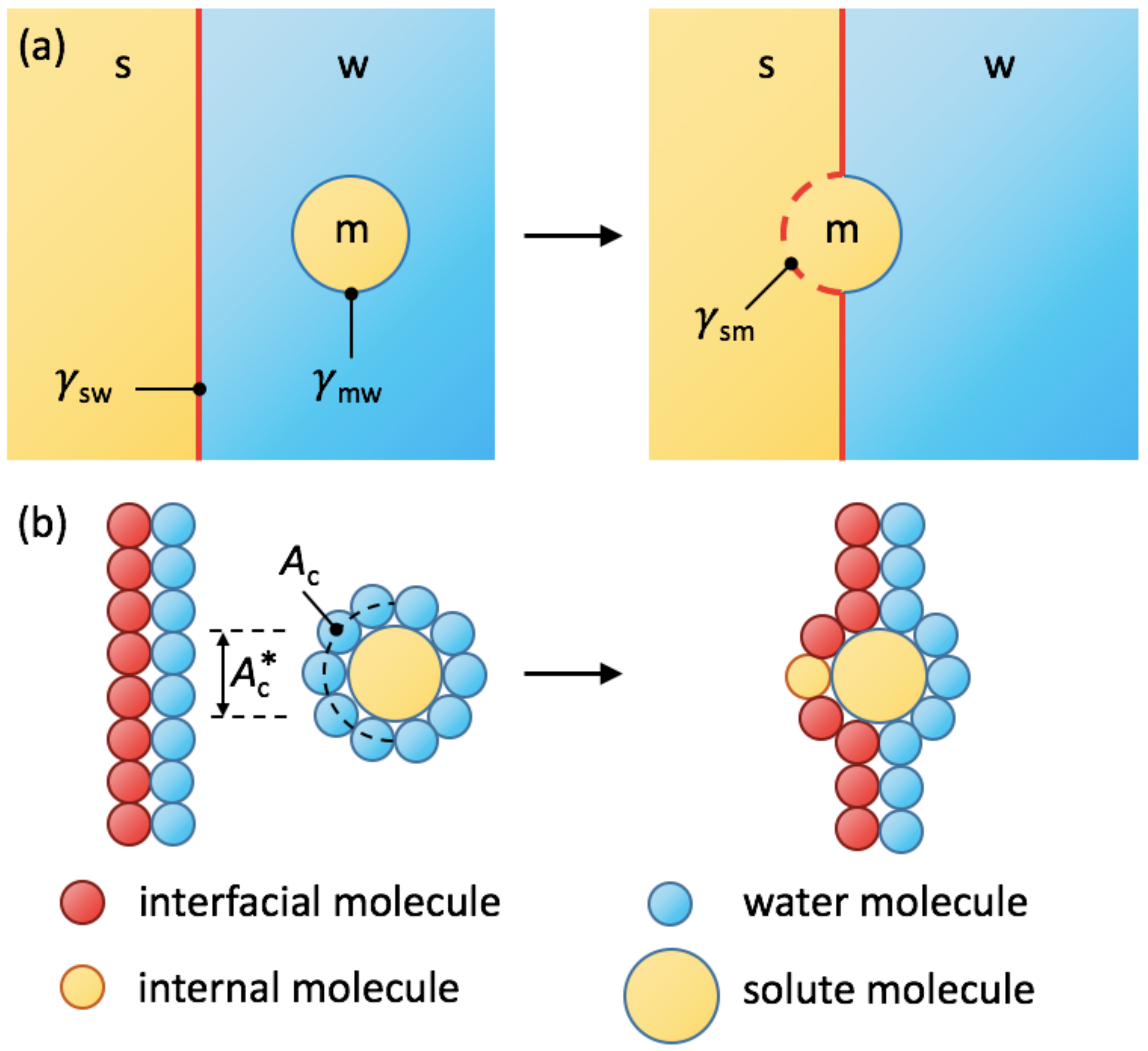}
	\caption{Schematic depiction of molecular adsorption to a soft surface. (a) Continuum picture: Adsorption is governed by surface tensions between the molecule (m), water (w), and the surface (s). (b) Molecular picture: The relevant areas are the bare cross sectional surface area of the molecule ($A_\trm{c}^*$) and the surface-accessible contact surface area ($A_\trm{c}$).
	\label{fig:schematics}}
\end{figure}

Since the adsorption increases with chain length, which is hydrophobic, the driving mechanism should be the hydrophobic effect~\cite{sendner2009interfacial}.
In order to at least qualitatively explain the observed relation, we resort to a continuum description of adsorption, schematically depicted in \Fig{fig:schematics}a: a surfactant molecule (m) adsorbs from bulk water (w) to the soft surface (s) by partially penetrating inside. 
The free energy of this adsorption scenario is composed of two contributions. 
Upon adsorption, the surfactant molecule forms direct contact with the surface of area $A_\trm{c}$. In doing so, the water molecules in this area of the surfactant molecule had to be removed. The corresponding free energy change is $-A_\trm{c}\gamma_\trm{mw}$, where $\gamma_\trm{mw}$ is the molecule--water surface tension.
The other contribution comes from new contacts between the molecule and the surface. However, even though the overall contact area with the surface is $A_\trm{c}$, the surface area with the OH head groups is equal to the cross-sectional area of the molecule $A_\trm{c}^*$.
The surplus $A_\trm{c}-A_\trm{c}^*$ comes from the hydrocarbon groups hitherto buried inside the surface that are now exposed to the surfactant (see \Fig{fig:schematics}b for illustration). Because the surfactant molecule is predominantly also hydrocarbon (an alkyl chain), the surface surplus does not contribute to the excess surface free energy. The free energy contribution due to the new contacts is therefore $A_\trm{c}^*(\gamma_\trm{sm}-\gamma_\trm{sw})$, where $\gamma_\trm{sm}$ and $\gamma_\trm{sw}$ are solid--molecule and solid--water surface tensions, respectively.
Summing up both contributions gives the adsorption free energy of the surfactant molecule in the continuum, macroscopic picture as
\begin{equation}
    \Delta G_\trm{s}=A_\trm{c}^*( \gamma_\trm{sm}-\gamma_\trm{sw})- A_\trm{c}\gamma_\trm{mw}
    \label{eq:dG}
\end{equation}

\Fig{fig:schematics}b outlines the essential molecular rearrangements during the adsorption.  The effective cross-sectional area $A_\trm{c}^*$, which is the area of the removed water molecules from the surface, is best described by the cross section of the bare molecule. If we approximate the molecule by a sphere (\ie, $A_\trm{m}=4\pi R_\trm{m}^2$ and $A_\trm{c}^*=\pi R_\trm{m}^2$, where $R_\trm{m}$ is its radius), then the cross-sectional area is $A_\trm{c}^*=\frac{1}{4}A_\trm{m}$.
In the other extreme limit, in which the molecule is considered as an infinitely long cylinder (\ie, $A_\trm{m}=2\pi R_\trm{m} L$ and $A_\trm{c}^*=2R_\trm{m}L$, where $R_\trm{m}$ is the radius and $L$ the length of the cylinder), the relation becomes $A_\trm{c}^*=\pi^{-1}A_\trm{m}$. In realistic cases of finite rod-like molecules (such as alcohols in our case), the the ratio $A_\trm{c}^*/A_\trm{m}$ lies somewhere between the two extremes of $1/4=0.25$ and $1/\pi\approx 0.32$, which is a rather narrow interval. Since the continuum approach for describing molecular details is very approximate, we will assume the spherical approximation in the forthcoming analysis. 

Before proceeding with \Eq~\ref{eq:dG}, we have to be aware that applying macroscopic concepts of interfacial surface at the molecular level is in general a delicate move. 
Nonetheless, some problems can be, at least qualitatively, formally resolved by identifying effective molecular surface areas and curvature (\ie, Tolman) corrections to surface tensions~\cite{tanford1979interfacial, ashbaugh2006colloquium}.
Such an analysis reaches, however, far beyond the scope of this study. Therefore, we will use the above continuum equation only to extract the dependence of adsorption on the contact angle. The latter is related to removal of water from the flat area of the solid, whose surface is flat (requiring no curvature corrections) and whose surface tension is macroscopically well defined.

The solid--water surface tension $\gamma_\trm{sw}$ is the only quantity in \Eq~\ref{eq:dG} that depends on the contact angle. The dependence is provided by the Young equation of a water droplet on the surface, 
\begin{equation}
\gamma_\trm{sw}=\gamma_\trm{sv}-\gamma\cos\theta
\label{eq:gamma_sw}
\end{equation}
where $\gamma_\trm{sv}$ is the solid--vapor surface tension.
Equation~\ref{eq:dG} now expresses as
\begin{equation}
    \Delta G_\trm{s}=\Delta G_\trm{s}^{(0)}+\frac 14 A_\trm{m} \gamma\cos\theta
    \label{eq:deltaGs}
\end{equation}
where the reference value $\Delta G_\trm{s}^{(0)}=A_\trm{c}^*( \gamma_\trm{sm}-\gamma_\trm{sv})- A_\trm{c}\gamma_\trm{mw}$ is the adsorption free energy to the surface with a vanishing a wetting coefficient, $\cos\theta=0$ (\ie, for $\theta=90^\circ$). The above equation nicely demonstrates the modulation of the adsorption free energy with the contact angle.


From a known $\Delta G_\trm{s}$, the adsorption coefficient to the surface can be estimated as 
\begin{equation}
K_\trm{s}=b_\trm{s}\,\rme^{- \beta\Delta G_\trm{s}}    
\label{eq:Ks}
\end{equation}
where $\beta=1/\kB T$ and $b_\trm{s}$ is a fitting parameter. 
Using \Eq~\ref{eq:deltaGs}, the dependence of the adsorption coefficient on $\cos\theta$ follows as
\begin{equation}
    K_\trm{s}= K_\trm{s}^{(0)}\exp\Bigl(-\frac{1}{4}\beta A_\trm{m}\gamma\cos\theta\Bigr)
    \label{eq:acostheta}
\end{equation}
where the reference $K_\trm{s}^{(0)}=b_\trm{s}\exp(-\beta \Delta G_\trm{s}^{(0)})$ is the adsorption coefficient for the surface with vanishing wetting coefficient. 
As seen in \Fig{fig:gamma_gc_a_xi}a, the agreement between the MD data and \Eq~\ref{eq:acostheta} (with $K_\trm{s}^{(0)}$ as a fitting parameter to the middle data points) is reasonably good, particularly in the hydrophobic regime ($\cos\theta<0$). 
For hydrophilic cases ($\cos\theta>0$), agreement becomes worse, especially for smaller molecules such as methanol, which exhibit weak adsorption.
One reason for the poorer agreement is that in weakly-adsorbing cases (\ie, small $K_\trm{s}$), the molecule penetrates less into the surface (see Fig.~S-3 in the SI), thus $A_\trm{c}^*$ is smaller than $A_\trm{m}/4$. 

The next relevant question is, how does the reference adsorption coefficient $K_\trm{s}^{(0)}$ depend on the molecular surface area.
In \Fig{fig:gamma_gc_a_xi}b, we plot the relation between $K_\trm{s}^{(0)}$ and the molecular surface area $A_\trm{m}$.
The result can be easily understood using \Eq~\ref{eq:dG}, which suggests that the adsorption free energy is proportional to the molecular surface area, $\Delta G_\trm{s}^{(0)}=-\tilde \gamma_\trm{s}A_\trm{m}$, where the proportionality coefficient $\tilde \gamma_\trm{s}$ can be considered as an effective molecular surface tension for adsorption~\cite{tanford1979interfacial,ashbaugh2006colloquium}.
For the reference adsorption coefficient we can thus write
\begin{equation}
    K_\trm{s}^{(0)}=b_\trm{s}\rme^{\beta \tilde \gamma_\trm{s} A_\trm{m}}
    \label{eq:KsAm}
\end{equation}
and likewise, for the adsorption coefficient at the water--vapor interface
\begin{equation}
    K_\trm{v}=b_\trm{v}\rme^{\beta \tilde \gamma_\trm{v} A_\trm{m}}
    \label{eq:KvAm}
\end{equation}
The above two equations fit the MD data points in \Fig{fig:gamma_gc_a_xi}b very well, with $b_i$ and $\tilde \gamma_i$ ($i=\trm{v}, \trm{s}$) used as fitting parameters.


It is insightful to look at the correlation between the adsorption coefficients to both interfaces  $K_\trm{s}$ and $K_\trm{v}$, as plotted in \Fig{fig:gamma_gc_a_xi}c. 
The two coefficients are very well correlated for a given surface contact angle, implying that the better a molecule adsorbs onto the water--vapor interface, the better it adsorbs onto the solid surface.
This correlation stems primarily from the linear dependence of adsorption energies on molecular surface area.
Using \Eqs~\ref{eq:acostheta}--\ref{eq:KvAm} and eliminating $A_\trm{m}$, we come up with the following analytic relation
\begin{equation}
    \ln \frac{K_\trm{s}}{b_\trm{s}}=\frac{\tilde \gamma_\trm{s}-\tfrac 14 \gamma\cos\theta}{\tilde \gamma_\trm{v}}\,\ln \frac{K_\trm{v}}{b_\trm{v}}
    \label{eq:logK}
\end{equation}
which demonstrates that, indeed, the logarithms of the two adsorption coefficients are linearly related, with a prefactor that linearly decreases with $\cos\theta$.
Using the fitted coefficients form \Fig{fig:gamma_gc_a_xi}b, we plot the predictions of \Eq~\ref{eq:logK} in \Fig{fig:gamma_gc_a_xi}c as solid lines. Even though the agreement is not perfect, the slope is nicely captured by the prefactor of \Eq~\ref{eq:logK}, at least for the larger two alcohols.

From the correlation plot, we conclude that the adsorption to the water--vapor interface is always stronger than to the polar solid surfaces with contact angles below $\theta\approx97^\circ$ --- the data lie below the diagonal symmetry line. Moreover, the ratio $K_\trm{s}/K_\trm{v}$ becomes progressively smaller with an increasing $K_\trm{v}$ (\ie, molecular size).
In contrast, the hydrophobic surfaces with contact angles above $\theta\approx120^\circ$ outdo the water--vapor interface in adsorption, at least for not too large and too strongly adsorbing molecules.
Qualitatively same trends were experimentally observed on hydrophobic and mildly hydrophilic surfaces~\cite{JanczukJCIS2006,BinksSM2010surfButt}.


\subsection*{Surfactant effect on the wetting contact angle}

In the end, we take a look at a scenario where the adsorption to the water--vapor and a solid surface compete with each other --- a sessile water droplet containing surfactants.
A neat water droplet deposited on a solid surface forms the contact angle $\theta$ with the surface, given by the Young equation (\Eq~\ref{eq:gamma_sw}). When surfactant is introduced into the droplet, it adsorbs to both interfaces, solid--water and water--vapor, thereby reducing their surface tensions, which become
dependent on the surfactant concentration (\ie, $\gamma_\trm{sw}(c_0)$ and $\gamma(c_0)$).
In principle, less soluble surfactants can also adsorb at the solid--vapor interface~\cite{bera2021antisurfactant}, which does, however, not occur in our case (see sec.~S-3 of the SI). Consequently, the solid--vapor surfaces tension, $\gamma_\trm{sv}$, remains unaffected.
The Young equation of the surfactant-laden droplet then reads~\cite{bera2016surfactant, sun2021ion}
\begin{equation} \label{eq:deltatheta_c0}
\cos (\theta+\Delta\theta) = \frac{\gamma_\trm{sv} - \gamma_\trm{sw}(c_0)}{\gamma(c_0)}
\end{equation}
where $\theta$ is the contact angle of the neat (surfactant-free) water droplet and $\Delta \theta$ is the change of the contact angle due to the surfactant.
For small changes in contact angle ($\Delta\theta\ll 1$), the above equation simplifies to 
\begin{equation}
\Delta\theta\simeq\frac{\Delta \gamma_\trm{sw}+\Delta\gamma \cos\theta}{\gamma\sin\theta}    
\label{eq:Delta_theta}
\end{equation}

In the linear adsorption regime, in which Henry's law and \Eq~\ref{eq:gamma_lin} apply, the expression further simplifies to
\begin{equation}\label{eq:deltatheta_c0_lin}
    \Delta\theta\simeq-\kB T\,\frac{K_\trm{s}+K_\trm{v}\cos\theta}{\gamma\sin\theta}\, c_0
\end{equation}
which is a modification of the Lucassen--Reynders equation~\cite{lucassen1963contact}.

\begin{figure*}[t!]
  \centering
    \includegraphics[width=0.32\textwidth]{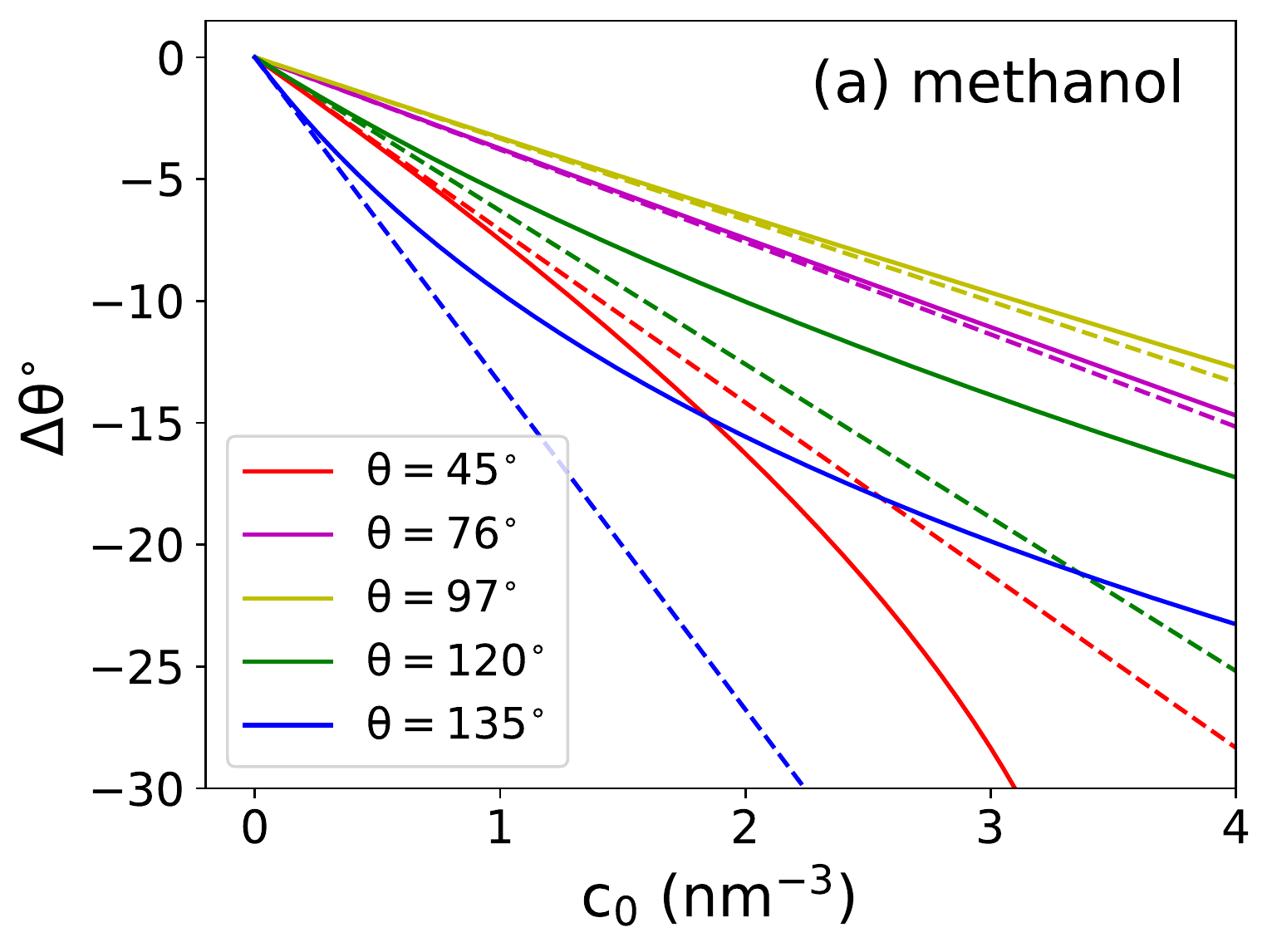}
    \includegraphics[width=0.31\textwidth]{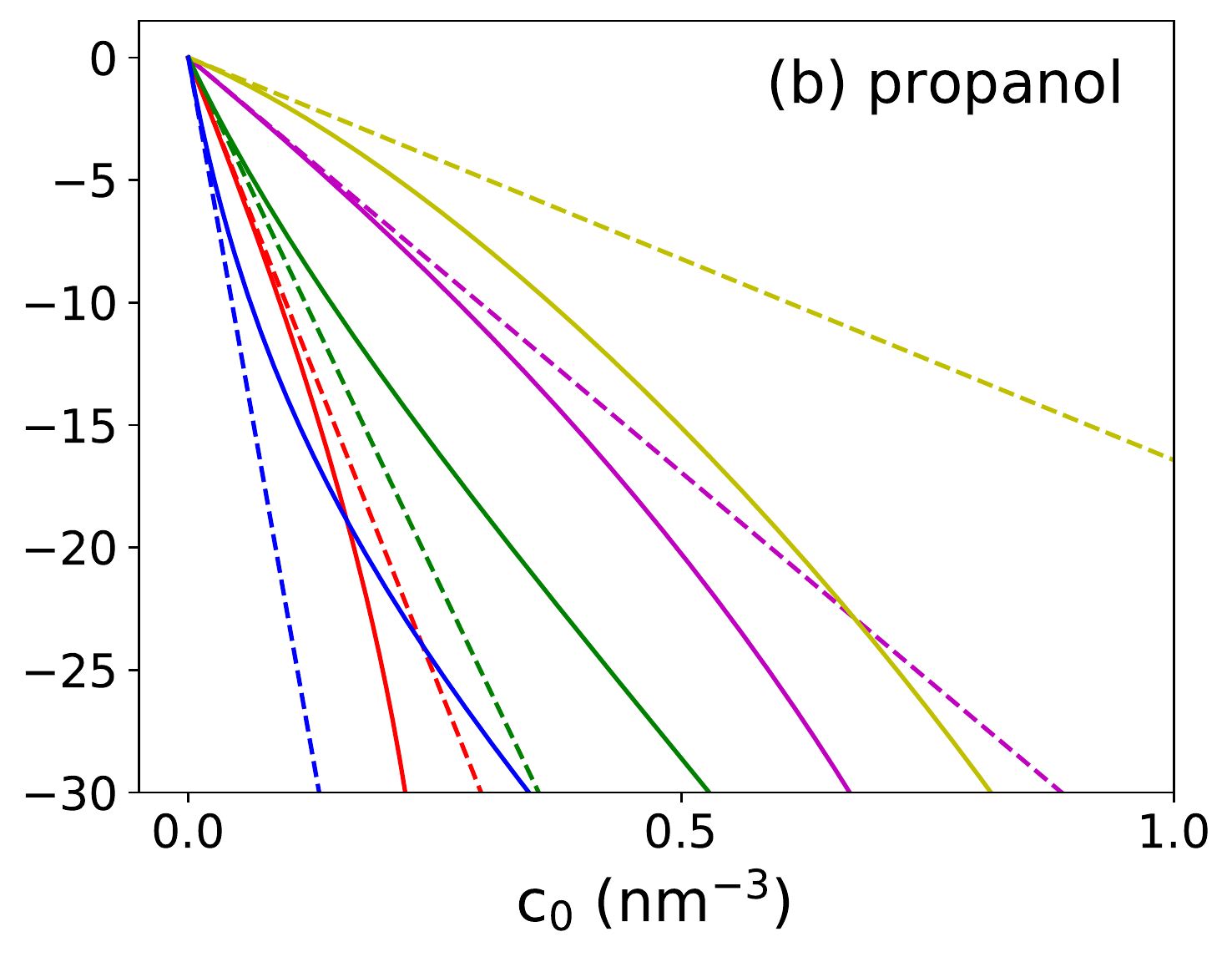}
   \includegraphics[width=0.30\textwidth]{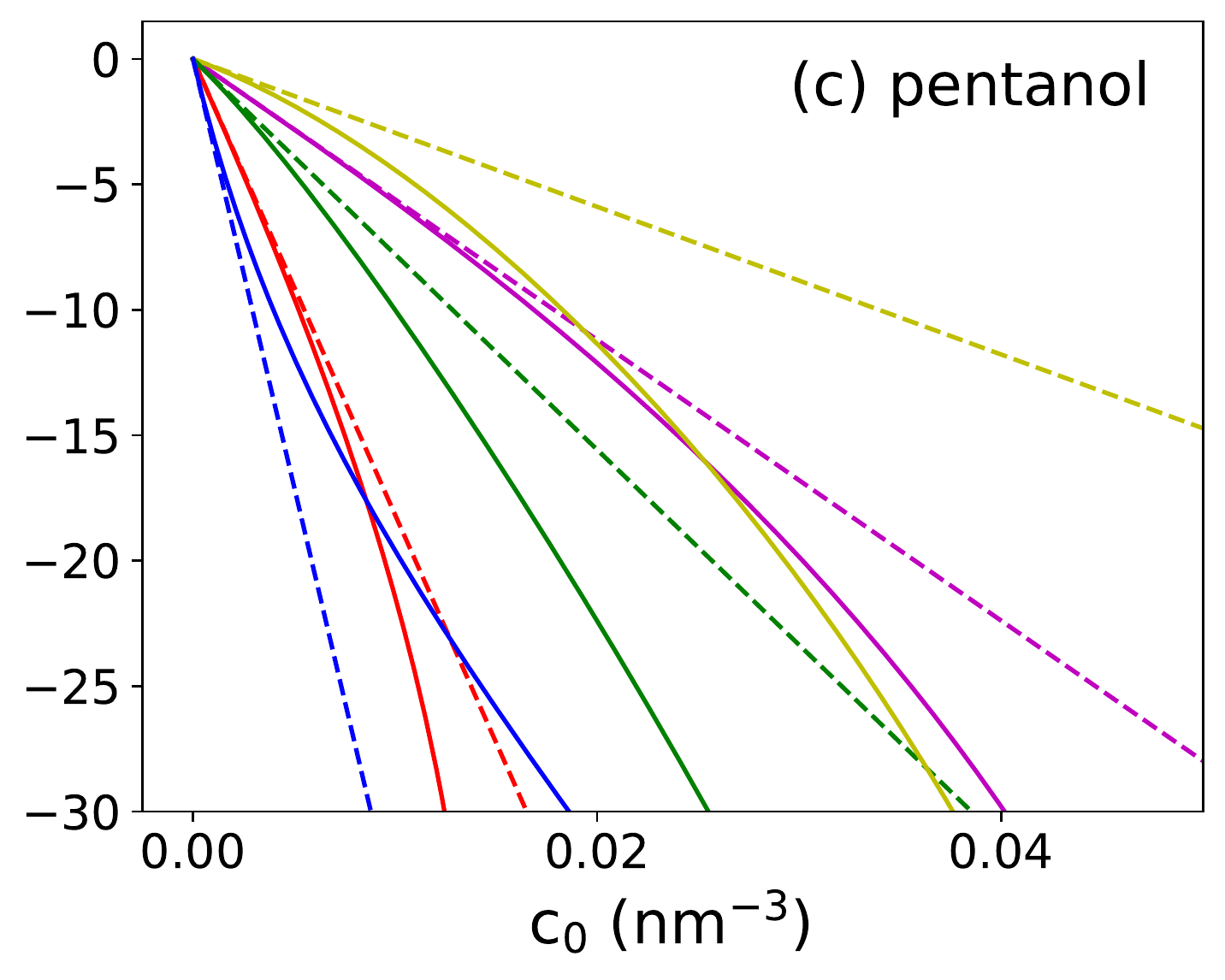}
\caption{
Change of the contact angle $\Delta \theta$ due to surfactant adsorption as a function of surfactant concentration for (a) methanol, propanol, and (c) pentanol on surfaces with different contact angles. Solid lines are predictions of \Eq~\eqref{eq:deltatheta_c0} and dashed lines are low-concentration predictions given by \Eq~\eqref{eq:deltatheta_c0_lin}.}
\label{fig:Dtheta_vs_c0}
\end{figure*}


In \Fig{fig:Dtheta_vs_c0}, we show the predictions of the contact angle change for all three alcohols and for different surface hydrophilicities, based on \Eq~\ref{eq:deltatheta_c0} (solid lines) and its linearized version, \Eq~\ref{eq:deltatheta_c0_lin} (dashed lines).
In \Eq~\ref{eq:deltatheta_c0}, we used \Eq~\ref{eq:gamma_vs_gamma} for calculating the surface tension reduction of both interfaces. 
We see that in all cases, the contact angle $\theta$ monotonically decreases with the bulk surfactant concentration in the droplet. That is, adding surfactant enhances wetting.
The $\Delta\theta$ vs.\ $c_0$ relation is altogether linear at first, as predicted by \Eq~\ref{eq:deltatheta_c0_lin}, and becomes nonlinear at higher concentrations: Sublinear on hydrophobic surfaces and superlinear on hydrophilic ones. 

\begin{figure}[t!]
  \centering
  \includegraphics[width=0.45\textwidth]{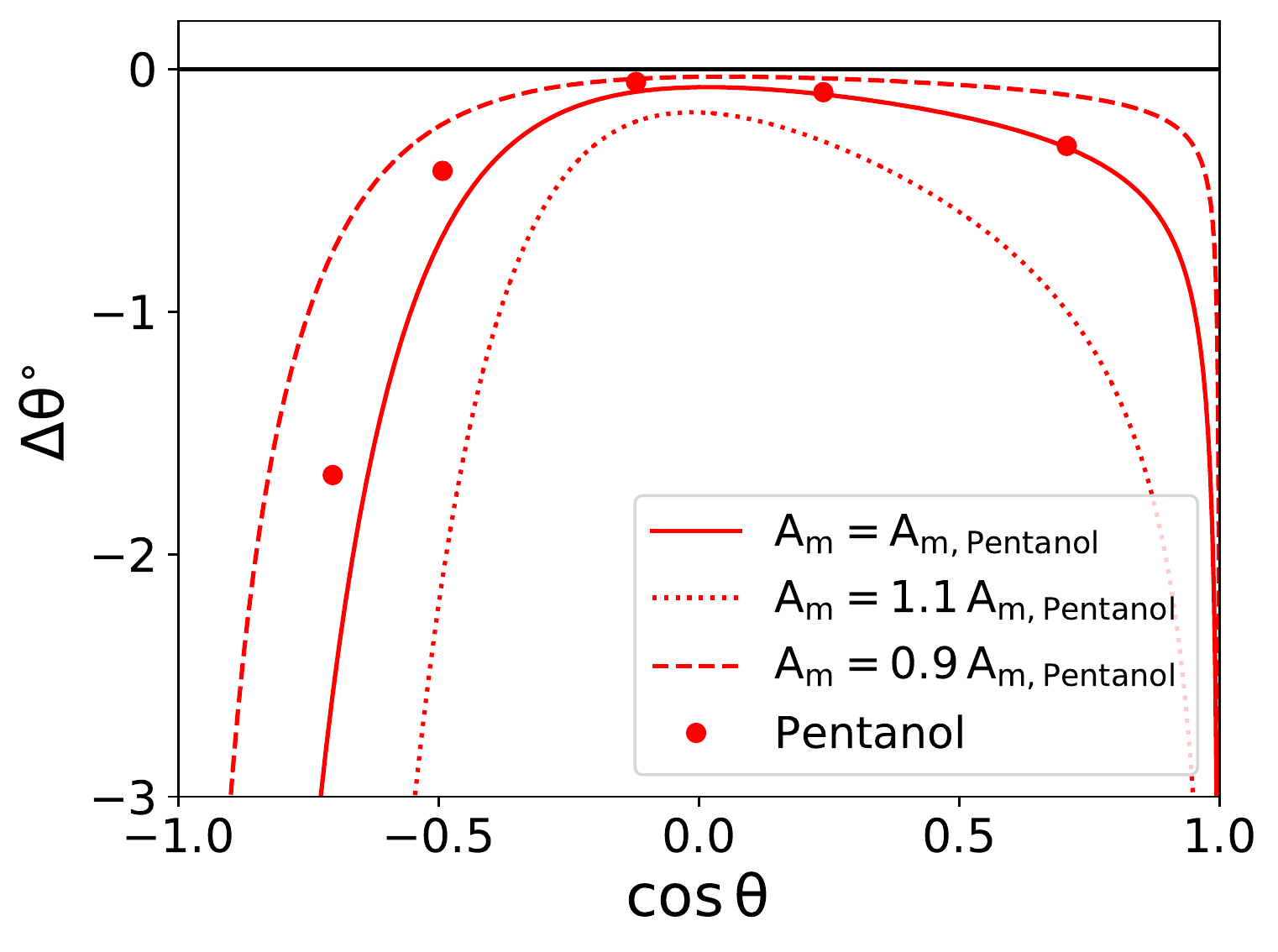}
\caption{Change in the water contact angle as a function of wetting coefficient for $c_0 = 0.01\,  \mbox{nm}^{-3}$ of added pentanol based on \Eq~\eqref{eq:deltatheta_c0_lin}.
The symbols are obtained by using $K_\trm{v}$ and $K_\trm{s}^{(0)}$ from the simulations of pentanol. The lines are obtained using the predictions of \Eqs~\eqref{eq:acostheta}, \eqref{eq:KsAm}, and~\eqref{eq:KvAm} for  $K_\trm{v}$ and $K_\trm{s}$ and three different values for the molecular surface area $A_\trm{m}$.}
\label{fig:Dtheta_vs_costheta}
\end{figure}

Interestingly, the change in contact angle drastically and nonmonotonically depends on the surface hydrophilicity, given by $\cos\theta$, as even better demonstrated in \Fig{fig:Dtheta_vs_costheta}.
The nonmonotonicity results from the competition between the adsorptions onto the water--vapor and solid--water interfaces of the droplet and is encoded in the numerator of \Eq~\ref{eq:deltatheta_c0_lin}, reading $K_\trm{s}(\theta)+K_\trm{v}\cos\theta$.
On considerably hydrophilic surfaces (small $\theta$), the adsorption of surfactants onto the surface is negligible (\ie, $K_\trm{s}\ll K_\trm{v}$) and thus the surfactant effect is dominated by the adsorption onto the water--vapor interface, dictated by the term $K_\trm{v}\cos\theta$ in \Eq~\ref{eq:deltatheta_c0_lin}. In this regime, the change in the contact angle scales as $\Delta\theta\propto -\cot\theta$. The effect of surfactant becomes extremely large for small contact angles, and it even diverges as the surface approaches the complete wetting regime ($\theta\to0^\circ$). 
This means that already low concentrations of surfactant in a low-contact angle droplet can easily push the droplet into the complete wetting regime. This observation also suggests that measurements of small contact angles are particularly challenging because of potential contamination of aqueous systems with surface-active molecules~\cite{CHANG19951, an2015wetting, diaz2022water}.

With increasing hydrophobicity (increasing $\theta$), the surface adsorption coefficient $K_\trm{s}$ rapidly increases (see \Fig{fig:gamma_gc_a_xi}b and \Eq~\ref{eq:acostheta}) and eventually exceeds $K_\trm{v}$. Thus, $|\Delta\theta|$ starts dramatically rising with the surface hydrophobicity.
Our analysis also shows that surfaces with contact angles around $\theta=90^\circ$ are the least sensitive to wetting alterations due to surfactants as compared to hydrophilic or hydrophobic surfaces.

Markedly, the net effect of adding simple alcohols to water is always to decrease the contact angle of the droplet ($\Delta\theta<0$), even though this is not strictly imposed by \Eq~\ref{eq:deltatheta_c0_lin}. Most experimental studies show that surfactants decrease the contact angle of aqueous solutions on hydrophobic surfaces~\cite{VelardeJCIS2000, JanczukJCIS2006, BinksSM2010surfButt}.
Theoretically, the effect could be positive ($\Delta\theta>0$) for hydrophobic surfaces (for which $\cos\theta<0$) if the adsorption onto the surface remains small, such that $K_\trm{s}>-K_\trm{v}\cos\theta$, which is, however, not the case in our systems.

\section{Conclusions}
We investigated the adsorption of short-chained alcohols (simple surfactants) onto two types of interfaces, the water--vapor and solid--water. The solid surface included various levels of polarity, which manifested in a broad span of water contact angles. 
We have analyzed how the adsorption depends on the surface contact angle. Adsorption properties calculated from MD simulations follow very well the Langmuir adsorption isotherm for most of the studied cases, except for some smaller deviations in the case of pentanol. We have found that the adsorption coefficient onto the solid surface scales roughly exponentially with the molecular surface area and the surface wetting coefficient (\Eq~\ref{eq:acostheta}). Based on the continuum-level description, this dependence arises from the free energy of removing water molecules from the contact area of the surface with the adsorbing surfactant molecule. 
In addition, we have shown a unique and practical correlation between the water--vapor and the water--solid adsorptions (\Fig{fig:gamma_gc_a_xi}c), which can also assist in estimating the solid--water adsorption based on the water--vapor adsorption.

The competition between the adsorption at the solid--water and water--vapor interface manifests by far the most directly in the contact angles of water droplets containing surfactants. Our theoretical analysis suggests that the short-chained alcohols in all our cases reduce the contact angle and, with that, enhance wetting. The wetting enhancement depends drastically and non-monotonically on the wetting coefficient. We have found high sensitivity of the contact angle on surfactant concentration on very hydrophilic surfaces, which can easily cross the complete wetting regime. A strong influence of surfactants is also observed for very hydrophobic surfaces, owing to strong surfactant adsorption to the surface. In contrast, mildly polar surfaces, with contact angles around 90$^\circ$, are much less sensitive to wetting alterations.

The outcomes of this study are also relevant to further experimental efforts at the nanoscale, in which tuning the surfactant concentration is used to improve liquids' sticking or coverage on solids, such as in detergency, inkjet printing, or pesticide spraying. Surface-active pharmacological agents could be used to promote liquid spreading and treat or prevent bubble obstruction of blood flow~\cite{eckmann2001wetting}. Surfactant uptake is also crucial for self-cleaning processes of hydrophobic surfaces by water drops~\cite{ButtSAdv2020sc}. 
Finally, making surface-active molecules charged brings about numerous electrochemical phenomena, which reflect, for instance, in zeta potential, nanobubble stability, and the Jones--Ray effect~\cite{uematsu2019impurity}, and which add a layer of complexity to wetting and would be interesting to investigate in future studies.


\subsection*{Acknowledgments}
We acknowledge funding from the Slovenian Research Agency ARRS (Research Core Funding No.\ P1-0055 and Research Grant No.\ J1-1701).

\bibliography{adsorption_at_the_nanoscale}


\end{document}


\pagenumbering{arabic}
\noindent

\parindent=0cm
\setlength\arraycolsep{2pt}

\twocolumn[	
\begin{@twocolumnfalse}
\oldmaketitle

\end{@twocolumnfalse}]

\section{Simulations of bulk properties of the solutions}\label{sec:suppl_bulk}

\begin{figure*}
  \centering
\includegraphics[width=0.32\textwidth]{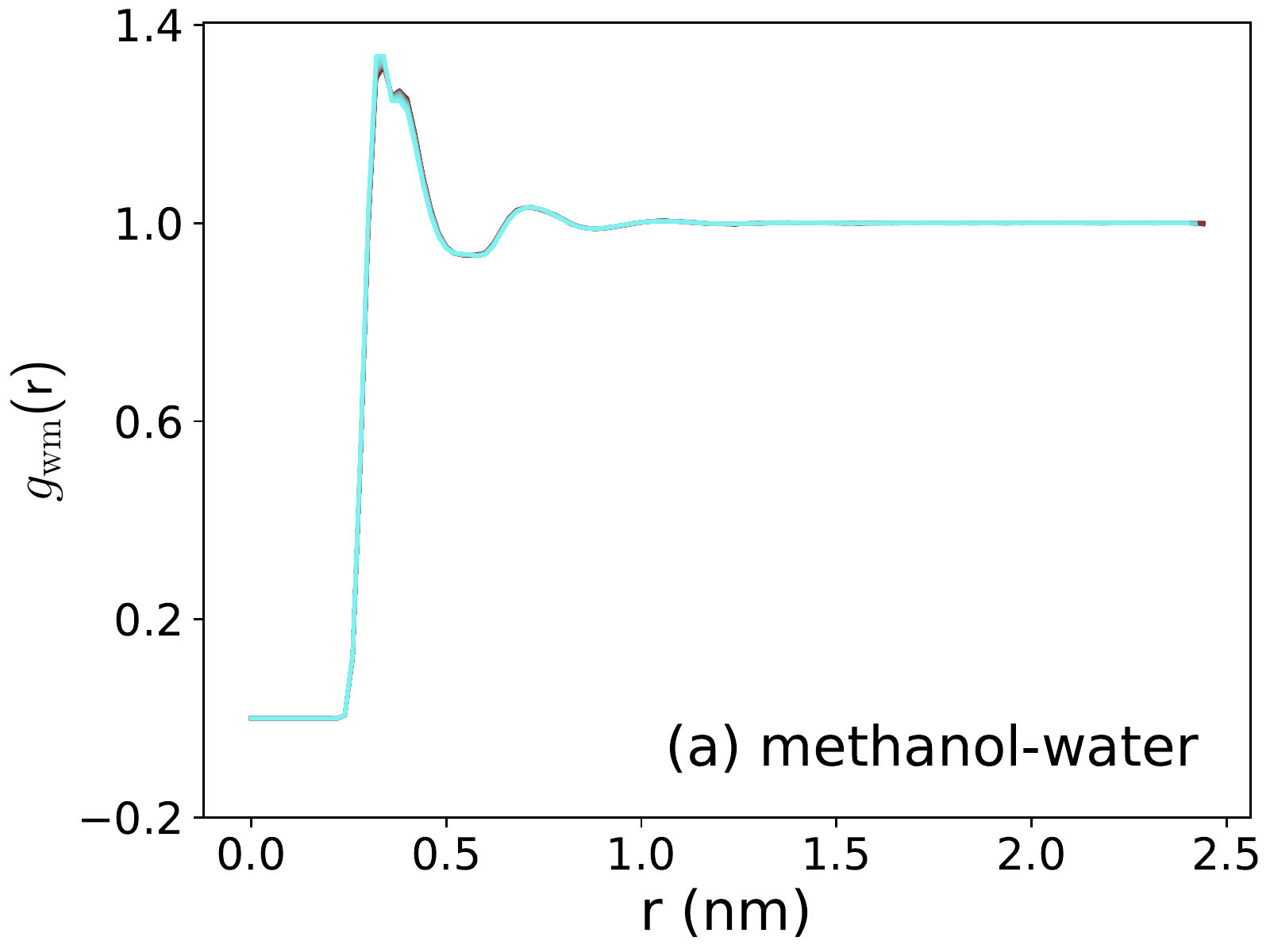}
\includegraphics[width=0.32\textwidth]{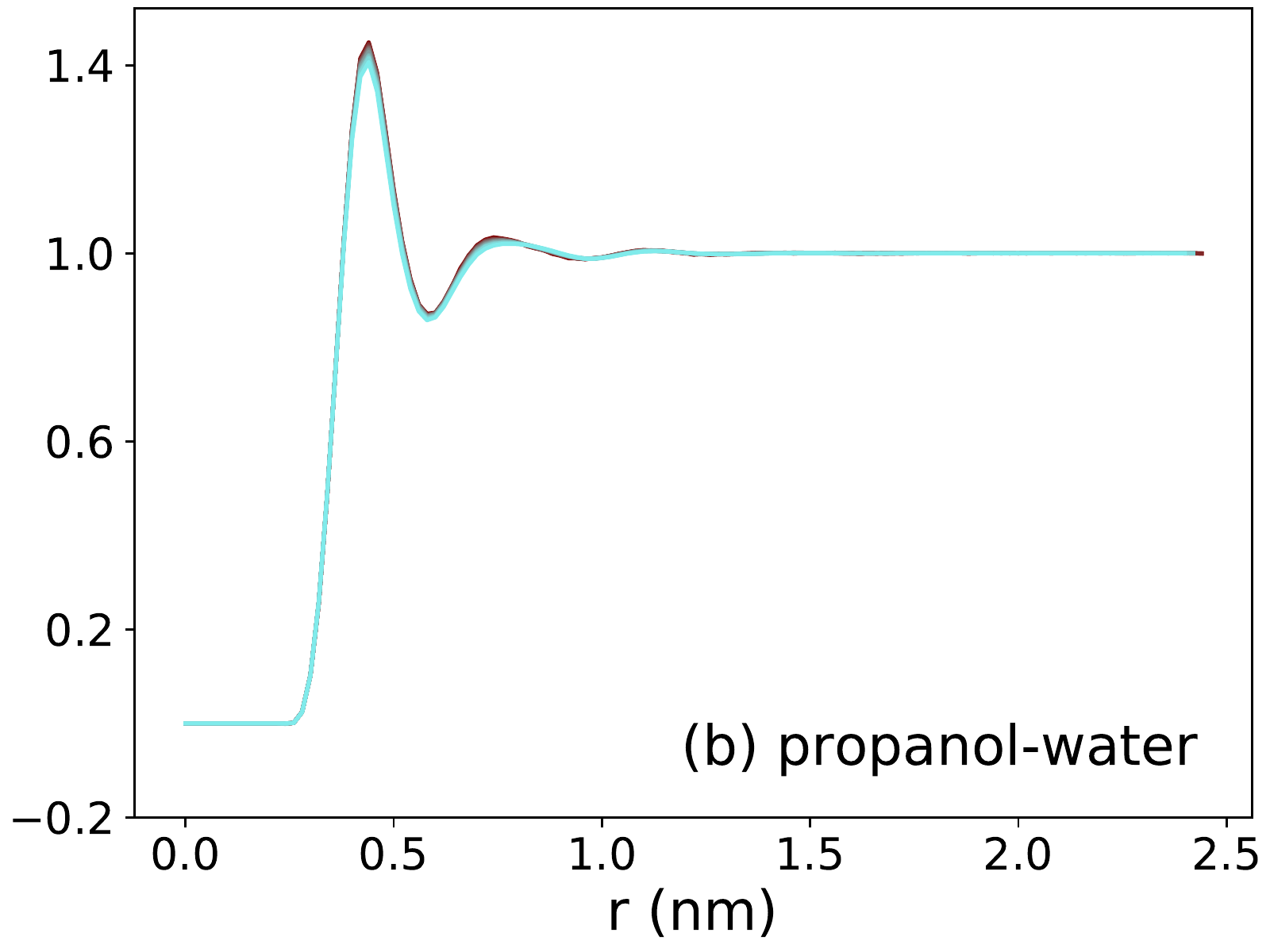}
\includegraphics[width=0.32\textwidth]{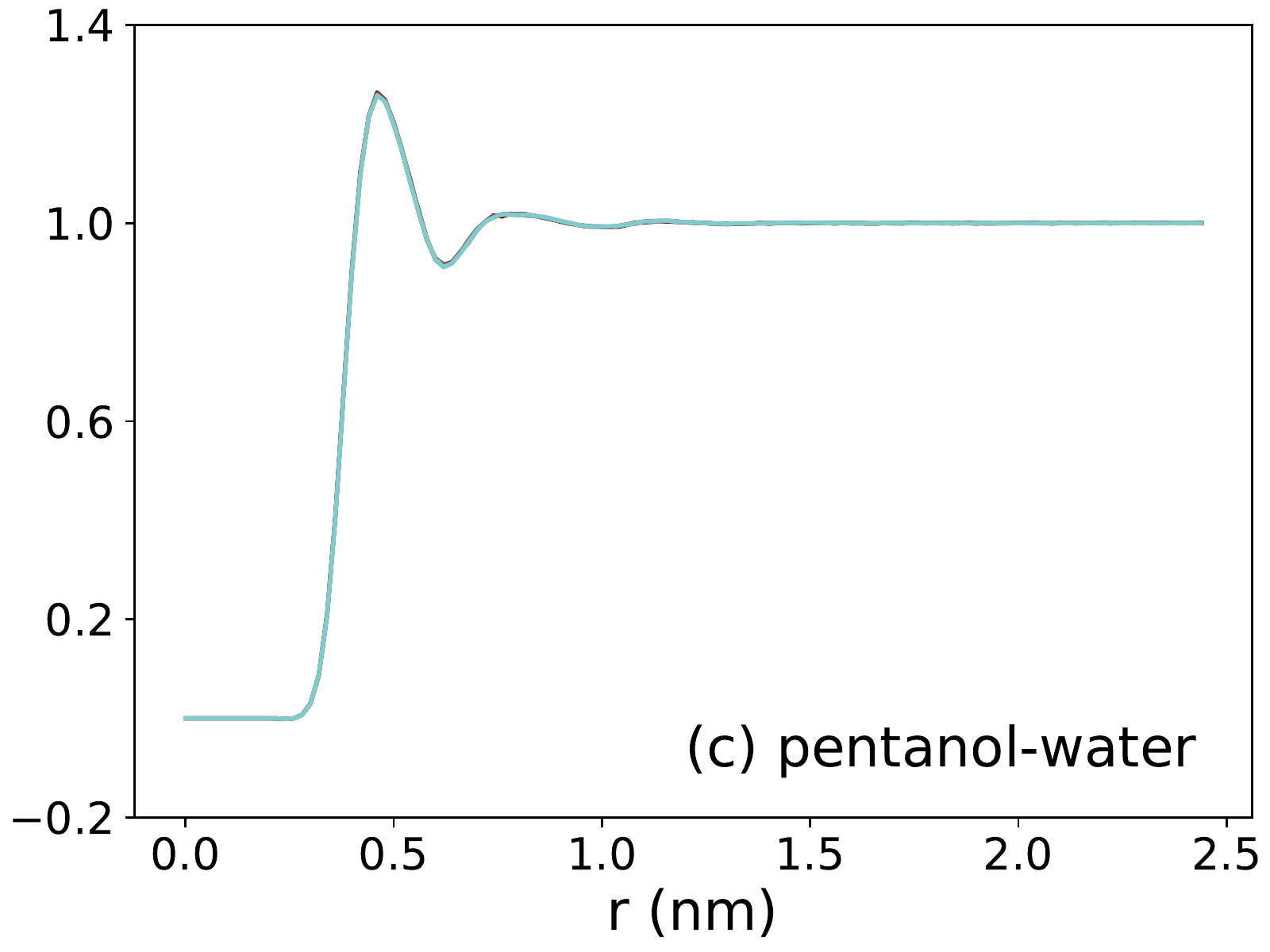}\\
\includegraphics[width=0.32\textwidth]{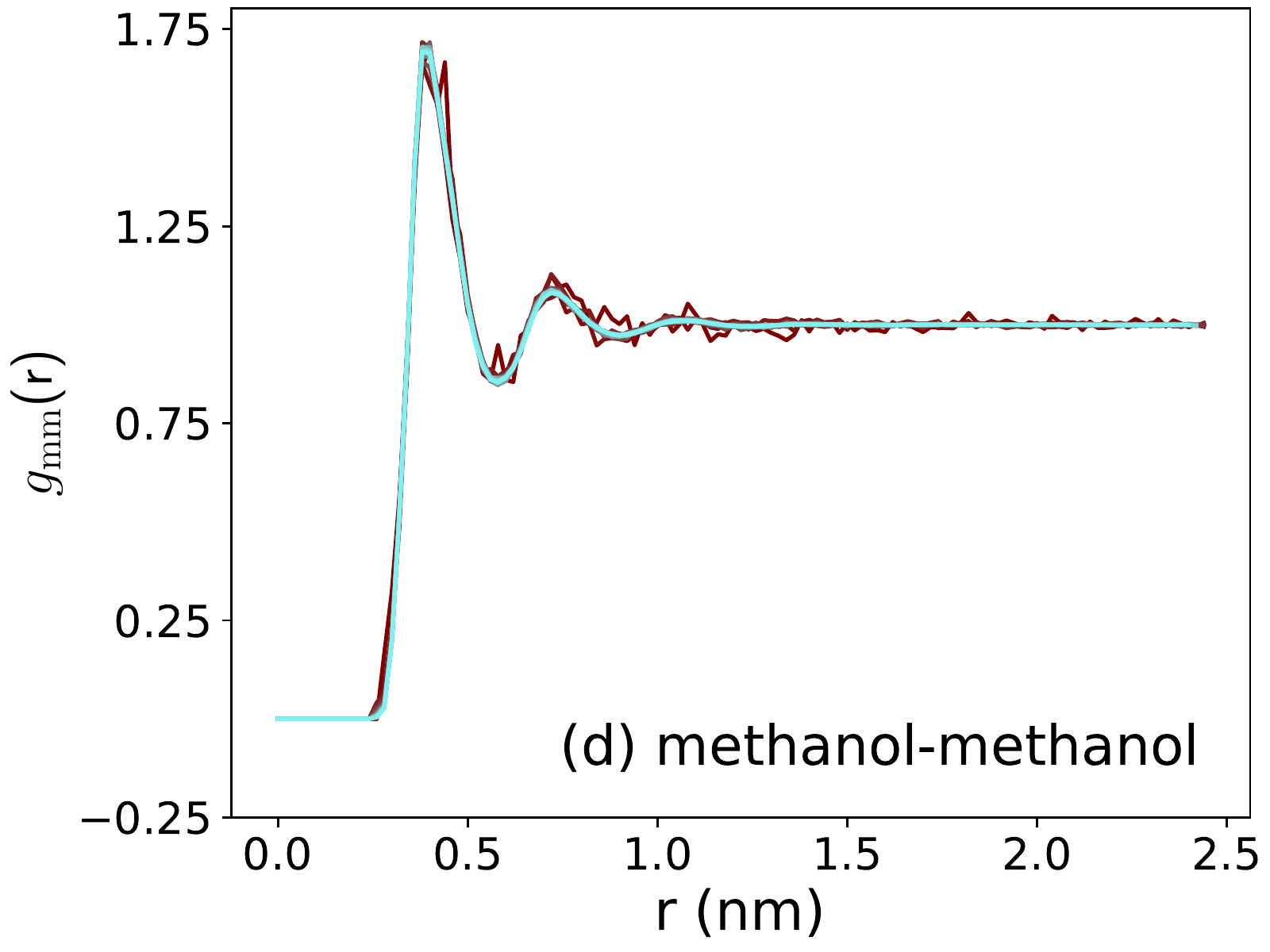}
\includegraphics[width=0.32\textwidth]{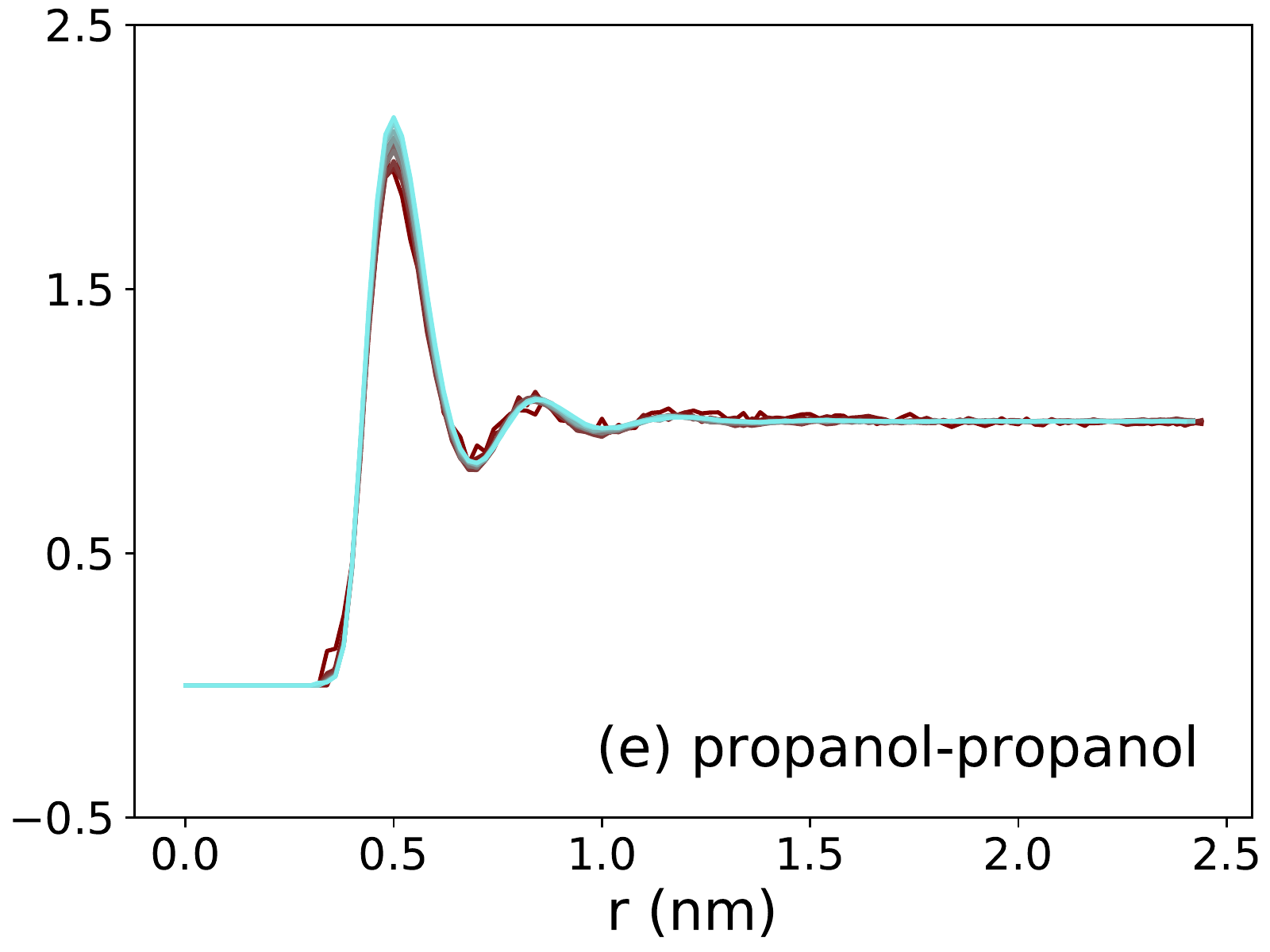}
\includegraphics[width=0.32\textwidth]{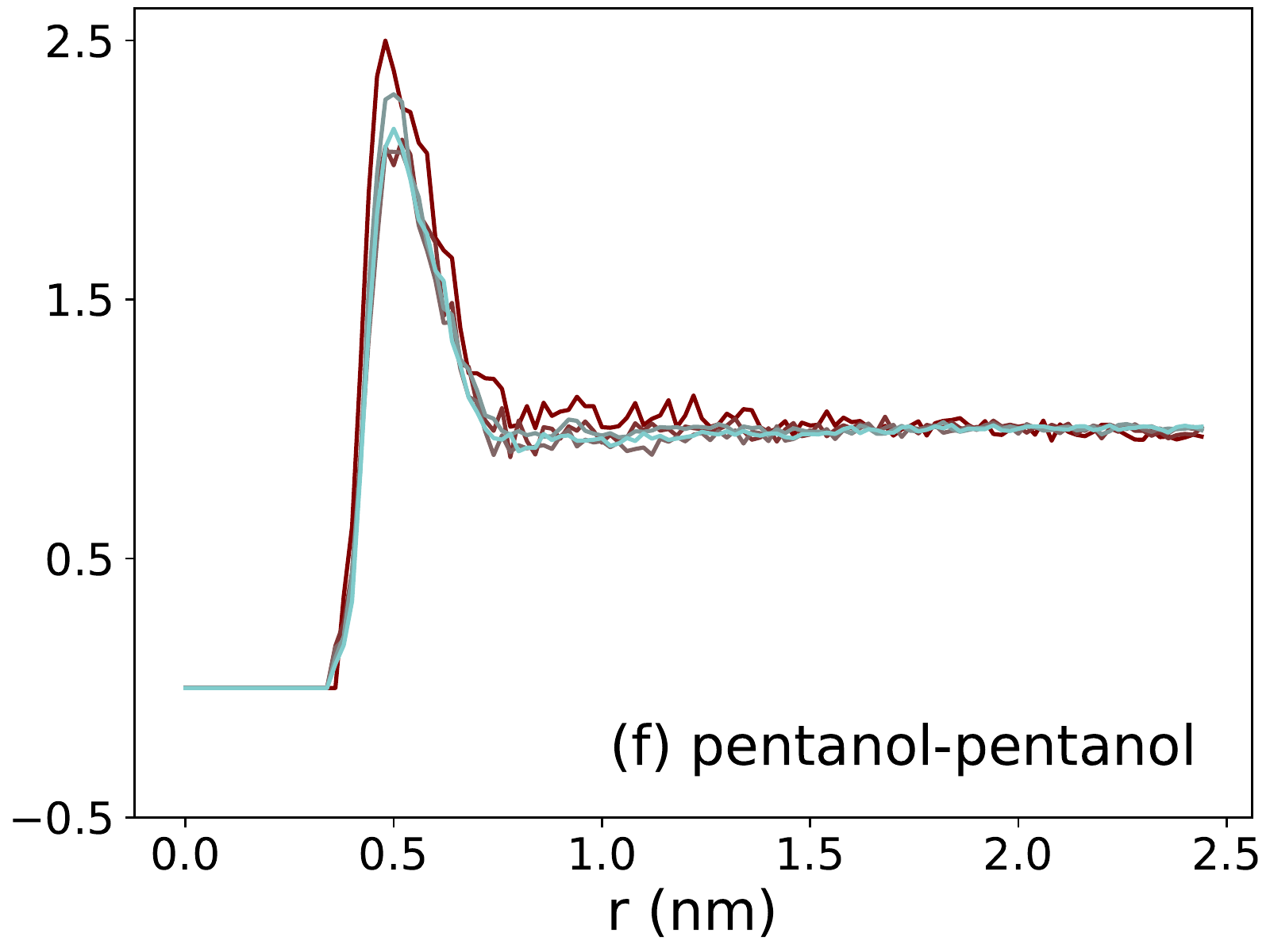}	\caption{Radial density functions between water and surfactant molecules (top row) and between the surfactant molecules themselves (bottom row). Different surfactant densities are shown in different colors. \label{fig:correlations_bulk}}
\end{figure*}

\begin{figure*}
  \centering
\includegraphics[width=0.32\textwidth]{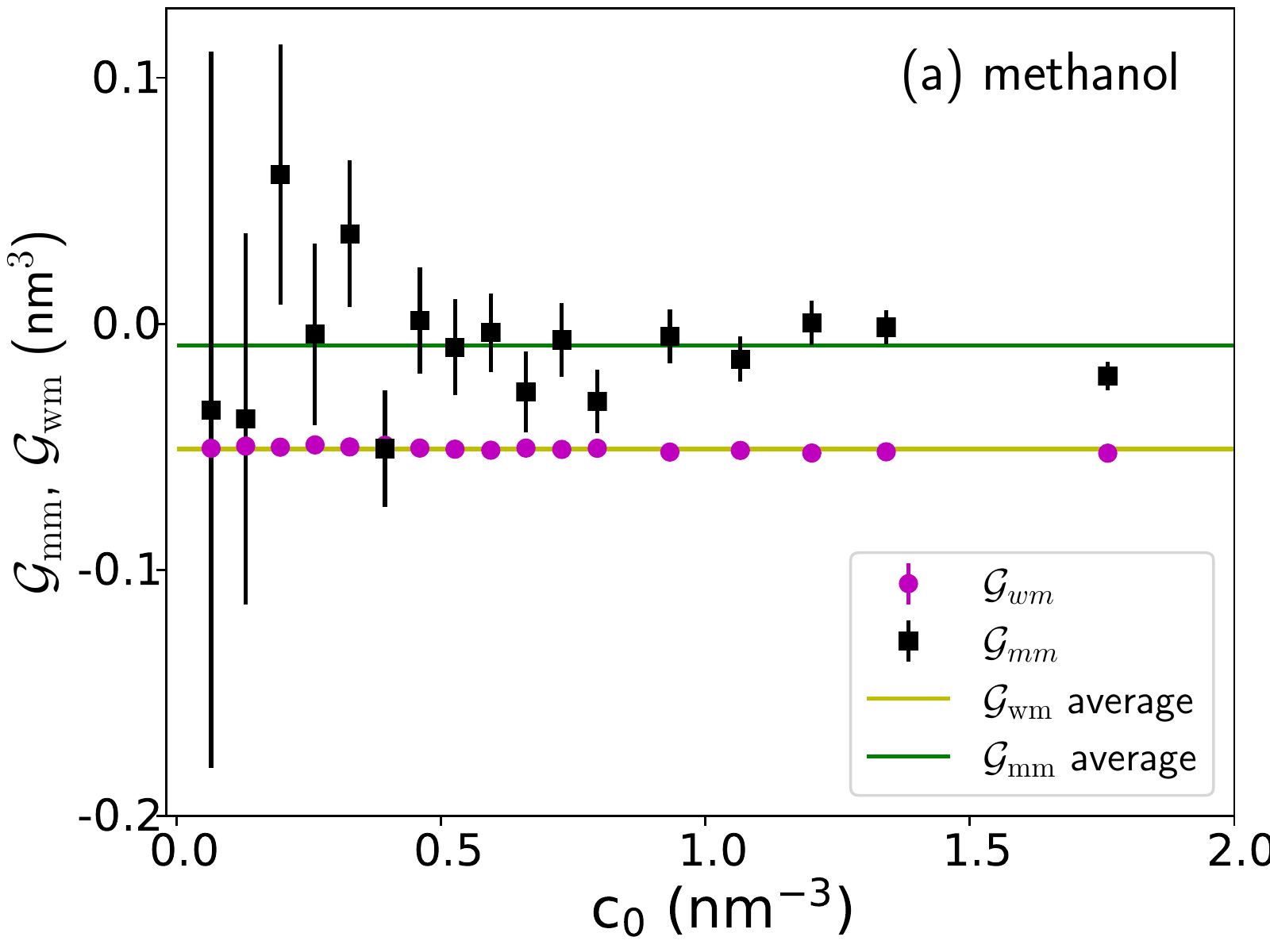}
\includegraphics[width=0.325\textwidth]{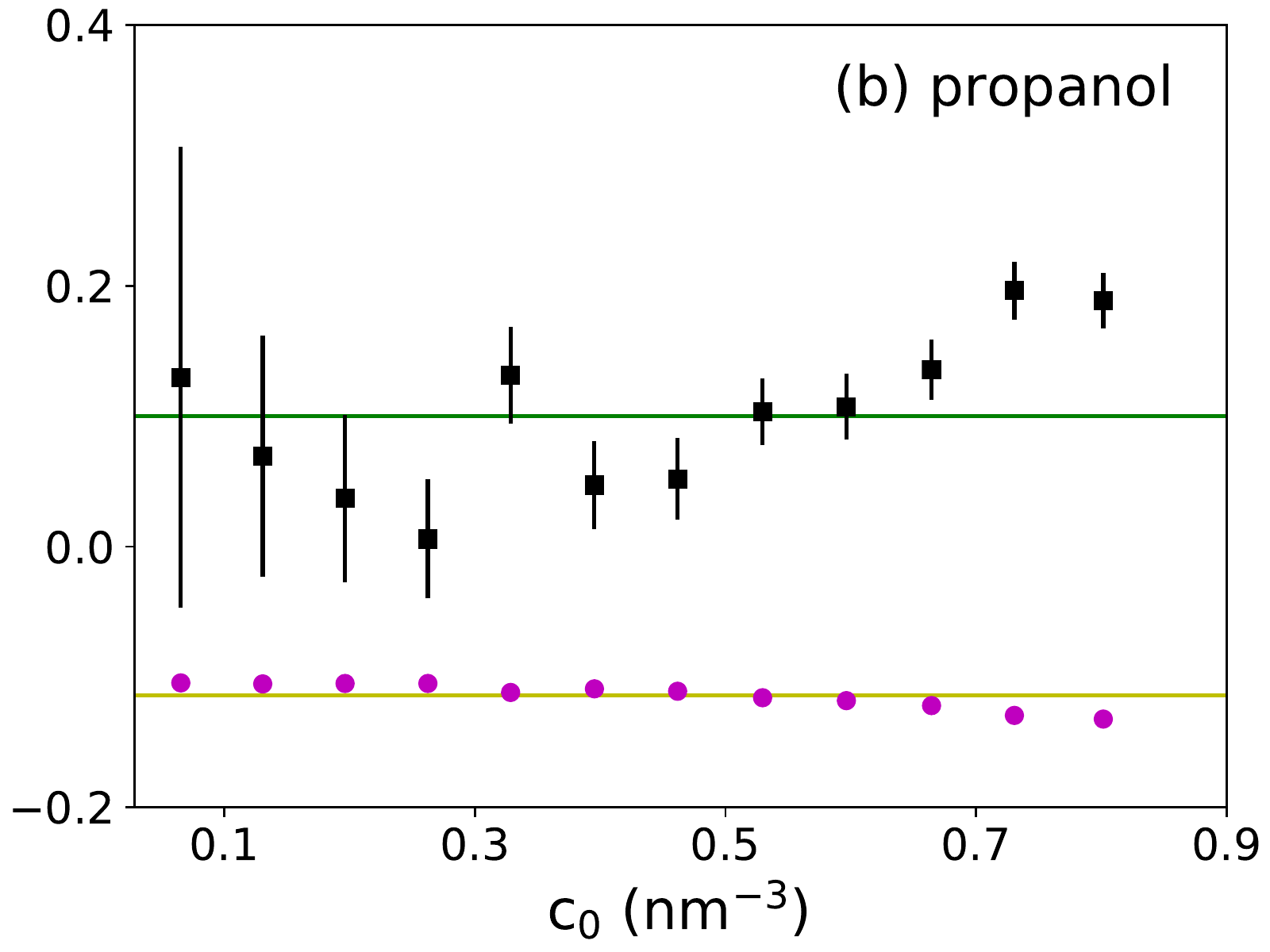}
\includegraphics[width=0.32\textwidth]{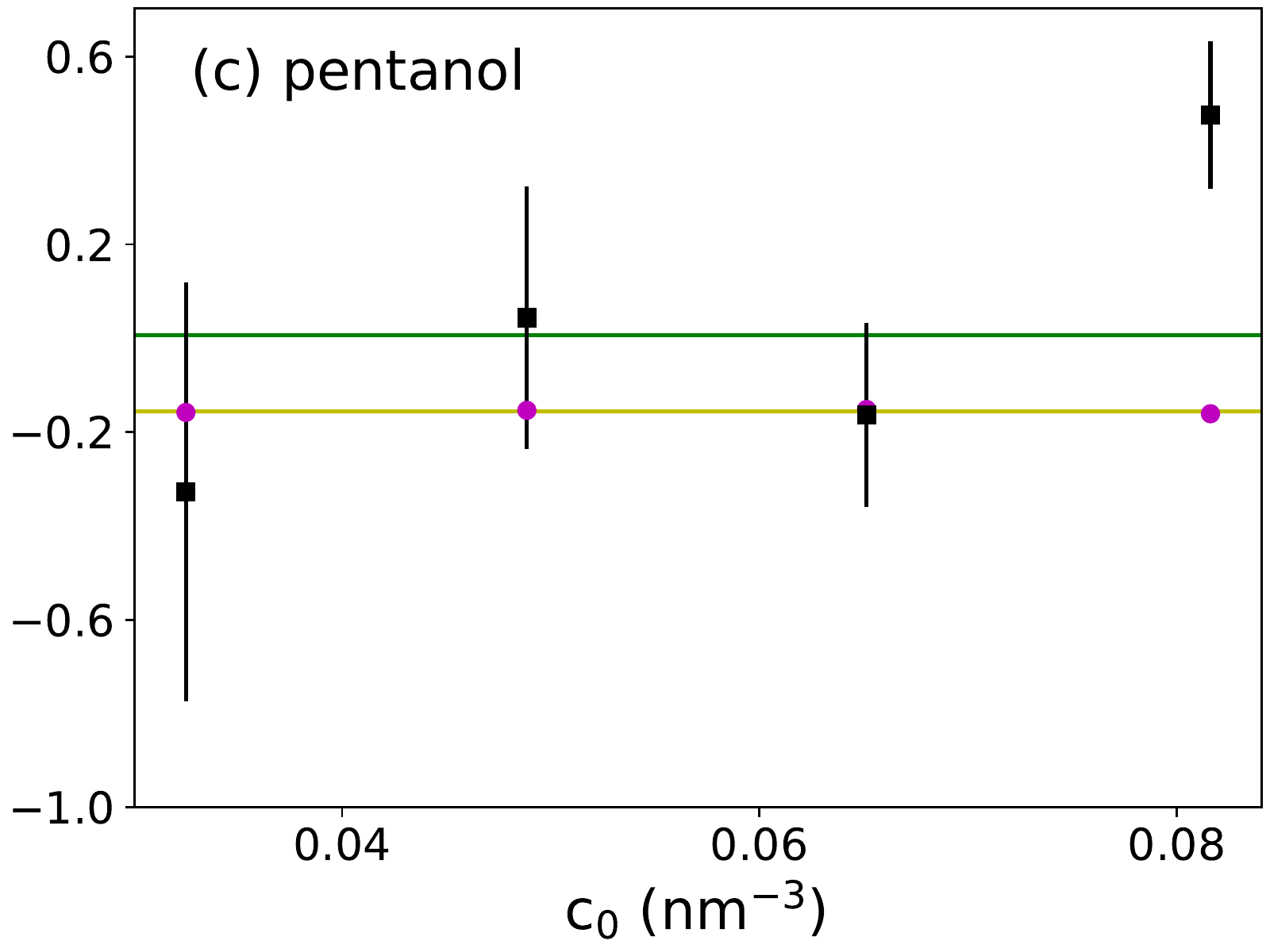}
	\caption{Parameters ${\cal G_\trm{wm}}$ and ${\cal G_\trm{mm}}$ calculated from  eq.~(5) in the main text from the data in \Fig{fig:correlations_bulk} as a function of surfactant concentration. The horizontal solid lines indicate the averaged values for each data set. \label{fig:G_vs_rho}}
\end{figure*}

To calculate the bulk properties necessary for the Kirkwood--Buff theory used in sec.~3.1, we simulate three independent realizations of a homogeneous system in a cubic simulation box of edge $5$~nm in the NPT ensemble for $100$~ns for each concentration of surfactant. 
From the simulations, we extract the radial density functions $g_\trm{mw}$ and $g_\trm{mm}$, shown in Fig.~\ref{fig:correlations_bulk}, and use them in eq.~(5) to calculate ${\cal G_\trm{mw}}$ and ${\cal G_\trm{mm}}$. 
The latter quantities are shown in Fig.~\ref{fig:G_vs_rho} as a function of the surfactant concentration. Both ${\cal G_\trm{mw}}$ and ${\cal G_\trm{mm}}$ are independent of the concentration at low concentrations (within the numerical incertitude), which justifies our approximation.

\section{Adsorption at a surface} \label{sec:suppl_SAM}
In~\Figs{fig:gama_c0_surf_meth}, \ref{fig:gama_c0_surf_prop}, and \ref{fig:gama_c0_surf_pent}, we show the comparison between MD simulations results and Langmuir~(2) and Henry~(3) isotherms  for the values of $\theta$ not shown in Fig.~6. 
The parameters of the Langmuir isotherm, $\Gamma_{\infty}$ and $k_c$, have been fitted, while the coefficient $K_\trm{s}$ for Henry's law has been obtained from the fit of the Langmuir isotherm as $K = \Gamma_{\infty} \, k_c$.
The outcome of these fits is shown in Figs.~7 and \ref{fig:gamma_vs_costheta}.

\begin{figure*}
  \centering
  \includegraphics[width=0.33\textwidth]{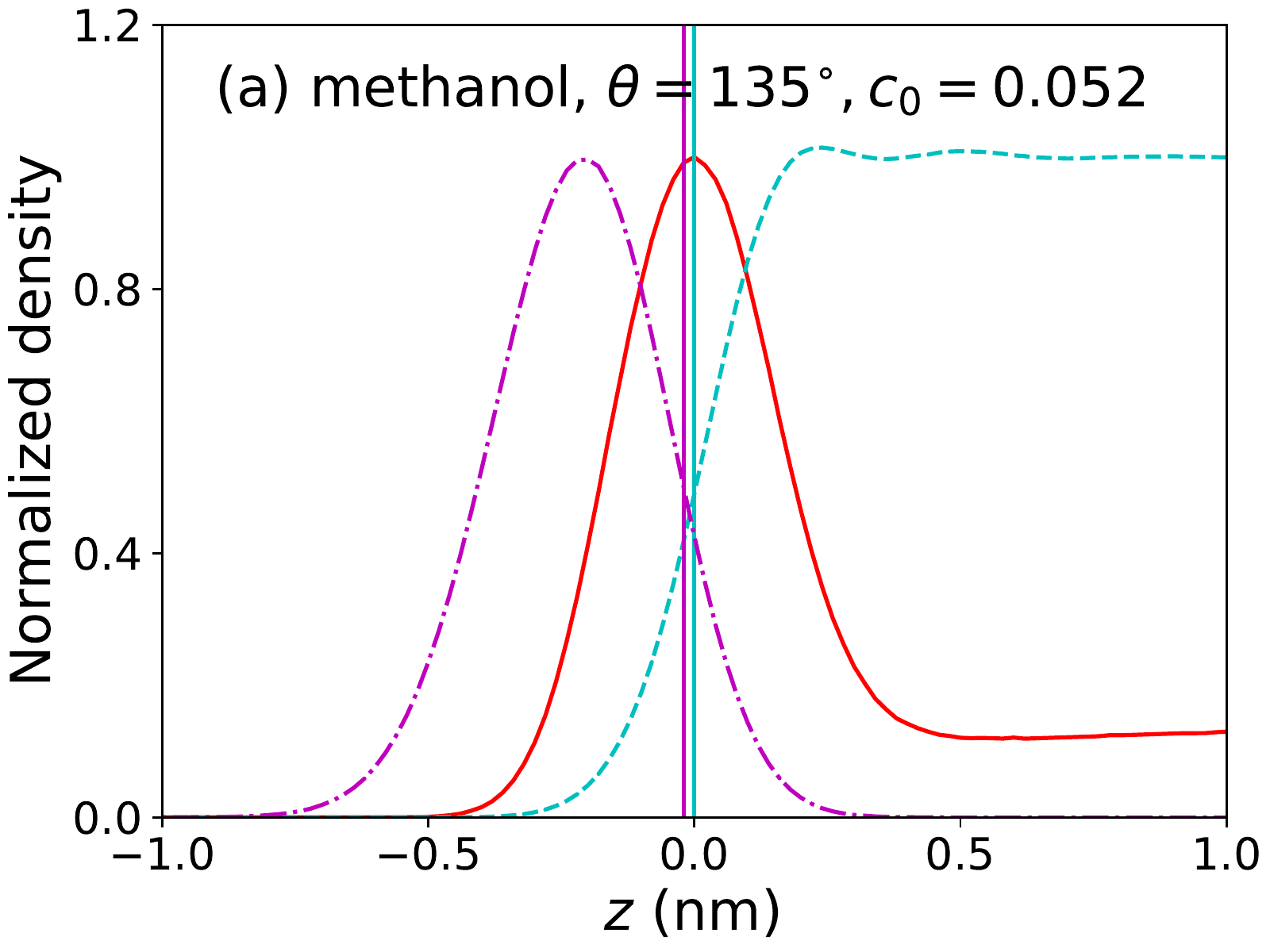}
\includegraphics[width=0.33\textwidth]{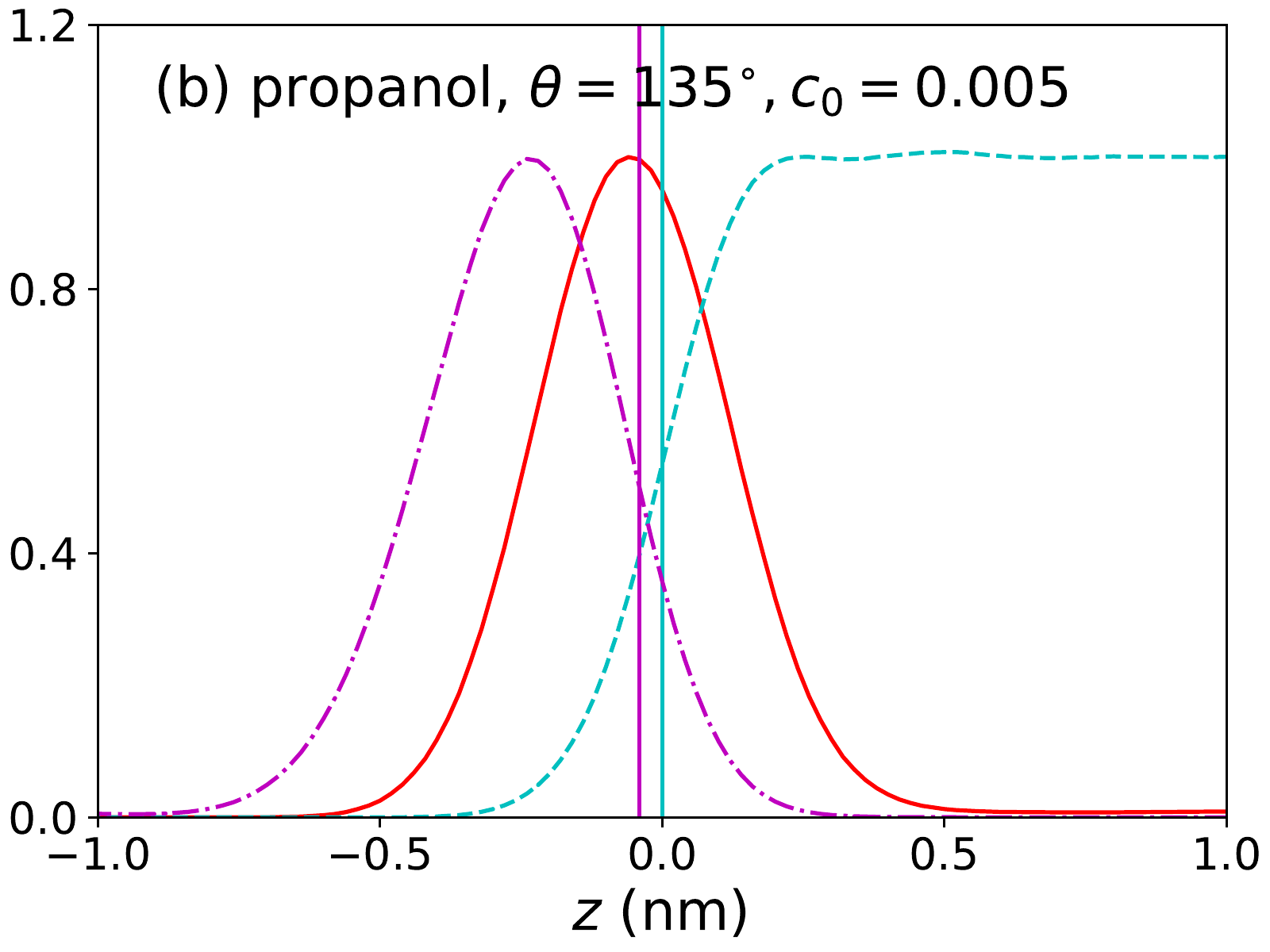}\\
  \includegraphics[width=0.33\textwidth]{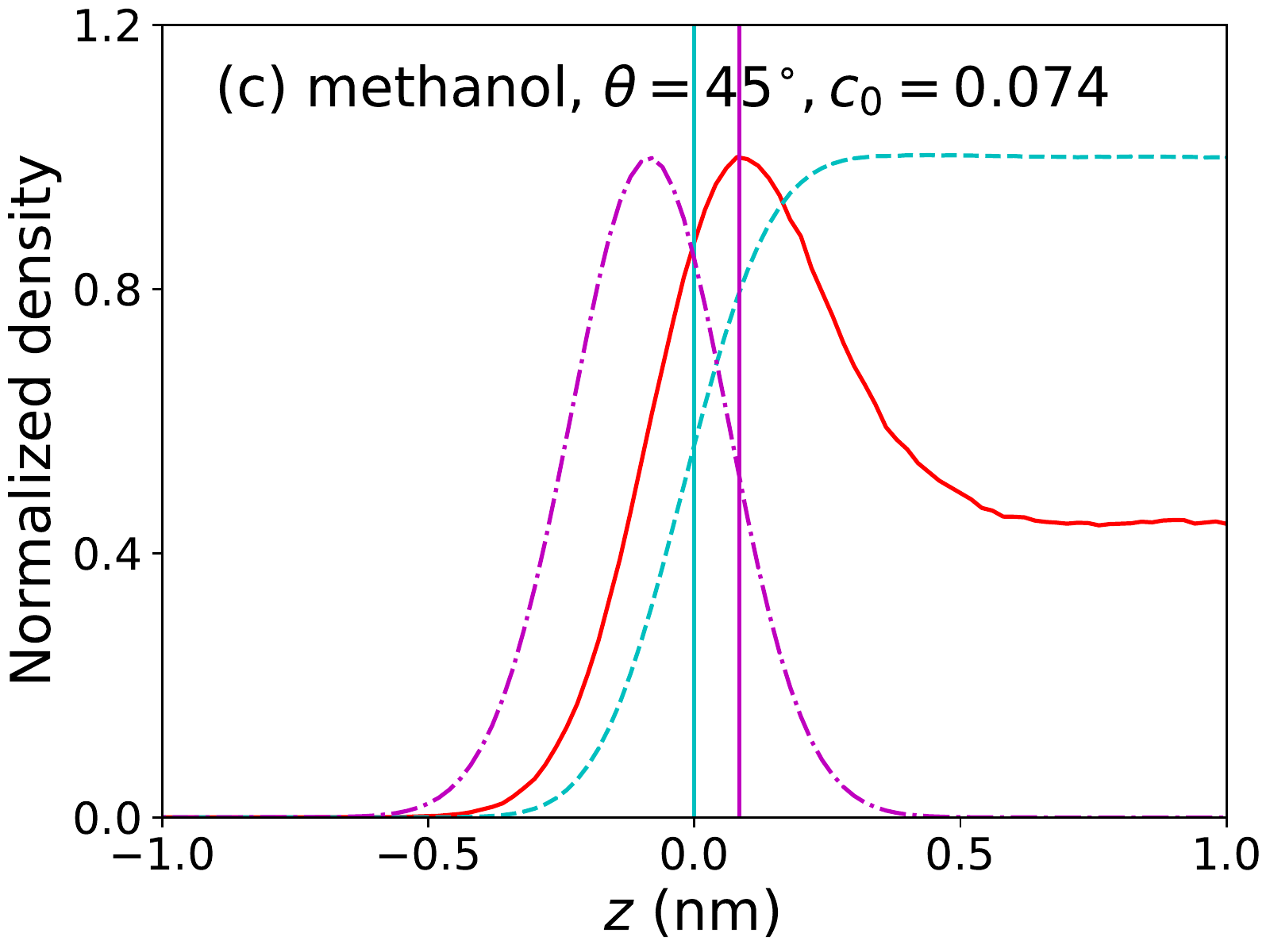}
\includegraphics[width=0.33\textwidth]{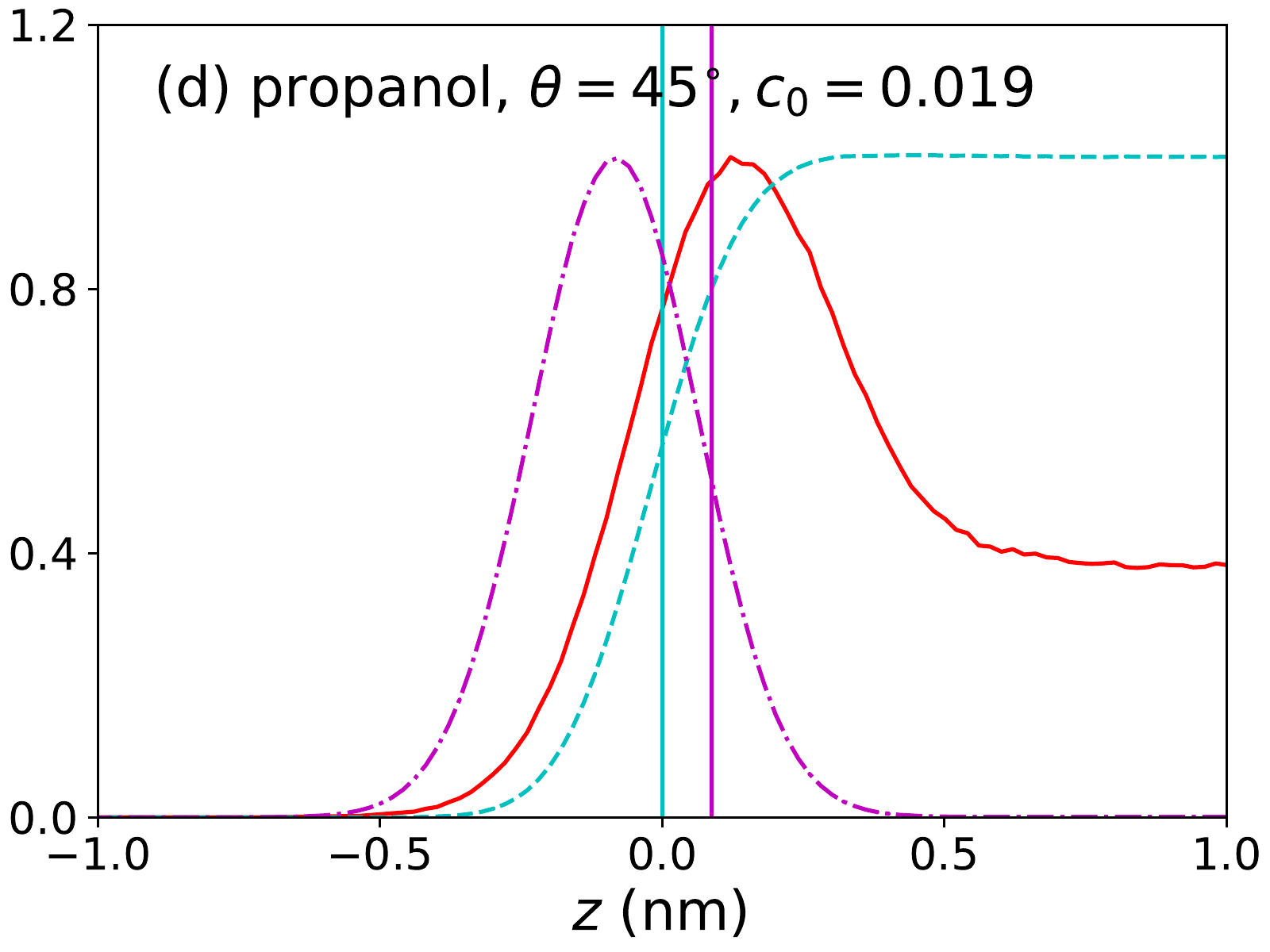}
\includegraphics[width=0.33\textwidth]{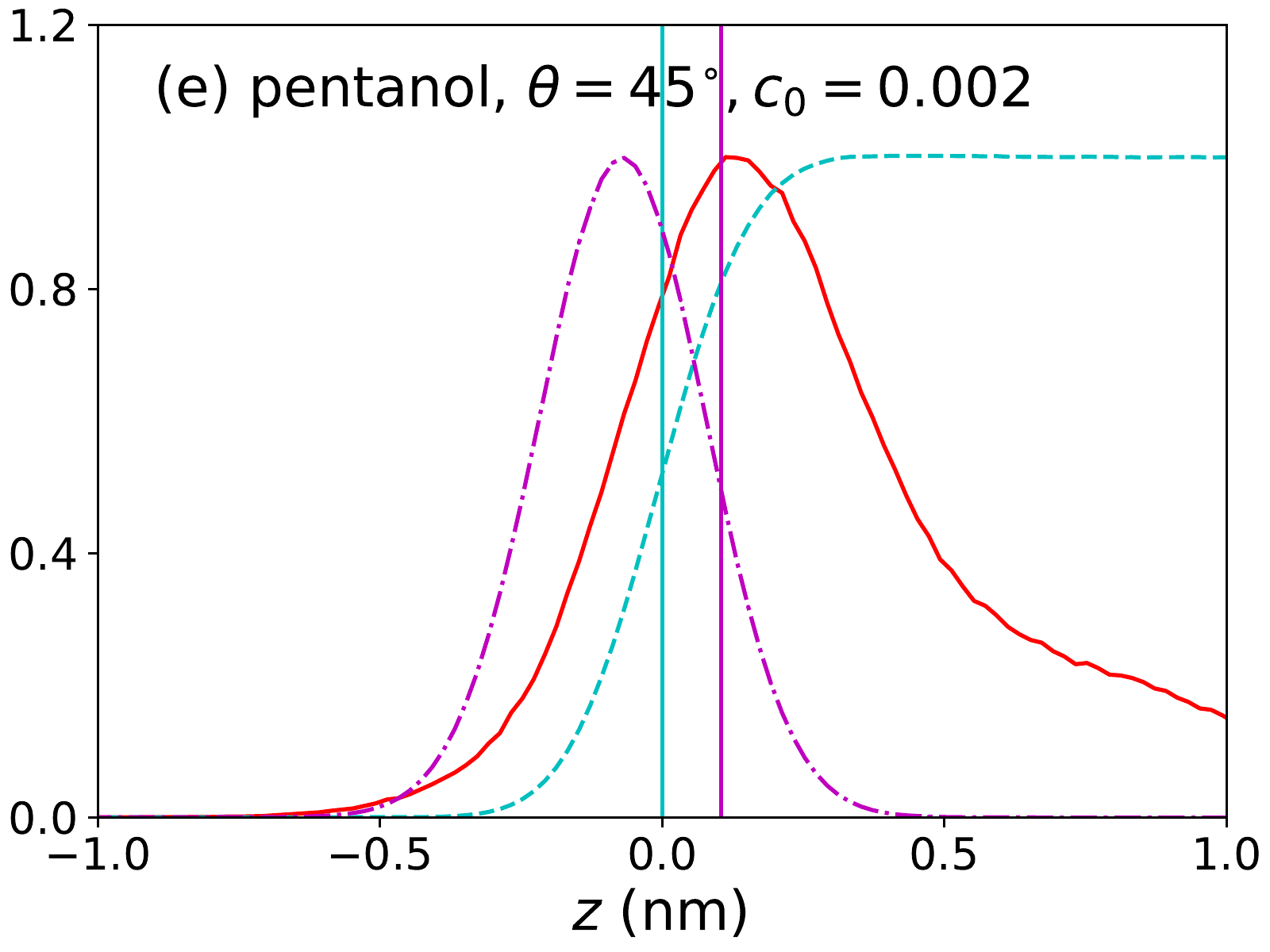}
	\caption{
	 Rescaled density profiles of surfactants (solid red lines), the surface OH groups (magenta dash-dotted lines), and water (cyan dashed lines). 
   	 Effective phase boundaries are depicted by the Gibbs dividing surface for water (cyan) and the position at half height on the water side of the OH group (magenta).
 \label{fig:norm_dens_SAM}}
\end{figure*}

\begin{figure*}
  \centering
  \includegraphics[width=0.4\textwidth]{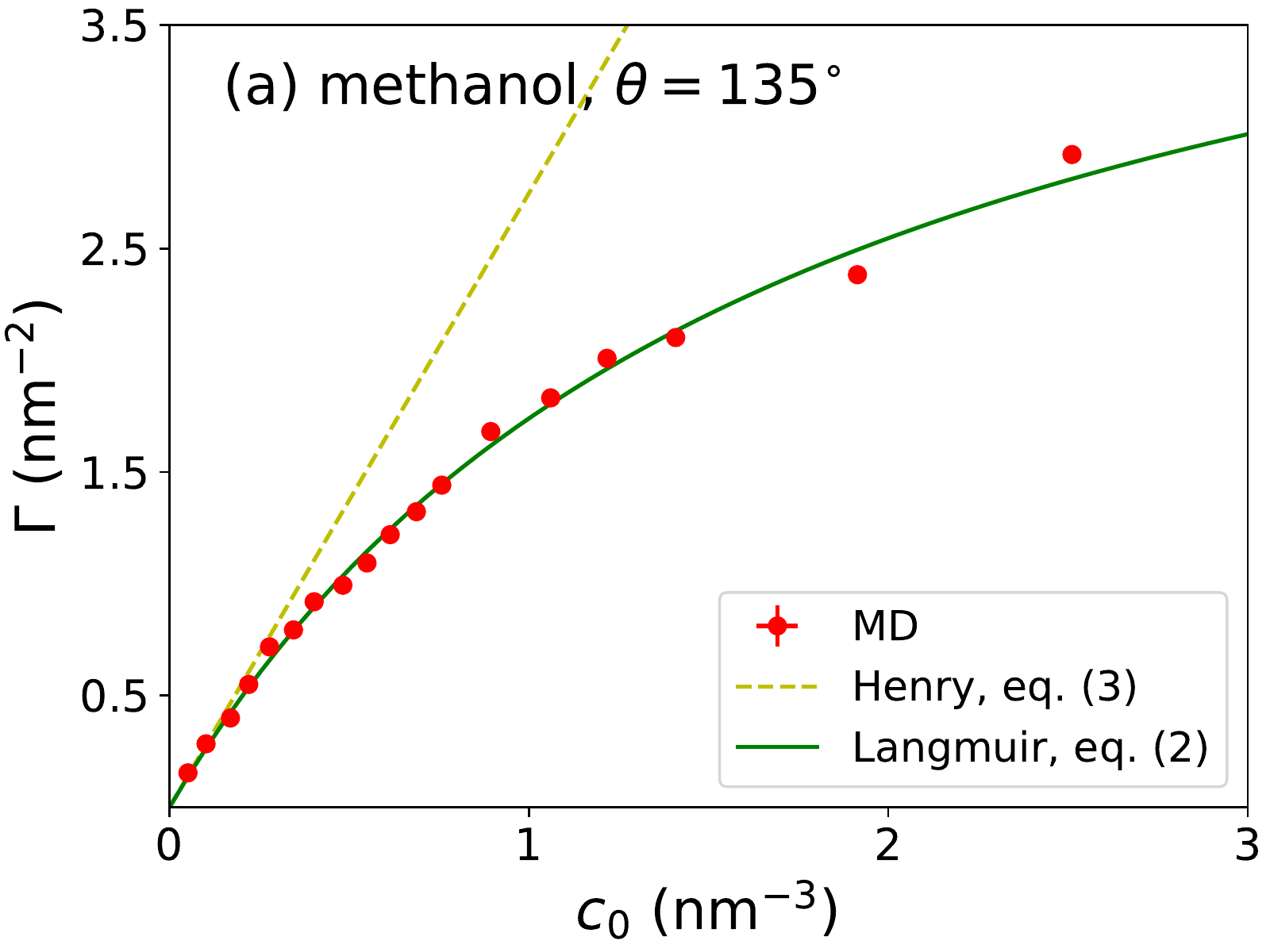}
\includegraphics[width=0.4\textwidth]{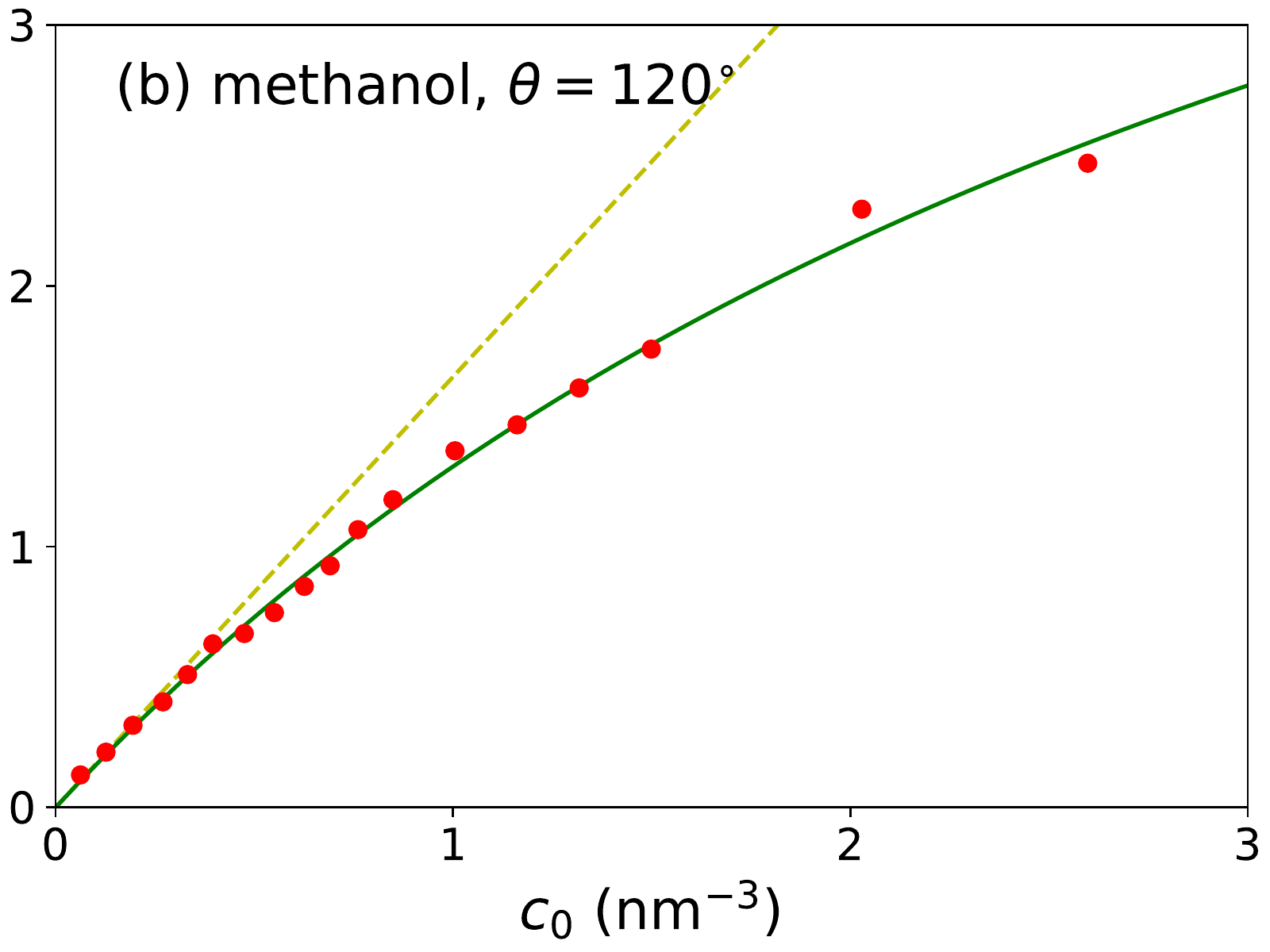}\\
\includegraphics[width=0.4\textwidth]{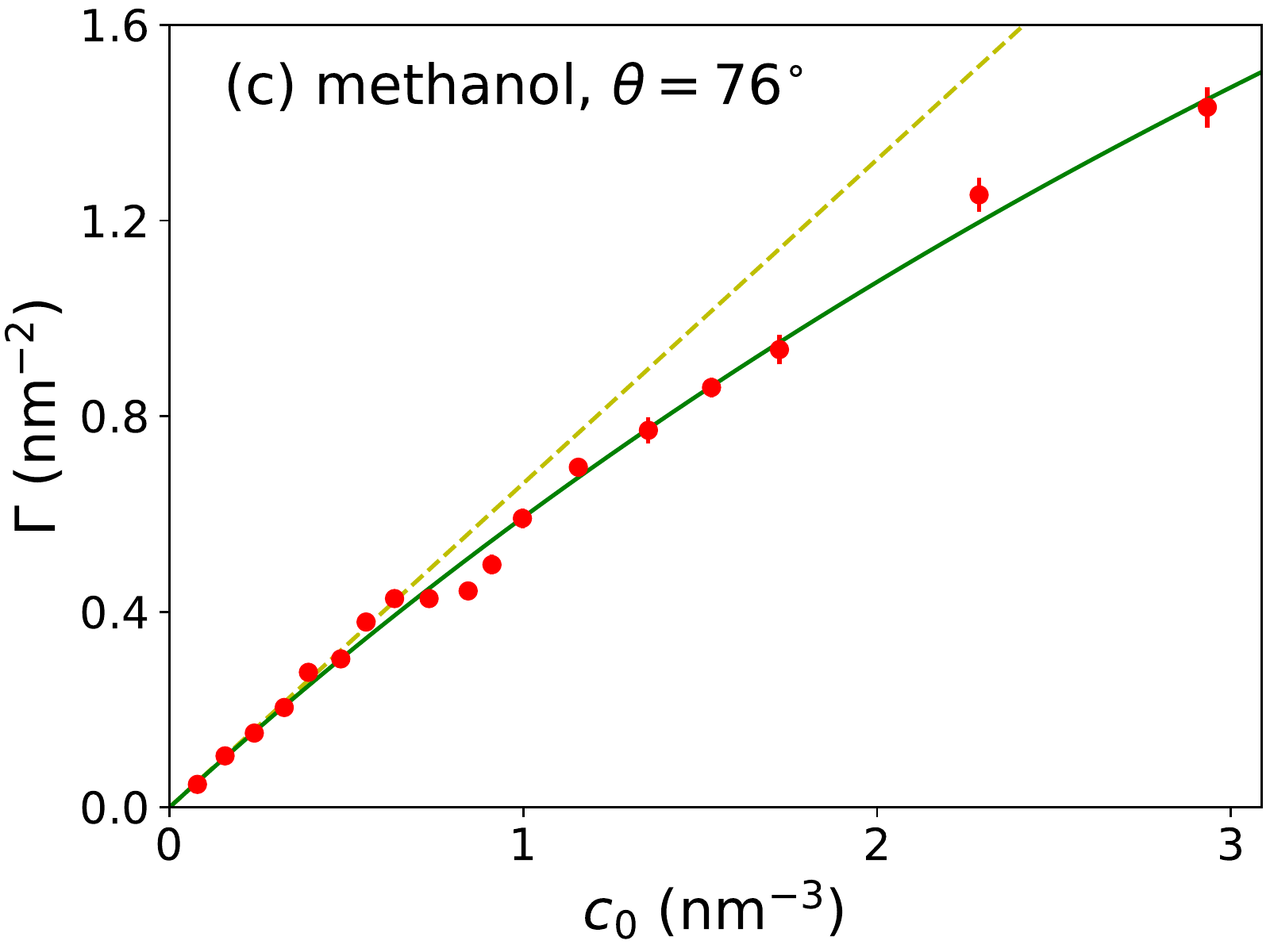}
\includegraphics[width=0.4\textwidth]{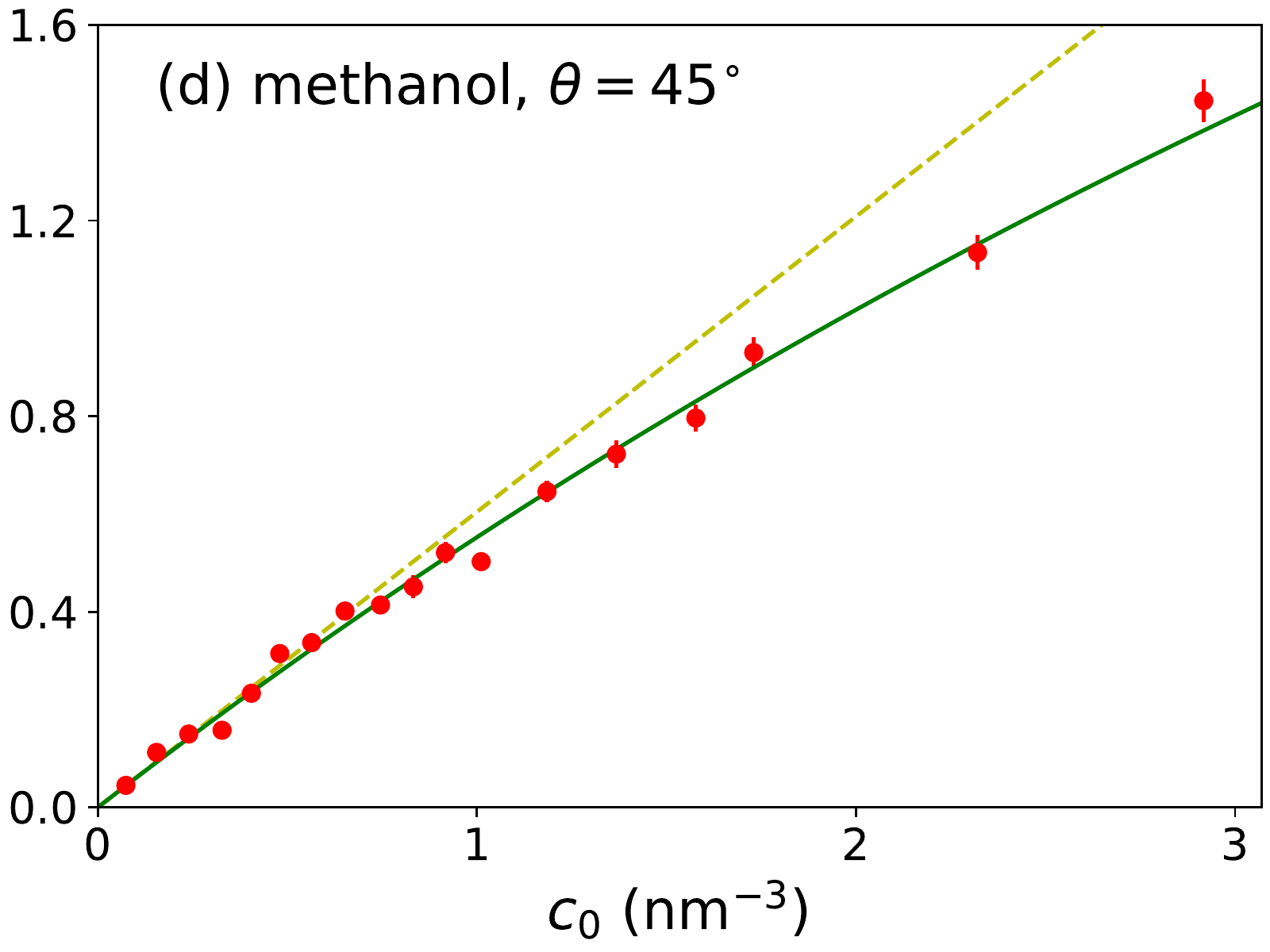}\\
	\caption{Adsorption $\Gamma$ at the solid--water interface as a function of surfactant concentration for various water contact angles. MD values are shown by red circles, whereas solid green lines show the fits of the Langmuir isotherm. Yellow dashed lines correspond to Henry's law (Eq.~3), for which the coefficient $K_\trm{v}$ is calculated from the Langmuir fit.\label{fig:gama_c0_surf_meth}
	}
\end{figure*}

\begin{figure*}
  \centering
  \includegraphics[width=0.4\textwidth]{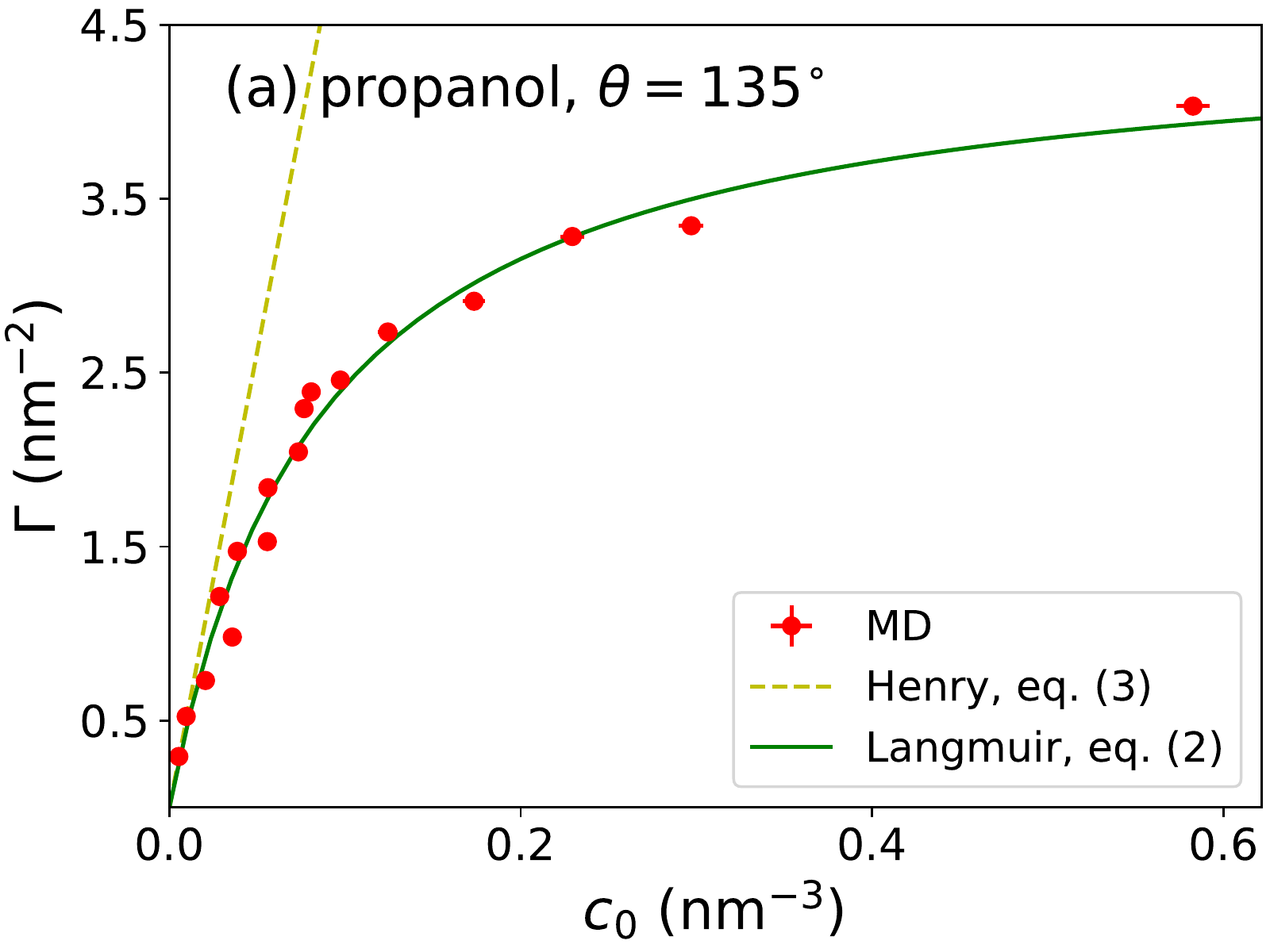}\hspace{3ex}
\includegraphics[width=0.39\textwidth]{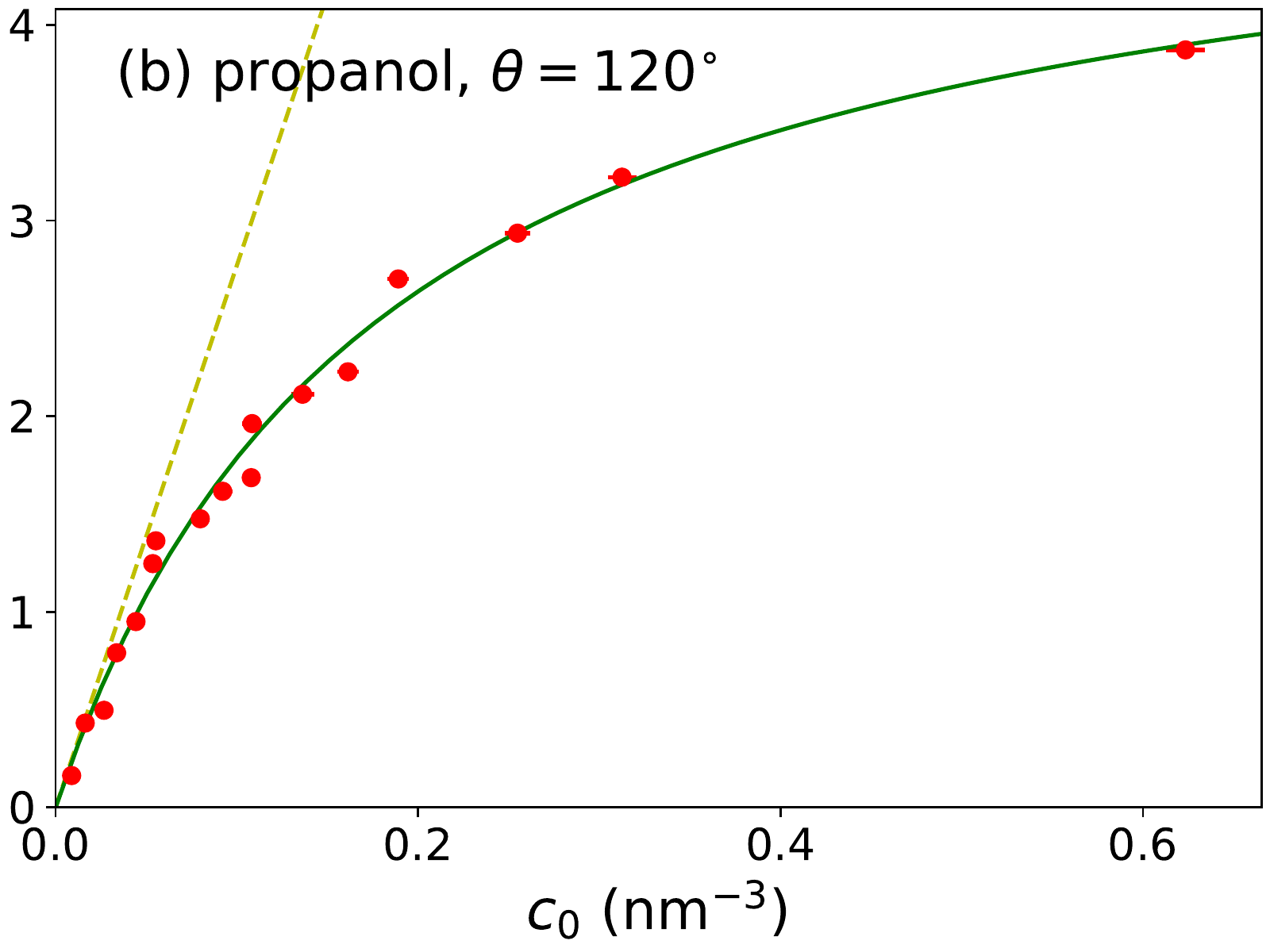}\\
\includegraphics[width=0.4\textwidth]{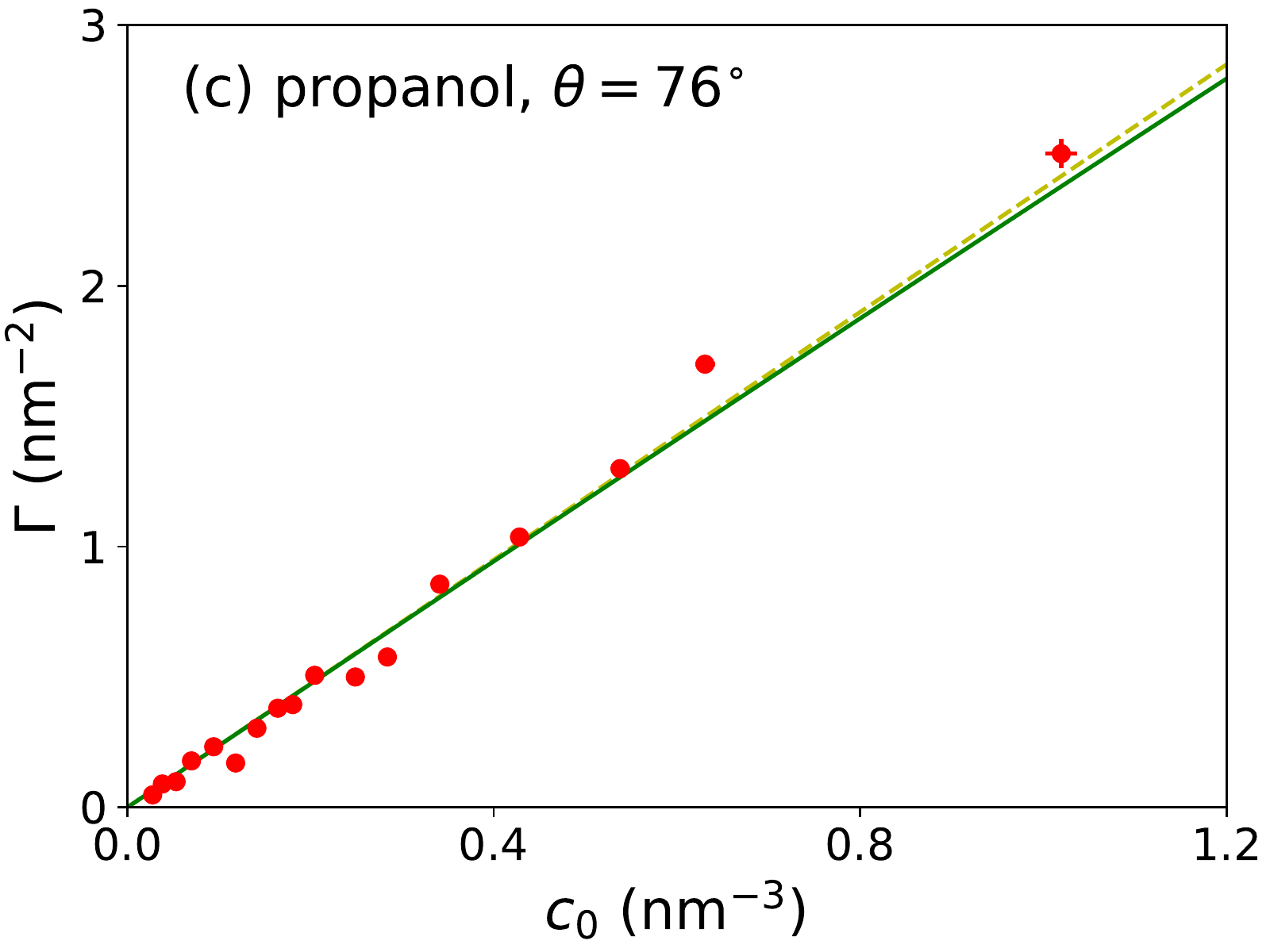}
\includegraphics[width=0.4\textwidth]{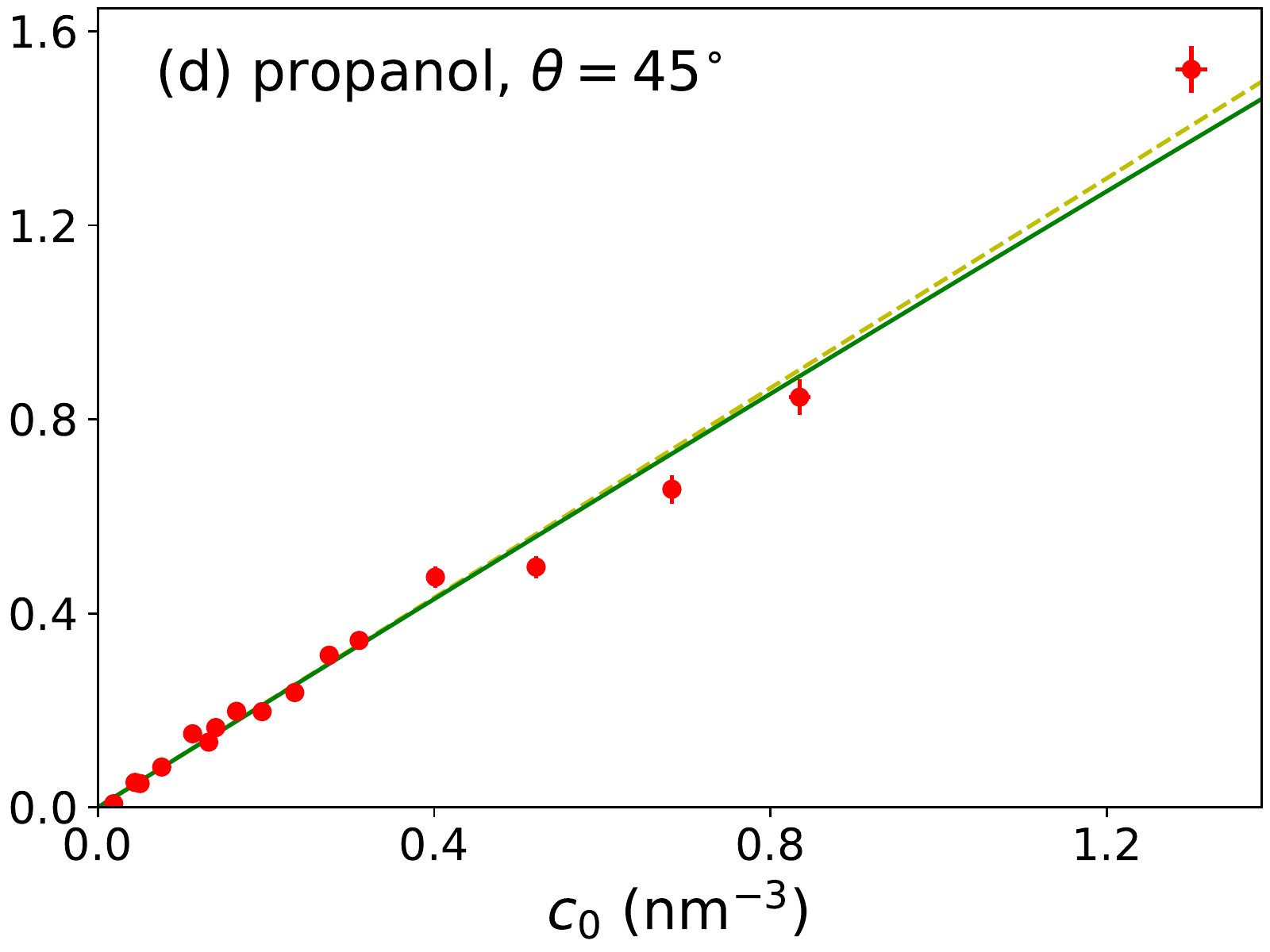}\\
	\caption{
	Same as Fig.~\ref{fig:gama_c0_surf_meth} but for propanol.
	\label{fig:gama_c0_surf_prop}}
\end{figure*}

\begin{figure*}
  \centering
  \includegraphics[width=0.4\textwidth]{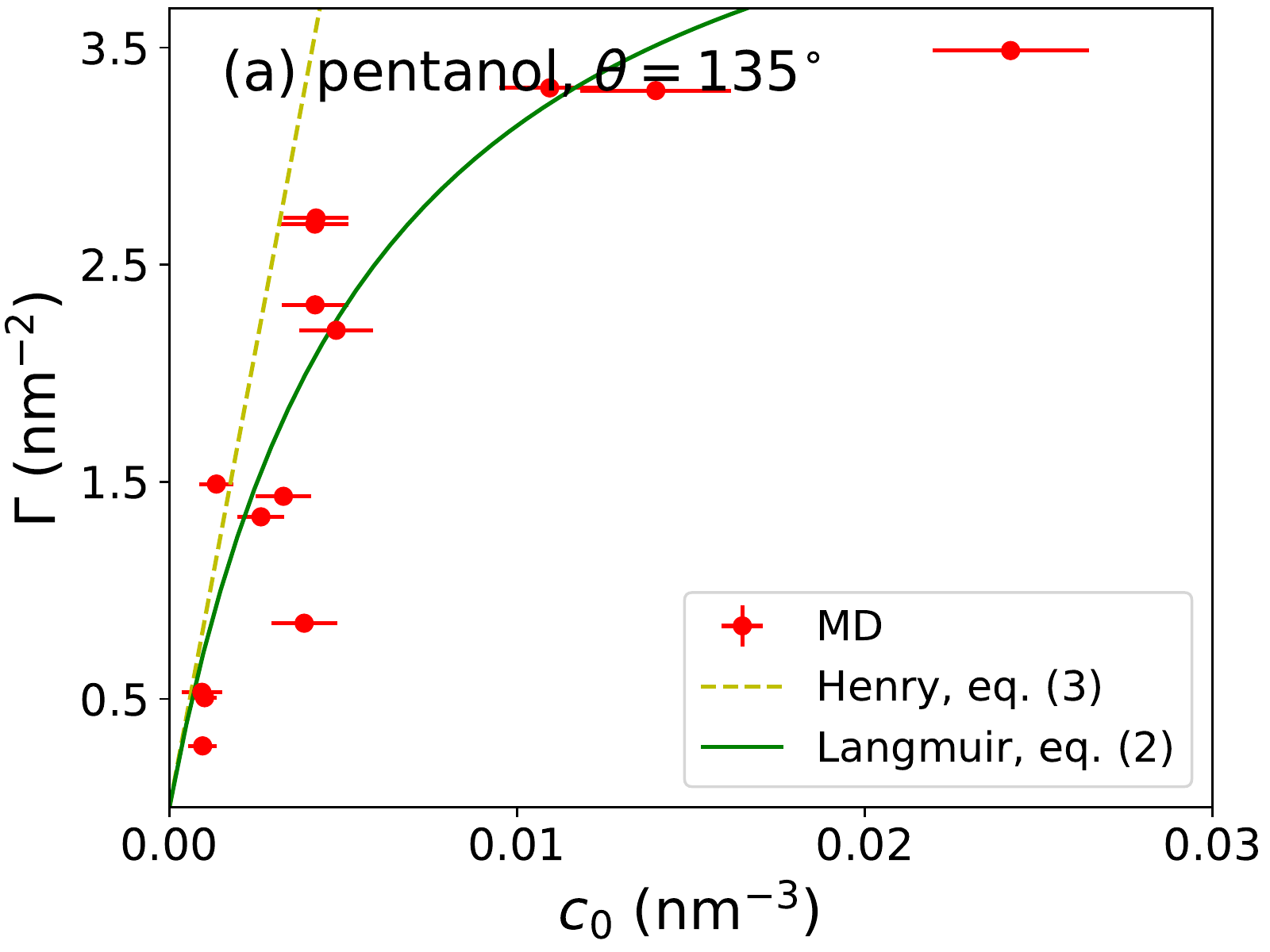}
\includegraphics[width=0.4\textwidth]{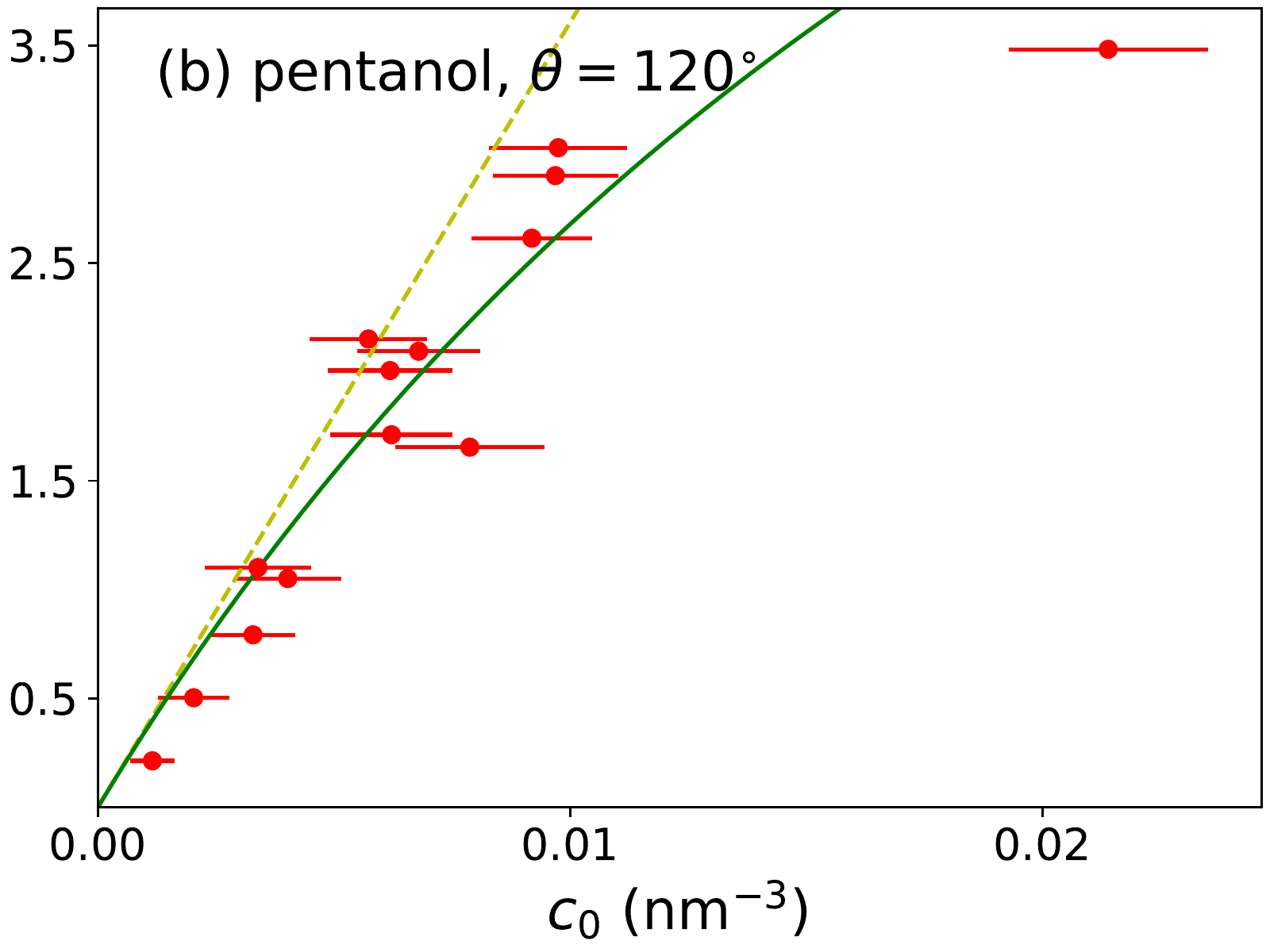}\\
\includegraphics[width=0.4\textwidth]{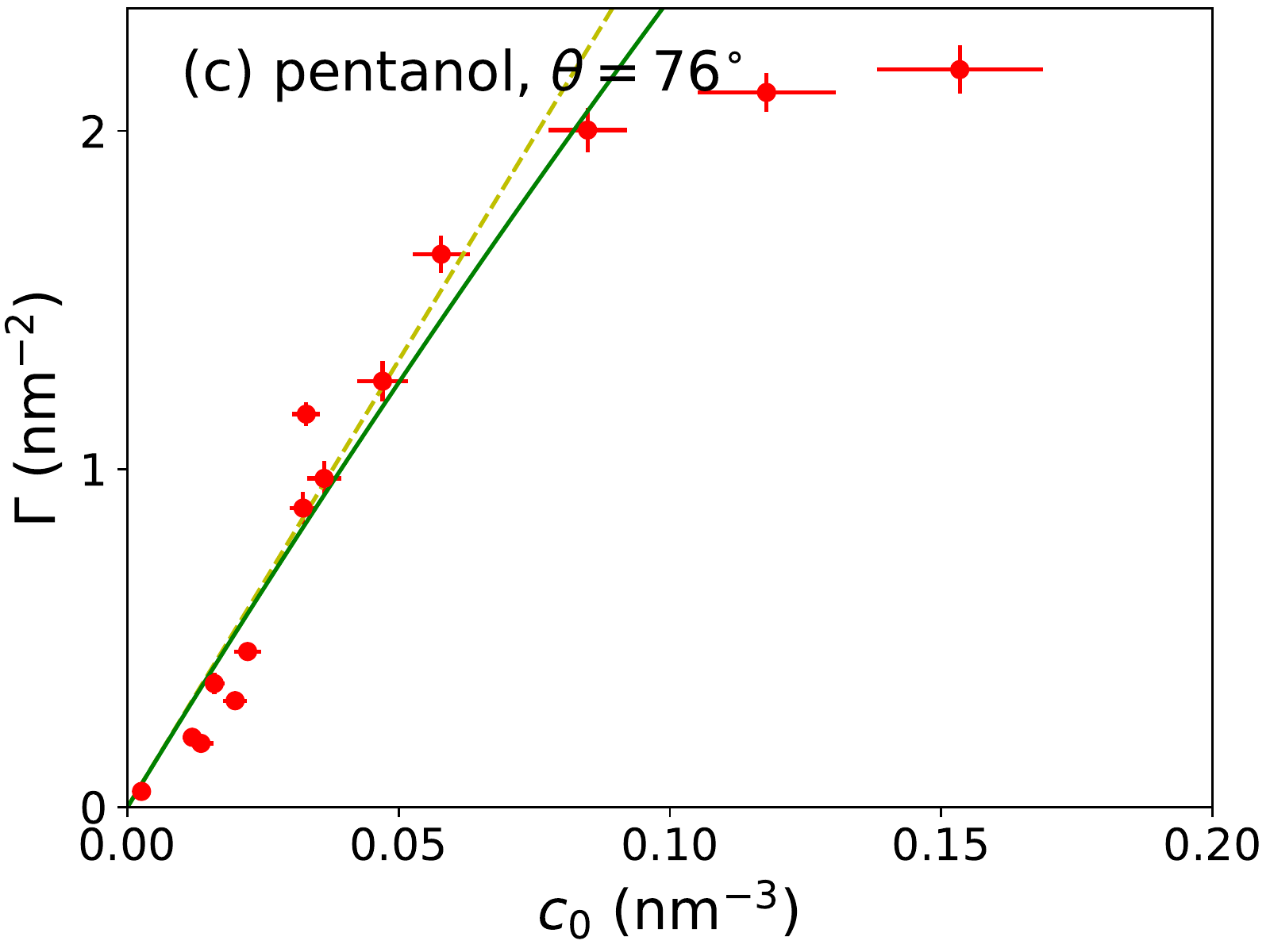}
\includegraphics[width=0.4\textwidth]{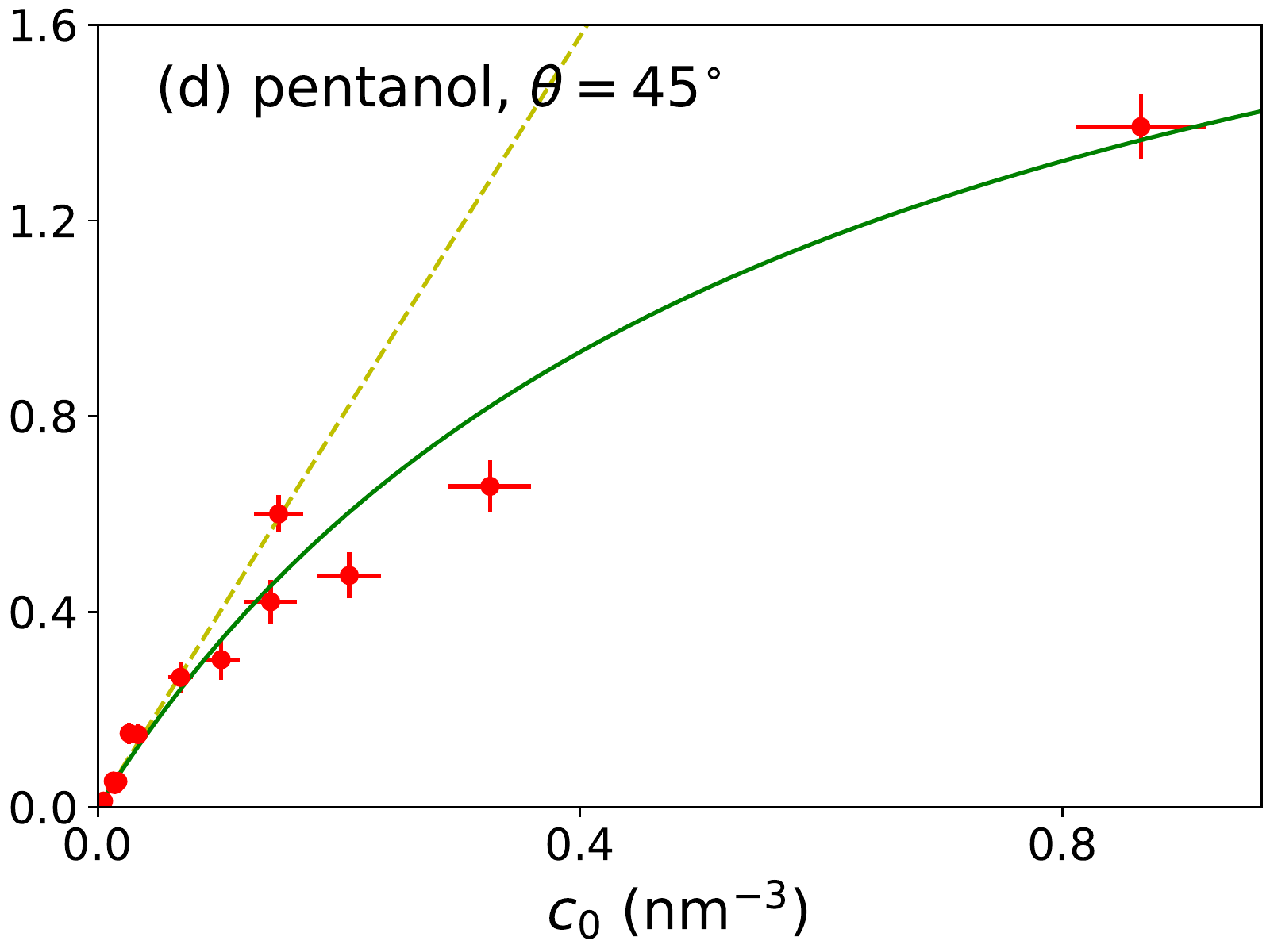}\\
	\caption{
	Same as Fig.~\ref{fig:gama_c0_surf_meth} but for pentanol.
	\label{fig:gama_c0_surf_pent}
	}
\end{figure*}

\begin{figure*}[t!]
  \centering
   \includegraphics[width=0.4\textwidth]{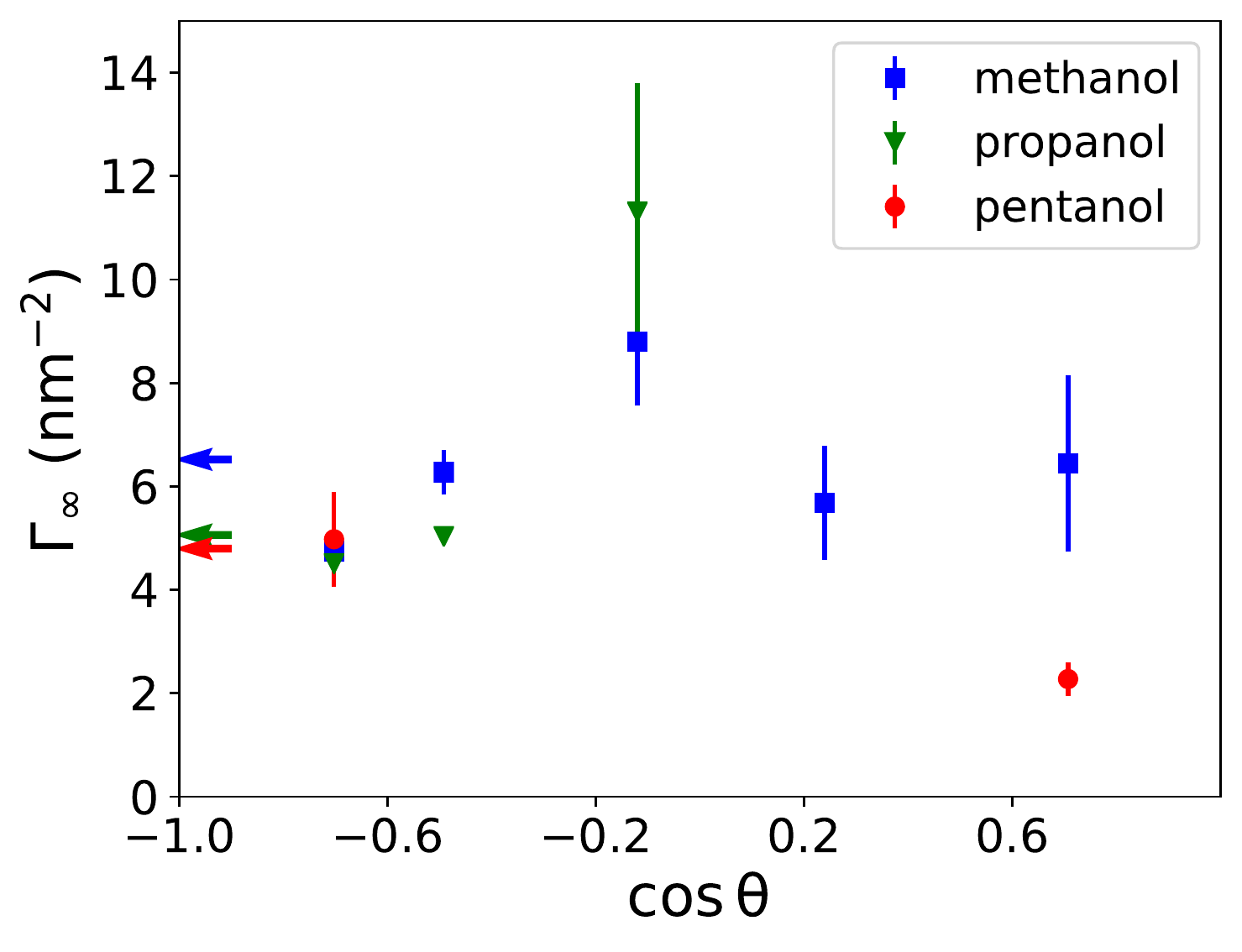}
\caption{
The fitted saturation value $\Gamma_\infty$ from the Langmuir isotherm {\em versus} the surface wetting coefficient for the three different alcohols. The arrows represent $\Gamma_\infty$  for the water--vapor interface.
	\label{fig:gamma_vs_costheta}}
\end{figure*}

\begin{figure*}[t!]
  \centering
   \includegraphics[width=0.4\textwidth]{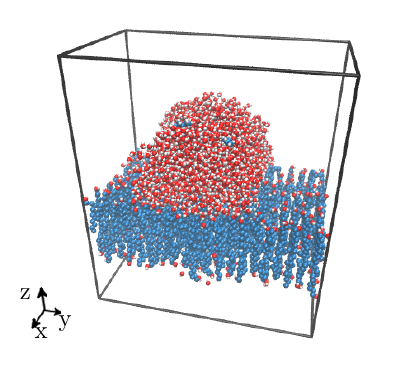}
\caption{Simulation snapshot of a cylindrical surfactant-containing water droplet on a planar solid surface, used to estimate the amount of the surfactant at the solid--vapor interface.
	\label{fig:wat_prop8_drop}}
\end{figure*}

\begin{figure*}[t!]
  \centering
	    \includegraphics[width=0.32\textwidth]{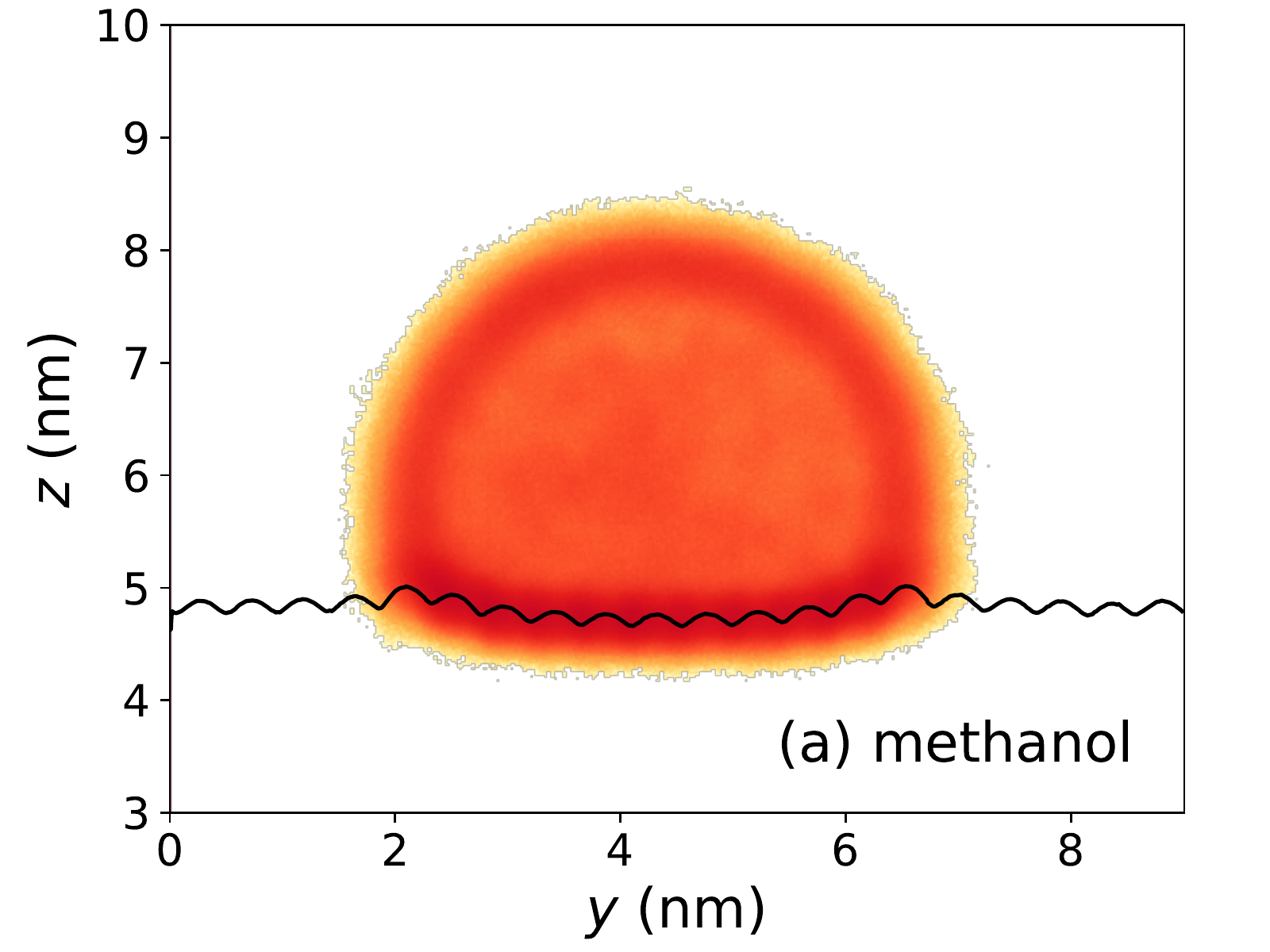}
	    \includegraphics[width=0.32\textwidth]{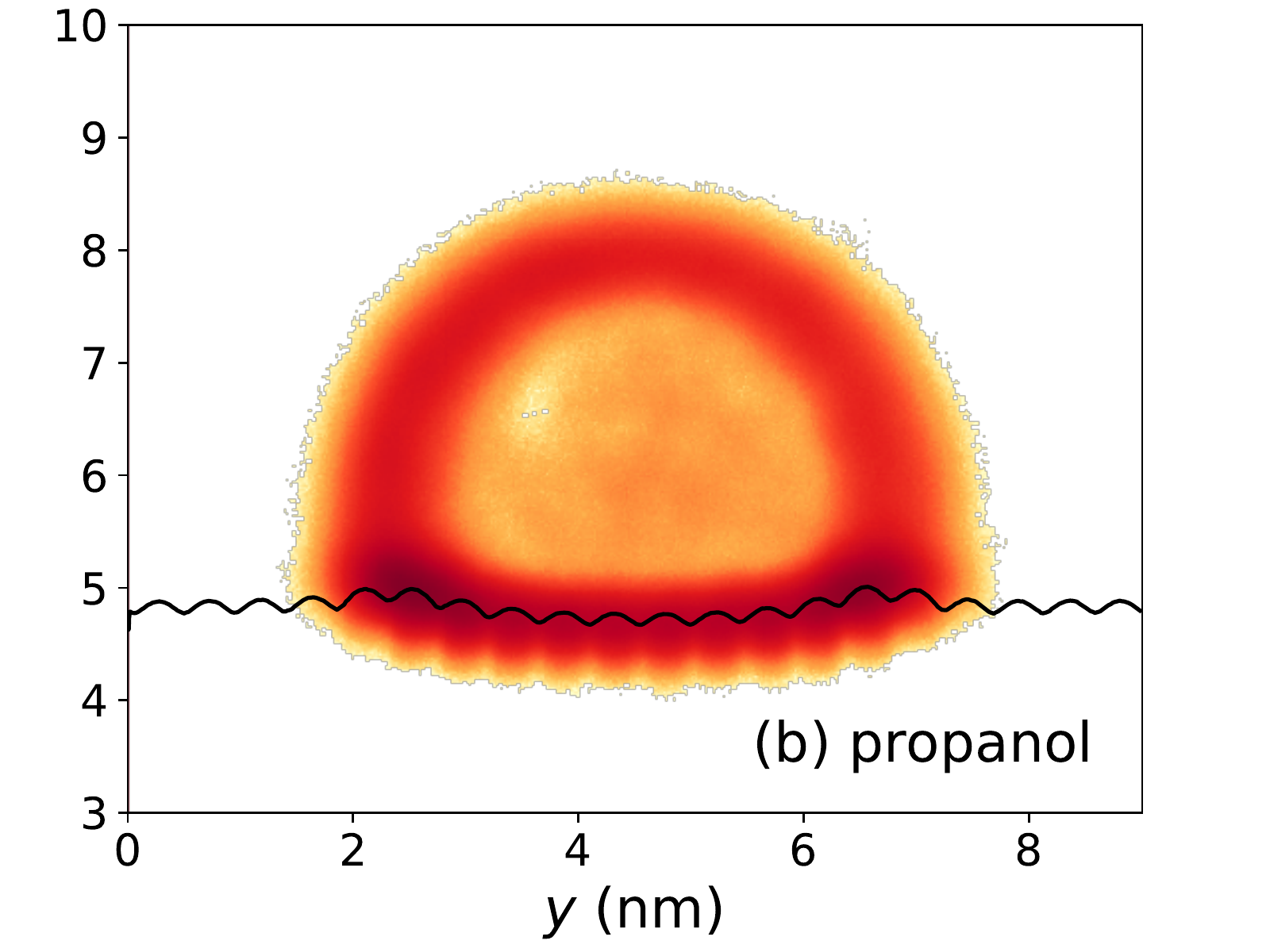}
	    \includegraphics[width=0.32\textwidth]{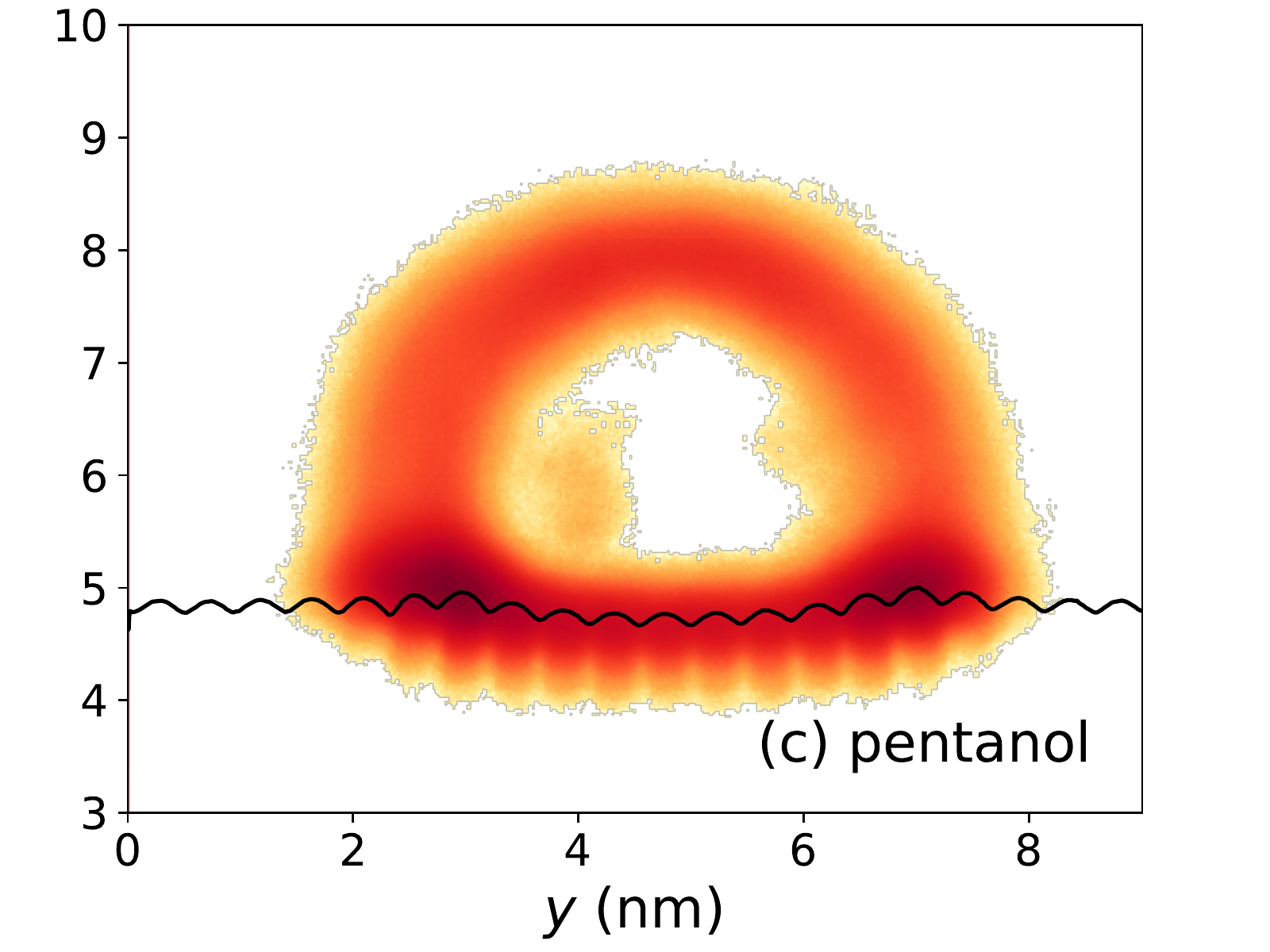}
\caption{
Side-view surfactant density plots (in logarithmic scale) of cylindrical water droplets for (a) methanol, (b) propanol, and (c) pentanol. The color scale runs from white (low densities) to red (large densities). In all three cases, the solid surfaces has a contact angle of $\theta = 120^{\circ}$. The black lines are the dividing surfaces of the OH groups of the surface.}
	\label{fig:droplet}
\end{figure*}

\section{Estimating surfactant adsorption at the solid--vapor interface}

To check that there is no adsorption at the solid--vapor interface, we run simulations of a cylindrical surfactant-containing water droplet on the solid surface. The surface has the same features as in the main simulations, except that the size along the $y$ direction is twice as large. The droplet is periodically replicated along the $x$ direction, as shown in \Fig{fig:wat_prop8_drop}. 

Averaging the densities over time and the $x$ direction, we obtain the $yz$-resolved surfactant density profile, which is shown in Fig.~\ref{fig:droplet}. From the obtained density plots of all three surfactants, it is possible to see that the density at the solid--vapor interface remains zero, which justifies our assumption.